\documentclass[final]{svjour2}
\usepackage{graphicx}
\usepackage{amssymb}
\usepackage{mathptmx}
\usepackage[numbers]{natbib}
\usepackage{dcolumn}
\usepackage{color}
\usepackage{bm}
\makeatletter
\begin{document}

%%%%%%%%%%%%%%%%%%%%%%%%%%%%%%%%%%%%%% AUTHORS %%%%%%%%%%%%%%%%%%%%%%%%%
\author{E.V. Kozik \and B.V. Svistunov}
\institute{E.V. Kozik \at Theoretische Physik, ETH Z\"urich, 8093 Z\"urich, Switzerland
\and
B.V. Svistunov \at Department of Physics, University of Massachusetts, Amherst, MA 01003, USA \\
Russian Research Center ``Kurchatov Institute'', 123182 Moscow, Russia}

%%%%%%%%%%%%%%%%%%%%%%%%%%%%%%%%%%%%%%%%%%%%%%%%%%%%%%%%%%%%%%%%%%%%%%%%%%%%%%

\title{Theory of decay of superfluid turbulence in the low-temperature limit}

%%%%%%%%%%%%%%%%%%%%%%%%%%%%%%%%%%%%%%%%%%%%%%%%%%%%%%%%%%%%%%%%%%%%%%%%%%%%%%
\date{XX.XX.2009}
\maketitle
\begin{abstract}
We review the theory of relaxational kinetics of superfluid turbulence---a tangle of quantized vortex lines---in the limit of very low temperatures when
the motion of vortices is conservative. While certain important aspects of the decay kinetics  depend on whether the tangle is non-structured, like the one corresponding to the Kibble-Zurek picture, or essentially polarized, like the one that emulates  the Richardson-Kolmogorov regime of classical turbulence, there are common fundamental features. In both cases, there exists an asymptotic range in the wavenumber space
where  the energy flux is supported by the cascade of Kelvin waves (kelvons)---precessing distortions propagating along the vortex filaments.

At large enough wavenumbers, the Kelvin-wave cascade is supported by three-kelvon elastic scattering. At zero temperature, the dissipative cutoff of the Kelvin-wave cascade is due to the emission of phonons, in which an elementary process converts two kelvons with almost opposite momenta into one bulk phonon.

Along with the standard set of conservation laws, a crucial role in the theory of low-temperature vortex dynamics is played by the fact of integrability of the local induction approximation (LIA) controlled by the parameter $\Lambda = \ln (\lambda/a_0)$, with $\lambda$ the characteristic kelvon wavelength  and $a_0$ the vortex core radius. While excluding a straightforward onset of the pure three-kelvon cascade, the integrability of LIA does not plug the cascade because of the natural availability of the kinetic channels associated with vortex line reconnections.

We argue that the crossover from  Richardson-Kolmogorov to the
Kelvin-wave cascade is due to eventual dominance of local induction of a single line over the collective induction of polarized eddies, which causes the breakdown of classical-fluid regime and gives rise to a reconnection-driven inertial range.

\end{abstract}

\maketitle

%%%%%%%%%%%%%%%%%%%%%%%%%%%%%%%%%
\section{Introduction}
\label{sec:intro}

Superfluid turbulence (ST), also known as quantum/quantized turbulence, is a tangle of quantized vortex lines in a superfluid \cite{Donnelly,Cambridge_workshop, Vinen_Niemela, Vinen06}.
In the last one and a half decade, and especially in the recent years, the problem of {\it zero-point} ST  has evolved into a really hot subfield of low-temperature physics \cite{Sv95, Nore, Davis, Vinen2000, Tsubota00, Vinen2001, Kivotides, Vinen_2003, KS_04, KS_05, KS_05_vortex_phonon, Bradley, sim_boll, Ladik, Golov, Lvov, KS_crossover, Lvov2, KS_scan, Golov2, Barenghi, vibr_review}. In this work, we summarize theoretical developments of the two authors on the theory of  decay of ST at $T=0$, previously published in a number of short papers/letters \cite{KS_04, KS_05, KS_05_vortex_phonon, KS_crossover, KS_scan}. To render the discussion self-contained, we also review the analysis of Ref.~\cite{Sv95}, where a number of conceptually important facts about zero-point ST has been revealed.

 Superfluid turbulence (ST) can be created in a number of ways: (i) in the counter-flow of normal and superfluid components \cite{Vinen, Schwartz}  (ii) by vibrating objects \cite{Davis, Bradley, sim_boll, vibr_review}, (iii) as a result of macroscopic motion of a superfluid (referred to as \textit{quasi-classical} turbulence), in which case ST can mimic, at large enough
length scales, classical-fluid turbulence \cite{Bradley, Maurer,Nore,Stalp, Oregon, Ladik, Golov, Golov2}, (iv) in the process of (strongly) non-equilibrium   Bose-Einstein condensation \cite{Berloff,BEC_expt}, in which case it is a manifestation of  generic Kibble-Zurek effect \cite{Kibble-Zurek}.

During about  four decades---from mid fifties till mid nineties---ST turbulence was intensively studied in the context of
counter-flow of normal and superfluid components. An impressive success  has been achieved in the field, with the prominent contributions by Vinen---the equation for qualitative description of growth/
decay kinetics (Vinen's equation), extensive experimental studies \cite{Vinen},---and pioneering microscopic simulations of vortex line dynamics by Schwartz \cite{Schwartz}. The counter-flow setup naturally implies  finite density of the normal component, and thus a relatively simple---by comparison with the $T=0$ case---relaxation mechanism.  The normal component exerts a drag force on a vortex filament.
As a result, the vortex line length decreases. For an illustration, consider a vortex  circle of the radius $R$. In the absence of normal component, the ring moves with a constant velocity $V\equiv V(R)$,  the radius $R$ remaining constant.  With the drag force, the ring collapses, obeying a simple law $\dot{R} \propto - \alpha V(R)$, where $\alpha$ is the dimensionless friction coefficient  measuring the drag force  in the units of Magnus  force. Similarly, the drag causes the decay of Kelvin waves---precessing distortions on the vortex filaments---and the value of $\alpha^{-1}$ gives the number of revolutions a distortion makes before its amplitude  significantly decreases. At $\alpha \sim 1$, the decay scenario of a (non-structured) vortex tangle is as follows. The vortex lines reconnect producing the distortions (Kelvin waves), these distortions decay due to the drag force thereby reducing the total line length and rendering the tangle more and more dilute.

\begin{figure}
\includegraphics[width = 1.0\columnwidth,keepaspectratio=true]{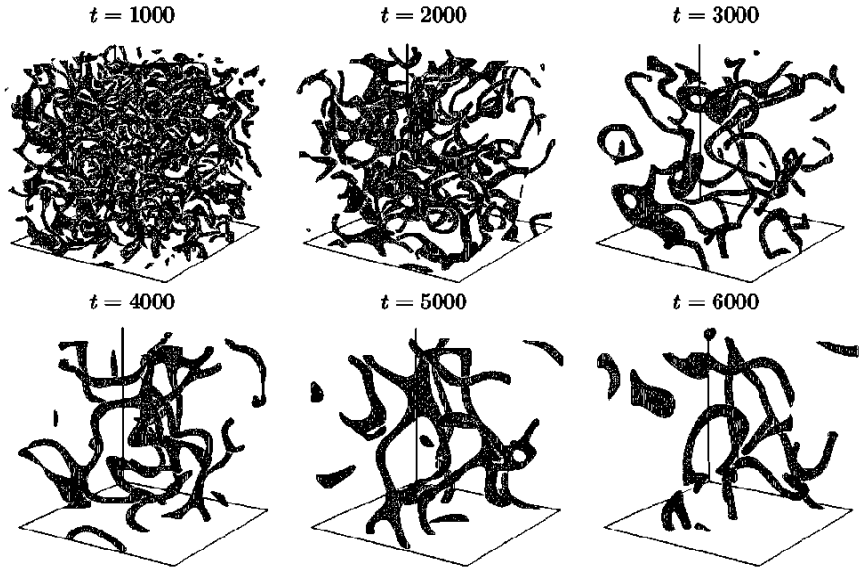}
\caption{Formation of superfluid turbulence in the process of strongly non-equilibrium kinetics of Bose-Einstein condensation governed by Gross-Pitaevskii equation, as simulated in Ref.~\cite{Berloff}.}
\label{fig:BEC_kin}
\end{figure}

In contrast to the counter-flow setup, where the case $\alpha \ll 1$ can be considered rather specific,  for the non-equilibrium process of Bose-Einstein condensation in a weakly interacting gas, the regime $\alpha \ll 1$ is quite characteristic in view of the following circumstance. The condensation kinetics is a classical-field process the essential part of which is quantitatively captured by the time dependent Gross-Pitaevskii equation (a.k.a. non-linear Schr\"odinger equation). Hence, the most natural setup is when the initial condition for the process features large occupation numbers for the bosons, and the whole process can be accurately described by the Gross-Pitaevskii equation from the very outset (see Fig.~\ref{fig:BEC_kin} for an illustration). In the idealized situation of the classical field, the final temperature asymptotically approaches zero in view of the ultraviolet catastrophe.
%%%%%%%%%%%%%% FIGURE simulation of BEC kinetics %%%%%%%%%%%%%
Another situation where the regime $\alpha \ll 1$ occurs naturally is the quasi-classical turbulence, which, unlike counterflow turbulence, is due to a macroscopic turbulent motion and is generated by purely classical means at arbitrary temperatures without the help of mutual friction.

At $\alpha \ll 1$ ($\alpha \propto T^5$ at $T \to 0$, see Ref.~\cite{Iordanskii}), dynamics of vortex lines in a significant range of length scales become essentially conservative. In this case, relaxation of turbulent motion in non-linear systems usually involves certain types of {\it cascades}. In both classical and superfluid turbulence the key role is played by the cascade of energy in the wavenumber space towards high wavenumbers. A realization of this peculiar relaxation regime is due to the following conditions satisfied by the system: (i) $\lambda_\mathrm{en} \gg \lambda_\mathrm{cutoff}$---a substantial separation of the energy-containing scale $\lambda_\mathrm{en}$ from the scale $\lambda_\mathrm{cutoff}$ where the dissipation becomes appreciable, which defines the cascade \textit{inertial range}
$\lambda_\mathrm{en} > \lambda > \lambda_\mathrm{cutoff}$, (ii) the kinetics are local in the wavenumber space, i.e. the energy exchange is mainly between adjacent length scales, and (iii) the
``collisional'' kinetic time $\tau_{\rm coll} (\lambda)$---the
time between elementary events of energy exchange at a certain
scale $\lambda$---gets progressively shorter down the scales. These requirements determine the main qualitative features of the cascade: The decay is governed by the slowest kinetics at
$\lambda_\mathrm{en}$, where the energy flux $\varepsilon$ in the wavenumber space is formed, while the faster kinetic processes at
shorter scales are able to instantly adjust to this flux
supporting the transfer of energy towards $\lambda_\mathrm{cutoff}$, where it is dissipated into heat. Thus, the cascade is a (quasi-)steady-state regime in which the energy flux $\varepsilon$
is constant through the length scales and the variation of
$\varepsilon$ in time happens on the longest time scale $\sim \tau_{\rm coll} (\lambda_\mathrm{en})$.

With vortex lines, one can think of quite a number of different cascades. Perhaps the most obvious one is the Richardson-Kolmogorov cascade of eddies, which can be realized in a superfluid (even at $T=0$) due to its ability to emulate the classical-fluid turbulent motion by the motion of a polarized vortex-line tangle. Clearly, this type of the cascade is fundamentally impossible in the non-structured Kibble-Zurek-type tangle. For quite a long time it was believed that non-structured ST decays at $T=0$ through the Feynman's cascade of vortex rings \cite{Feynman}. Feynman conjectured that  reconnections of the vortex lines produce vortex rings, with subsequent decay of each ring into a pair of smaller rings, and so forth. It can be shown, however, that this conjecture is inconsistent with simultaneous conservation of energy and momentum (see Ref.~\cite{Sv95} and Sec.~\ref{sec:self-rec}). Another obvious option is a cascade of Kelvin waves
supported by non-linear interaction (scattering) of kelvons. While such a cascade is indeed possible (see Ref.~\cite{KS_04} and Sec.~\ref{sec:pure}), it is subject to a peculiar constraint on the maximal energy flux. The origin of this constraint is the approximate integrability of the vortex dynamics controlled by the large parameter
\begin{equation}
\Lambda = \ln (\lambda/a_0) ,
\label{Lambda}
\end{equation}
where $\lambda$ is the characteristic kelvon wavelength  and $a_0$ is the vortex core radius. If the parameter $\Lambda$ is large, then the leading term of the Kelvin-wave dynamics is given by the so-called local induction approximation (LIA) in which the velocity of an element of the vortex line is due to local differential properties of the line. The LIA dynamics turns out to be integrable, so that  mere non-linearity of the LIA equations of motion does not lead to  the Kelvin-wave cascade. This circumstance predetermines the existence of intermediate cascades that are necessary to transfer the energy  to the wavenumbers high enough for the pure Kelvin-wave cascade to take over. All the intermediate cascades have to involve reconnections, to lift the constraint imposed by the integrability of LIA. In the non-structured tangle, there is only one type of reconnection-induced cascade.  An elementary step
of this cascade at the wavelength $\lambda$ consists of two adjacent events: (i) emission of a  vortex ring with the radius $\sim \lambda$ by a local self-crossing of a kinky vortex line and (ii) re-absorbtion of the ring by the tangle. Both  events  are accompanied by  transferring  Kelvin-wave energy  to shorter wavelengths (but still on the order of $\lambda$), since  reconnections of the vortex loops produce Kelvin-wave structure with wavelength smaller than the loop radii. For the self-reconnection cascade to be efficient, the vortex line has to be kinky (loosely speaking, fractalized) at all relevant length scales. Due to the integrability of LIA, the  fractalization is provided by the cascade itself: In the absence of self-crossings at a given length scale, the amplitude of Kelvin waves keeps
growing due to self-crossings at larger length scales.  It is worth emphasizing that in contrast to Feynman's scenario, where
the vortex rings  are the energy carriers,  in the local self-crossing scenario, the rings play only a supportive role. In the process (i) a ring is  just a product of self-crossing and in the process (ii) the ring is just a cause of yet another reconnection event. Clearly, the crucial part is played by  Kelvin waves, which carry the energy, and reconnections, which directly promote the Kelvin-wave cascade.

In the polarized tangle emulating the Richardson-Kolmogorov classical-fluid cascade, there are two more varieties  of reconnection-supported Kelvin-wave cascades. One is when reconnections are between the bundles of quasi-parallel vortex lines, and the other one when the reconnections are between two neighboring lines in a bundle. In Sec.~\ref{sec:crossover}
we present the arguments \cite{KS_crossover} that in the theoretical limit of $\Lambda\to \infty$, all the three reconnection driven cascades are necessary to cross over from Richardson-Kolmogorov to pure Kelvin-wave regime, the crossover being a series of three distinct cascades: bundle driven, neighboring reconnection driven, and local self-crossing driven. The total extent of the crossover regime in the Kelvin wavenumber space is predicted to be  $\sim \Lambda/10$ decades. With realistic $\Lambda\lesssim 15$ (for $^4$He), this means that the crossover takes about one decade, within which one can hardly expect sharp distinctions between the three different regimes.

At $T=0$, the reconnection-supported cascade(s) ultimately crosses over to the pure Kelvin-wave cascade, and the latter is cut off
by  phonon emission \cite{Vinen2000, Vinen2001, KS_05_vortex_phonon}. The elementary process of phonon emission converts two kelvons with almost opposite momenta into one bulk phonon \cite{KS_05_vortex_phonon}.

The rest of the paper is organized as follows. In Sec.~\ref{sec:basic} we render basic equations and notions of vortex dynamics:
Biot-Savart equation and the LIA, Hasimoto representation for the LIA in terms of curvature and torsion, conservation of energy and momentum, Hamiltonian formalism for the
Kelvin waves, conservation of angular momentum, implying conservation of the total number of kelvons. We pay special attention to the
problem of {\it ultraviolet regularization} of the theory, crucial for an adequate treatment of non-local corrections to the LIA.
In Sec.~\ref{sec:pure} we develop a theory of the pure Kelvin-wave cascade (supported
by elastic scattering of three kelvons---the leading non-trivial scattering event consistent with conservation of energy, momentum,
and the total number of kelvons). We show that the leading term in the three-kelvon scattering amplitude does not contain local contributions, in accordance with the integrability of LIA. We then find the spectrum and energy flux of the pure cascade.
In Sec.~\ref{sec:self-rec} we analyze the scenario of local self-crossings, starting with the explanation why pure Feynman's cascade cannot be realized in view of conservation of momentum and energy. We show that this scenario implies fratalization of the vortex lines, and derive corresponding spectrum of Kelvin waves. We then discuss a simple Hamiltonian model \cite{Sv95} featuring a collapse-driven cascade in otherwise integrable system, and argue why the spectrum of the cascade in this model is similar to the one in the scenario of local self-crossings. In Sec.~\ref{sec:crossover} we address the problem of the crossover from Richardson-Kolmogorov to
Kelvin-wave cascade.  Sec.~\ref{sec:kelvon_phonon} is devoted to the theory of kelvon-phonon interaction. Starting from the hydrodynamical Lagrangian,
 we derive the Hamiltonian of kelvon-phonon interaction. The Hamiltonian allows us to straightforwardly formulate kinetics of the kelvon-phonon processes, and, in particular, find the cutoff momentum for the pure Kelvin-wave cascade.
In Sec.~\ref{sec:conclusions} we summarize the main qualitative aspects of the theory and put our findings in the context of experiment and simulations, including the ones that are still missing. We also formulate certain theoretical questions to be addressed in the future.  Finally,  we critically discuss the concept of bottleneck between Richardson-Kolmogorov and Kelvin-wave cascades, put forward recently in Ref.~\cite{Lvov} (see also Ref.~\cite{Lvov2}), which we disagree with on the basis
of our theoretical analysis.

%%%%%%%%%%%%%%%%%%%%%%%%%%%%%%%
\section{Basic relations}
\label{sec:basic}

\subsection{Biot-Savart law. Energy and momentum}

The zero-temperature hydrodynamics of a superfluid is nothing but the hydrodynamics of a classical ideal fluid with
quantized vorticity---the only possible rotational motion is in the form of vortex filaments with a fixed value of velocity circulation $\kappa$.  Therefore, one can apply the Kelvin-Helmholtz  theorem of classical hydrodynamics---stating that the vortices move with the local fluid velicity---to obtain a closed dynamic equation for vortex filaments in a superfluid.  Alternatively, one can start with the
hydrodynamic action for the complex-valued field and derive the  equation of vortex motion from the least-action principle. In the present section, we use the former approach,  readily yielding the answer.  [The latter approach will  be used in Sec.~\ref{subsec:hydro}, where it will prove crucial for describing the vortex-phonon interaction.]

If the vortex
lines are the only degrees of freedom excited in the fluid, and if the
typical curvature radius and interline separations are much larger than the
vortex core size, $a_0$, then the instant velocity field at distances much
larger than $a_0$ from the vortex lines is defined by the form of the vortex line
configuration. Indeed, away from the vortex core the density is practically constant and
the hydrodynamic continuity equation reduces to
\begin{equation}
\nabla \cdot {\bf v} = 0\; .
\label{div_0}
\end{equation}
Then, taking into account that $\nabla \times {\bf v} = 0$ everywhere except for the
vortex lines, and that for any (positively oriented) contour $\Gamma$ surrounding
one vortex line
\begin{equation}
\oint_{\Gamma} {\bf v} \cdot d {\bf l} = \kappa ,
\label{kappa}
\end{equation}
we note a direct analogy of our problem
with the magnetostatic problem of finding magnetic field produced by thin wires: Velocity field is identified with the magnetic field, and the absolute value
of the current of each wire is one and the same and is proportional to $\kappa$.
The result is given by the Biot-Savart formula
\begin{equation}
{\bf v} ({\bf r}) = { \kappa \over 4 \pi}
\int \frac{ d {\bf s} \times ({\bf r} - {\bf s})}
{|{\bf r} - {\bf s}|^3}  \; ,
\label{BS}
\end{equation}
where radius-vector ${\bf s}$ runs along all the vortex filaments.

According to Kelvin-Helmholtz  theorem, each piece of the vortex line should move
with a velocity corresponding to the velocity of the {\it net} motion of small
contour (of the size, say, of order $a_0$) surrounding this element. This fact is
very important, since the velocity field (\ref{BS}) is singular
at all points on the vortex line, and the Kelvin-Helmholtz theorem yields a simple
regularization prescription: Take a small circular contour with the center of the
circle at some vortex line point and the plane of the circle perpendicular to the
vortex line (at the point of intersection), and average the velocity field around
the contour---to eliminate rotational component of the contour motion.

As is clear from (\ref{BS}),
the singularity
of the velocity field at some point ${\bf s}={\bf s}_0$ in the vortex line comes from the integration over the close vicinity of the point ${\bf s}_0$. To isolate the singularity, we  expand the function
${\bf s}(\xi)$  around the point ${\bf s}_0={\bf s}(\xi_0)$:
\begin{equation}
{\bf s}={\bf s}_0 + {\bf s}'_{\xi} \, \xi +
{\bf s}''_{\xi \xi} \, \xi^2/2 + \ldots\;  .
\label{arc}
\end{equation}
Here  $\xi$ is the parameter of the line. It is convenient to use the
natural parameterization, that is to choose $\xi$ to be  the (algebraic) arc length measured from the point ${\bf s}_0$.
Substituting this expansion into (\ref{BS}),
we get
\begin{equation}
{\bf v}({\bf r} \to {\bf s}_0) \; = \;  {\kappa \over 8 \pi} \; {\bf s}'_{\xi} \times
{\bf s}''_{\xi \xi}   \, \int^{\xi_*}_{-\xi_*} {d \xi \over |\xi|} +
\mbox{regular~part}  ,
\label{BS1}
\end{equation}
where $\xi_*$ is some upper cutoff parameter on the order of the curvature radius at
the point ${\bf s}_0$. The integral in the right-hand side is divergent at
$\xi \to 0$. To regularize it we note (i) that, by its very origin,
the expression (\ref{BS}) is meaningful only at $|\xi| > a_0$, and (ii)
the part of the vortex line with $|\xi| \ll a_0$ does not give a
significant contribution to the net velocity on the contour of the
radius $\sim a_0$.
Hence, with a logarithmic accuracy we can adopt the regularization
$|\xi| > a_0$, that is
\begin{equation}
\int^{\xi_*}_{-\xi_*} {d \xi \over |\xi|} \; \to \; 2 \ln(\xi_* /a_0) \; .
\label{reg}
\end{equation}
Moreover, by fine-tuning the value of $a_0$ in accordance with a model-specific behavior at the distances $\sim a_0$,
the logarithmic accuracy of the regularization (\ref{reg}) can be improved to the accuracy $\sim a_0/\xi_*$.
We discuss this option in detail in Sec.~\ref{subsec:ham}, and utilize in Sec.~\ref {sec:pure}.

We thus arrive at the Biot-Savart equation of vortex line motion
\begin{equation}
\dot{{\bf s}} = { \kappa \over 4 \pi}
\int \frac{d {\bf s}_0 \times ({\bf s} - {\bf s}_0)}
{|{\bf s} - {\bf s}_0|^3}   ,
\label{Biot-Savart}
\end{equation}
with the integral regularized in accordance with (\ref{reg}).

Equation  (\ref{Biot-Savart}) has two constants of motion:
\begin{equation}
{\cal E} = \int \frac{d {\bf s} \cdot d {\bf s}_0}{|{\bf s} - {\bf s}_0|} \; ,
\label{E}
\end{equation}

\begin{equation}
\vec{{\cal P}} = \int {\bf s} \times d {\bf s} \; ,
\label{P}
\end{equation}
which (up to dimensional factors) are nothing than the energy and momentum, respectively.

%%%%%%%%%%%%%%%%%%%
\subsection{Local induction approximation (LIA)}

In the absence of significant enhancement of non-local interactions by polarization of the vortex tangle,
the regular part in (\ref{BS1}) is smaller than the
first term containing large logarithm.  In such cases it is often---but not always (!), see the theory of the pure Kelvin-wave cascade---safe to neglect the second term in (\ref{BS1})
and proceed within LIA:
\begin{equation}
\dot{\bf s}  =  \beta \; {\bf s}'_{\xi} \times
{\bf s}''_{\xi \xi}   ,
\label{LIA}
\end{equation}
where
\begin{equation}
\beta = {\kappa \over 4 \pi} \, \ln (R/a_0) \label{beta0} ,
\end{equation}
with  the typical curvature radius $R$ treated as a constant.
Recalling that the parameter $\xi$ in (\ref{BS1})  is the arc length, it is crucial that equation of motion (\ref{LIA}) is consistent with
this requirement. [Speaking generally, in  the course of evolution
 $\xi$ might deviate from the arc length.]  The consistency is established by directly checking that (\ref{LIA}) implies
 \begin{equation}
{ d \over  d t} \, \sqrt{ d{\bf s}\cdot d{\bf s}} = 0 .
\label{arc_t}
\end{equation}
From Eq.~(\ref{arc_t}) it trivially follows---by integrating over $\xi$---that the total
line length is conserved. The total line length in LIA plays the same role as the energy (\ref{E}) in
the genuine Biot-Savart equation. Indeed, from (\ref{E}) it is seen that within
the logarithmic accuracy ${\cal E}$ is proportional to $\beta$ times
line length. It is not difficult to also make sure that Eq.~(\ref{LIA}) conserves ${\cal P}$ (\ref{P}).

The standard constants of motion---the energy (line length),  momentum, and angular momentum---are not the only quantities
conserved by Eq.~(\ref{LIA}). For example, the integral of the square of the curvature radius is also conserved \cite{Betchov}:
\begin{equation}
{d\over dt} \int ({\bf s}''_{\xi \xi})^2   d \xi = 0 .
\label{invar}
\end{equation}
In the next section we will see that the constant of motion (\ref{invar}) is just one of the {\it infinite}
set of constants of motion implied by the integrability of LIA.

%%%%%%%%%%%%%%%%%%%
\subsection{Hasimoto representation. Integrability of LIA}
Betchov \cite{Betchov} revealed certain interesting properties of LIA, e.g., Eq.~(\ref{invar}), by re-writing Eq.~(\ref{LIA}) in terms of intrinsic
variables of the vortex line: curvature, $\zeta$, and torsion, $\tau$. Hasimoto \cite{Hasimoto} further advanced these ideas by discovering that for the complex variable $\psi(\xi,t) $, such that
\begin{equation}
\zeta = |\psi|, ~~~~~~~~~ \tau = {\partial \Phi\over \partial \xi}
\label{Hasimoto_repr}
\end{equation}
($\Phi$ is the phase of $\psi$) the LIA equation (\ref{LIA}) is equivalent to the non-linear Schr\"odinger equation (time is measured in units $\beta^{-1}$)
\begin{equation}
i{\partial \psi \over \partial  t}= -{\partial ^2\psi \over \partial  \xi^2} - {1\over 2} |\psi|^2\psi .
\label{NLSE}
\end{equation}
The one-dimensional non-linear Schr\"odinger equation is known to be an integrable system featuring and infinite number of additive constants of motion, the explicit form of which is given by \cite{Zakharov}
\begin{eqnarray}
I_n=\int_{-\infty}^{\infty} \varphi_n(\xi)d\xi , ~~~~~~~~~~~~~~~~~~~~~~~~~~\nonumber \\
\varphi_1={1\over 4}|\psi|^2 ,~~~~~~~~~~~~~~~~~~~~~~~~~~~~~~~~~~~ \\
 \varphi_{n+1}=\psi {d\over d\xi} (\varphi_n/\psi)+\sum_{n_1+n_2=n} \varphi_{n_1} \varphi_{n_2} . ~~~~\nonumber
\label{constants}
\end{eqnarray}
The integrability of LIA renders any non-trivial relaxation kinetics impossible, unless the reconnections are involved to break the conservation of $I_n$'s.

%%%%%%%%%%%%%%%%%%%%%%%%%%%
\subsection{Hamiltonian formalism}
\label{subsec:ham}
Suppose a vortex line can be  parametrized by its Cartesian  coordinates $x$ and $y$ as single-valued functions of the third coordinate $z$. In this case, the Biot-Savart law can be cast into a Hamiltonian form,  very convenient for  our purposes. First, introduce a vector $\vec{\rho}(z,t) = (x(z,t), y(z,t))$. In contrast to the earlier-defined vector ${\bf s}(\xi,t)$ that  follows the motion of the element of fluid containing the element of vortex line, the vector $\vec{\rho}(z_0,t)$ just defines the geometrical point of intersection of the vortex line with the plane $z=z_0$. With this distinction in mind, it is easy to relate the time derivatives of the two vectors:
\begin{equation}
{\partial \vec{\rho}\over \partial t} = {\partial {\bf s} \over \partial t} -
\left(\hat{z}\cdot {\partial {\bf s} \over \partial t}\right)\left(\hat{z} + {\partial\vec{\rho} \over \partial z}\right) .
\label{time der}
\end{equation}
Here $\hat{z}$ is the unit vector in the $z$-direction. Substituting for $\partial {\bf s}/ \partial t$ the right-hand side of Eq.~(\ref{Biot-Savart}), expressing then ${\bf s}$ in terms of $\vec{\rho}$, and finally replacing $\vec{\rho}$ with a
complex variable  $w(z,t)= x(z,t)+iy(z,t)$, we arrive at  the Hamiltonian equation of vortex line motion:
\begin{equation}
i\dot{w} = {\delta H\over \delta w^*} ,
\label{eq_motion}
\end{equation}
\begin{equation}
H= (\kappa/4\pi)  \int  {\left[ 1+{\rm Re} \,
w'^*(z_1)w'(z_2) \right]  dz_1 dz_2 \over \sqrt{ (z_1-z_2)^2+|w(z_1)-w(z_2)|^2}} .
\label{ham}
\end{equation}
The Hamiltonian (\ref{ham}) is singular at $z_1\to z_2$ and thus needs to be regularized.
To this end we introduce $r_*$ such that
\begin{equation}
a_0 \ll r_* \ll \lambda ,
\label{r_star}
\end{equation}
and write
\begin{equation}
H =H_{\rm loc}+H_{\rm n.l.}+ {\cal O}(r_*/\lambda),
\label{ham_sum}
\end{equation}
\begin{equation}
H_{\rm n.l.} = (\kappa/4\pi) \!\!\!\! \int\limits_{\left| {z_1  - z_2 } \right| > r_* } \! \!\!\!  {\left[ 1+{\rm Re} \,
w'^*(z_1)w'(z_2) \right]  dz_1 dz_2 \over \sqrt{ (z_1-z_2)^2+|w(z_1)-w(z_2)|^2}} \; ,
\label{ham_non_loc}
\end{equation}
\begin{equation}
H_{\rm loc} = 2\beta\int dz \sqrt{1+|w'(z)|^2}  ,
\label{ham_loc}
\end{equation}
\begin{equation}
\beta = (\kappa/ 2 \pi)\ln (r_*/a_*) .
\label{beta}
\end{equation}
Here  the value of $a_* \sim a_0$  is specially tuned to eliminate a factor of order unity in the logarithm.---The absence of such a factor in Eq.~(\ref{beta})  should not be confused with a lack  of control on  first sub-logarithmic corrections. In view of the first inequality in (\ref{r_star}), the Hamiltonian $H_{\rm n.l.}$ is of purely hydrodynamic nature, while the Hamiltonian $H_{\rm loc}$ takes care of {\it both}  hydrodynamic and microscopic (and system specific) features at distances smaller that $r_*$. In view of the condition $\lambda \gg a_0$, implied by  Eq.~(\ref{r_star}), the microscopic specifics of the system is completely absorbed by the proper  value of $a_*$. Indeed, whatever is the physics at the distances of the order of the vortex core radius, the leading contribution from these distances to the energy of a smooth vortex line should be directly proportional to the line length; and this is precisely what is expressed by Eq.~(\ref{ham_loc}).

The freedom of choosing a particular value of $r_*$ within the range (\ref{r_star}) can be used to introduce the local induction approximation by requiring that
\begin{equation}
\ln ( r_*/a_0) \gg \ln (\lambda / r_*) .
\label{r_star_LIA}
\end{equation}
In this case, the Hamiltonian (\ref{ham_sum}), with
\begin{equation}
\beta = (\kappa/ 2 \pi)\ln (\lambda/a_0) =\Lambda \kappa/ 2 \pi ,
\label{beta_appr}
\end{equation}
captures the leading---as long as the interline interactions are not relevant---logarithmic contribution, compared to which the non-local Hamiltonian (\ref{ham_non_loc}) can be omitted with logarithmic accuracy guaranteed by the parameter $\Lambda^{-1}\ll 1$.

Less obvious technical trick  is to {\it formally} set
\begin{equation}
r_* = a_*
\label{r_star_a_0}
\end{equation}
to  {\it nullify} the Hamiltonian $H_{\rm loc}$. Doing so might seem to violate the range of applicability of the essentially hydrodynamic Hamiltonian $H_{\rm n.l.}$ \cite{Laurie}.  Nevertheless, it is readily seen by inspection that under the condition $\lambda \gg a_0$ the resulting theory is equivalent to the theory (\ref{ham_sum})-(\ref{beta}) up
to negligibly small corrections of the order $a_0/\lambda$. The resulting Hamiltonian is
\begin{equation}
H_{\rm psd}=(\kappa/4\pi) \!\!\!\! \int\limits_{\left| {z_1  - z_2 } \right| > a_* } \! \!\!\!  {\left[ 1+{\rm Re} \,
w'^*(z_1)w'(z_2) \right]  dz_1 dz_2 \over \sqrt{ (z_1-z_2)^2+|w(z_1)-w(z_2)|^2}} .
\label{ham_pseudo}
\end{equation}
In direct analogy with the pseudo-potential method in the scattering theory, it has a status of a {\it pseudo}-Hamiltonian in the sense that, being a convenient but rather inadequate model at the scales $\sim a_0$, it accurately accounts for the long-wave motion of the vortex lines (including all sub-leading corrections to LIA coming from the distances $\sim a_0$) by an appropriate choice of $a_*$. In the next subsection, we describe an explicit procedure of extracting the value of $a_*$ from a model-specific dispersion relation of Kelvin waves.

%%%%%%%%%%%%%%%%%%%%%%%%%
\subsection{Kelvons. Angular momentum and the number of kelvons}

Suppose the amplitude of Kelvin waves (KW) is small enough, so that the following condition is satisfied:
\begin{equation}
\alpha(z_1,z_2) = {|w(z_1)-w(z_2)| \over |z_1-z_2|} \ll 1  .
\label{ALPHA}
\end{equation}
Linearizing (\ref{ham_pseudo}) up to the leading orders in the dimensionless function $\alpha(z_1,z_2)$, we obtain
\begin{equation}
H_0=(\kappa / 8\pi ) \!\!\!\! \int\limits_{\left| {z_1  - z_2 } \right| > a_* } \! \!\!\! {dz_1 dz_2 \over |z_1-z_2|} \left[ 2{\rm
Re} \, w'^*(z_1)w'(z_2)- \alpha^2 \right]  . \label{H0}
\end{equation}
The Hamiltonian $H_0$ describes the linear properties of KW. It is
diagonalized by the Fourier transformation $w(z) = L^{-1/2} \sum_k
w_k e^{ikz}$ ($L$ is the system size, periodic boundary conditions
are assumed):
\begin{equation}
 H_0=(\kappa / 4\pi) \sum_k   \,  \omega_k \, w_k^* w_k  , \label{H0_diag}
\end{equation}
yielding Kelvin's dispersion law
\begin{equation}
 \omega_k = {\kappa \over  4\pi} \left[ \ln { 1\over k a_*} + C_0 + {\cal O}\left( (k a_*)^2\right) \right]  k^2   ,
 \label{disp_law}
\end{equation}
with
\begin{eqnarray}
C_0=2\int_1^{\infty} {d x \over x} \left[ \cos x + (\cos x - 1) /x^2 \right]  + \nonumber \\
2\int_0^{1} {d x \over x} \left[ \cos x -1/2 + (\cos x - 1) /x^2 \right]
\approx -2.077 .
 \label{C_0}
\end{eqnarray}
Generally speaking, the term ${\cal O}\left( (k a_*)^2\right)$ should be omitted, since it goes beyond the
universality range of the pseudo-Hamiltonian (\ref{ham_pseudo}).---In each particular microscopic model this term
will be sensitive to the details of the physics at the length scale $\sim a_0$. [With logarithmic accuracy, the constant $C_0$ can be omitted as well, while the specific value of $a_*$ can be replaced with just and order-of-magnitude estimate $a_0$.]

The practical utility of Eq.~(\ref{disp_law}) with sub-logarithmic correction $C_0$  (\ref{C_0}) for a given specific microscopic model---{\it different} from the pseudo-Hamiltonian (\ref{ham_pseudo})---is as follows. Eqs.~(\ref{disp_law})-(\ref{C_0}) can be used to {\it calibrate} the value of $a_*$, and thus fix the pseudo-Hamiltonian, by separately solving for the KW spectrum in the given microscopic model and then casting the answer in the form (\ref{disp_law})-(\ref{C_0}). Likewise, for a realistic strongly-correlated system like $^4$He the appropriate value of $a_*$ in the pseudo-Hamiltonian (\ref{ham_pseudo}) can be calibrated by an experimentally measured kelvon dispersion.

Although the problem of KW cascade generated by decaying superfluid
turbulence is purely classical, it will be convenient to approach it
quantum mechanically---by introducing KW quanta, kelvons. In
accordance with the canonical quantization procedure, we
understand $w_k$ as the annihilation operator of the kelvon with
momentum $k$ and correspondingly treat $w(z)$ as a quantum field.
The Hamiltonian functional (\ref{ham_pseudo}) is {\it proportional} to the energy---with the coefficient $\kappa \rho/2$, where $\rho$ is the mass
density---but not {\it equal} to it. This means that if one
prefers to work with genuine Quantum Mechanics rather than a fake
one (for our purposes, the latter is also sufficient), using true
kelvon annihilation operators, $\hat{a}_k$, field operator
$\hat{w}$, and Hamiltonian, $\hat{H}$, he should take into account
proper dimensional coefficients: $\hat{H}=(\kappa \rho/2)H$,
$\hat{w}_k=\sqrt{{2 \hbar / \kappa \rho }}\, \hat{a}_k $. By
choosing the units $\hbar = \kappa = 1$, $\rho = 2$, we ignore
these coefficients until the final answers are obtained.

In the quantum approach, there naturally arises the notion of the
number of kelvons. This number is {\it conserved}
 in view of its global $U(1)$ symmetry of the Hamiltonian, $w
\to e^{i\varphi} w$ reflecting the rotational symmetry of the
problem. The rotational symmetry is also responsible for the conservation of the angular momentum component along the vortex line direction. Thus the number of kelvons is immediately related to the angular momentum component \cite{Epstein_Baym}: one kelvon carries a single (negative) quantum, $-\hbar$, of the angular momentum component along the vortex line relative to the macroscopic angular momentum of the rectilinear vortex line.

Another advantage of the quantum language in
weak-turbulence problems, which we intensively employ in this paper, is that the collision term of the kinetic
equation immediately follows from the Golden Rule for the
corresponding elementary processes.

%%%%%%%%%%%%%%%%%%%%%%%%%%%%%%%%%%%%%%%%%%%%%%%%%%%%%%%%%%%%%%%%%%%%%%
\section{Pure Kelvin-wave cascade}
\label{sec:pure}

\subsection{Qualitative analysis}

%We start by discussing the key qualitative features that lie in the basis of %the pure Kelvin-wave cascade scenario. It is instructive that these quite %general features are sufficient to dramatically narrow down the range of %possible solutions---in this subsection, we obtain the solution up to a %single parameter that contains all the relevant information about the %details of microscopic dynamics.

In the problem of low-temperature ST decay at the length scales smaller than the typical interline separation, the pure Kelvin-wave cascade on individual vortex lines is perhaps the most natural decay scenario one can think of. Indeed, the non-linear nature of Kelvin-wave dynamics should allow such a process in which the energy in the form of Kelvin waves is transferred from the long-wave energy-containing modes to the short wavelengths where the dissipation becomes efficient. Provided with a wide inertial range, one could also expect such a transfer of energy to be local in the wavenumber space, in which case the relaxation is likely to be due to a cascade.

A subtlety that is immediately clear is the complete absence of kinetics in the leading approximation, which stems from the aforementioned integrability of the LIA, Eq.~(\ref{LIA}). If it exists, the pure Kelvin-wave cascade must be entirely due to the non-local coupling between different vortex-line elements. Hence, being suppressed by the small parameter $\Lambda^{-1} \ll 1$ relative to the leading LIA dynamics, the pure KW cascade is an \textit{a priori} weak phenomenon. This fact provides us with an important consistency check of the results: the leading contribution $\propto \Lambda$---determined by the effective microscopic cutoff parameter $a_*$, if one uses the pseudo-Hamiltonian (\ref{ham_pseudo})---must completely drop out of the effective kelvon scattering amplitude.

The success of the Kolmogorov-type argumentation makes it very tempting to use a dimensional analysis of the dynamical equations of motion to immediately extract, e.g., energy spectra. Such a simple approach for Kelvin-wave turbulence, however, is doomed to failure. The reason is that here the scale invariance dictates the kinetics to be controlled by a purely geometrical \textit{dimensionless} parameter $\alpha_k=b_k k$, where $b_k$ is the typical KW amplitude at the wavenumber $k$. Thus, the solution could be only obtained from the corresponding kinetic equation, which reveals the proper combination of $\alpha_k$ and $k$ the kinetics are due to. Deriving and solving the kinetic equation will be our main goal in the this section.

Let us first determine the general structure of the collisional term. The key aspects that essentially fix this structure are (i) the conservation of the number of kelvons, discussed in Sec.~\ref{sec:basic}, (ii) the dimensionality of the problem, and (iii) weakness of non-linearities. The first circumstance means that the kinetics are entirely due to kelvon \textit{elastic} scattering since the elementary events of kelvon creation/annihilation are prohibited. In other words, the only allowed kinetic channel is energy-momentum exchange between kelvons. On the other hand, the energy-momentum conservation laws in one dimension make this exchange impossible in two-kelvon collisions: the process $(k_1,k_2) \to (k_3,k_4)$ is only allowed if
either $(k_3=k_1,~k_4=k_2)$, or $(k_3=k_2,~k_4=k_1)$ which does
not lead to any kinetics. Thus any non-trivial kinetics can be only due to collisions of three or more kelvons.

The third condition translates into the fact that the amplitudes of Kelvin waves $b_k$ in the pure cascade are necessarily small compared to their wavelengths, $\alpha_k=b_k k \ll 1$, at least at sufficiently large $k$. This is the central feature of this regime, which is due to the following. The cascade solution with the largest possible amplitudes, $\alpha_k \approx 1$, corresponds to the regime driven by self-reconnections, as described in Sec.~\ref{sec:self-rec}, the regime where the non-linear effects are negligible altogether. The weakness of the purely non-linear kinetics makes the amplitude at a scale $k$ rise (due to the energy supplied from the larger length scales) until $b_k$ is of order $1/k$ and the self-reconnection can happen causing the energy transfer to a smaller scale. If non-linear processes are appreciable, smaller amplitudes $b_k$ are sufficient to sustain the energy flux. Thus, the pure cascade spectrum, which due to the scale invariance has a general power-law form $\alpha_k =b_k k \propto k^{\beta}$, must be constrained by $\beta \leq 0$. Apart from the marginal $\beta=0$, the latter requirement guarantees that a theory built on $\alpha_k \ll 1$ becomes asymptotically exact at hight wavenumbers. Correspondingly, self-reconnections, being exponentially dependant on $\alpha_k$, necessarily cease in the purely non-linear cascade.

The smallness of the Kelvin-wave amplitudes in the particle language means that the many-kelvon collisions are rare events and we can confine ourselves to the leading three-kelvon processes in the kinetic term.

These simple considerations already substantially limit the freedom in writing the collision term. In fact, the only missing ingredient, which needs to be calculated, is the wavenumber dependence of the effective kelvon scattering vertex. Postponing the final issue of finding this dependence until the next section, let us analyze the kinetic equation in a general form. Written in terms
of averaged over the statistical ensemble kelvon occupation
numbers $n_k = \langle a^{\dagger}_k a_k \rangle $, the kinetic equation is given by
\begin{equation}
\dot{n}_1 = \frac{1}{(3-1)! \; 3!} \sum_{k_2, \ldots , k_6} \left(
W_{4,5,6}^{1,2,3}-W_{1,2,3}^{4,5,6} \right) \; . \label{KE_gen}
\end{equation}
Here $W_{1,2,3}^{4,5,6}$ is the probability per unit time of the
elementary three-kelvon scattering event $(k_1,k_2,k_3) \to (k_4,k_5,k_6)$, and the combinatorial factor compensates multiple counting of the same scattering event. The three-kelvon  effective interaction Hamiltonian has the general form
\begin{equation}
H_{\rm int} =
\sum_{k_1, \ldots , k_6} \delta(\Delta k) \,
\tilde{V}_{1,2,3}^{4,5,6} \, a^{\dagger}_6 a^{\dagger}_5
a^{\dagger}_4 a^{ }_3 a^{ }_2 a^{ }_1 , \label{H_int_gen}
\end{equation}
where the effective vertex
$\tilde{V}$ is symmetrized with respect to the
corresponding momenta permutations, $\delta(k)$ is understood
as the discrete $\delta_{k, 0}$, and  $\Delta k =
k_1+k_2+k_3-k_4-k_5-k_6$. The probabilities $W_{1,2,3}^{4,5,6}$ are then straightforwardly given by the Fermi Golden Rule applied to the Hamiltonian (\ref{H_int_gen}):
\begin{eqnarray}
W_{1,2,3}^{4,5,6}=2 \pi  |(3!)^2 \tilde{V}_{1,2,3}^{4,5,6}|^2
f_{1,2,3}^{4,5,6} \, \delta(\Delta \omega) \delta(\Delta k), \nonumber \\
f_{1,2,3}^{4,5,6}\;=\;n_1n_2n_3(n_4+1)(n_5+1)(n_6+1), \label{FGR} \\
\Delta\omega \; = \; \omega_1+\omega_2+\omega_3-\omega_4-\omega_5-\omega_6 .\nonumber
\end{eqnarray}
Here the combinatorial factor $(3!)^2$ accounts for the addition of
equivalent amplitudes. We are of course interested in the classical-field limit of the Eq.~(\ref{FGR}), which is obtained by taking $n_k \gg 1$ and retaining only the leading in $n_k$ terms. This procedure finally yields
\begin{eqnarray}
\dot{n}_1 \! \! =  \!   216 \pi \! \! \!  \! \sum_{k_2, \ldots ,
k_6} \! \! \!  \! |\tilde{V}_{1,2,3}^{4,5,6}|^2  \, \delta(\Delta
\omega) \, \delta(\Delta k) \left( \tilde{f}_{4,5,6}^{1,2,3} \! -
\! \tilde{f}_{1,2,3}^{4,5,6} \right )  , \label{KE} \\
\tilde{f}_{1,2,3}^{4,5,6}=n_1n_2n_3(n_4n_5+n_4n_6+n_5n_6). \nonumber
\end{eqnarray}

The kinetic equation (\ref{KE}) supports
an energy cascade \cite{Zakharov}, provided two conditions
are met: (i) the kinetic time is getting progressively smaller
(vanishes) in the limit of large wavenumbers, (ii) the collision
term is local in the wavenumber space, that is the
relevant scattering events are only those where all the kelvon
momenta are of the same order of magnitude. In the following subsection, we make sure that both conditions are satisfied: the condition (i) can be checked by a dimensional estimate, provided (ii) is true. The
condition (ii) is verified numerically. Under these conditions one can establish the cascade spectrum by a simple dimensional analysis of the kinetic equation.

The locality of the collision term implies that the integral in (\ref{KE}) builds up around $(k_2, \ldots ,
k_6) \sim k_1$, which dramatically simplifies the kinetic equation:
\begin{equation}
\dot{n}_k \propto k^5 \cdot |\tilde{V}|^2 \cdot \omega_k^{-1} \cdot k^{-1}
\cdot n_k^5, \label{KE_est}
\end{equation}
where the factors go in the order of the appearance of
corresponding terms in (\ref{KE}). At $k_1 \sim \ldots \sim k_6
\sim k$ we have $|V| \propto k^\nu$ and
\begin{equation}
\dot{n}_k \;  \propto \; \omega^{-1}_k \, n^5_k \, k^{4+2\nu} \;
.\label{n_dot}
\end{equation}
The energy flux (per unit vortex line length), $\theta_k$, at the
momentum scale $k$ is defined as
\begin{equation}
\theta_k \; = \;  L^{-1} \sum_{k' < k } \, \omega_{k'} \,
\dot{n}_{k'} \; , \label{flux_def}
\end{equation}
implying the estimate $\theta_k \sim k \dot{n}_k \omega_k $.
Combined with (\ref{n_dot}), this yields $\theta_k \sim n_k^5
k^{5+2\nu}$, and the cascade requirement that $\theta_k $ be actually
$k$-independent leads to the spectrum
\begin{equation}
n_k \propto \langle \hat{w}^{\dagger}_k \hat{w}_k \rangle = A \, k^{-(5+2\nu)/5} \; . \label{spect_nu}
\end{equation}
The value of the spectrum amplitude $A$ in (\ref{spect_nu}) controls the energy flux that the cascade transports. The relation between $\theta$ and $A$ is
\begin{equation}
\theta = C_{\theta} \kappa^3 \rho \, A^5 \; . \label{rel}
\end{equation}
The dimensionless coefficient $C_{\theta}$ in this formula can be obtained, e.g., by a straightforward numerical calculation, as it was done by the authors in Ref.~\cite{KS_04}. However, the calculation of Ref.~\cite{KS_04} in view of a large relative error allowed to obtain only the order of magnitude of $C_{\theta}$, which, as we explain in the next subsection, can be further questioned due to an erroneous omission of an order-one term in the effective vertex.

\subsection{Quantitative analysis}

The central problem of this section is the derivation of the effective vertex $V_{1,2,3}^{4,5,6}$, which defines the effective kelvon interaction Hamiltonian (\ref{H_int_gen}) ($\tilde{V}$ is obtained from $V$ by symmetrization with respect to corresponding momenta permutations). We start with the pseudo-Hamiltonian (\ref{ham_pseudo}). Our fundamental requirement that the amplitude of KW turbulence is small as compared to the wavelength, $\alpha_k \ll 1$, is formulated by Eq.~(\ref{ALPHA}). This allows us to expand (\ref{ham_pseudo}) in powers of $\alpha(z_1,z_2) \ll 1$:
$H=E_0+H_0+H_1+H_2+ \ldots $ ($E_0$ is just a number and will be
ignored). The term $H_0$ is given by Eq. (\ref{H0}). As we demonstrated in the previous subsection, it describes the linear properties of the Kelvin waves, in particular it determines the Kelvin-wave dispersion law, Eq.~(\ref{disp_law}). The higher-order terms are responsible for interactions between kelvons. The terms that will prove relevant are
\begin{equation}
H_1=\frac{\kappa }{32\pi}  \!\!\!\! \int\limits_{\left| {z_1  - z_2 } \right| > a_* } \! \!\!\!  {dz_1 dz_2 \over |z_1-z_2|} [ 3
\alpha^4 - 4 \alpha^2 {\rm Re} \, w'^*(z_1)w'(z_2) ] \; ,
\label{H1}
\end{equation}
and
\begin{equation}
H_2=\frac{\kappa }{64\pi}  \!\!\!\! \int\limits_{\left| {z_1  - z_2 } \right| > a_* } \! \!\!\!  {dz_1 dz_2 \over |z_1-z_2|} [ 6
\alpha^4 {\rm Re} \, w'^*(z_1)w'(z_2) - 5 \alpha^6 ] \; .
\label{H2}
\end{equation}

\begin{figure}[htb]
\includegraphics[width = 0.8\columnwidth,keepaspectratio=true]{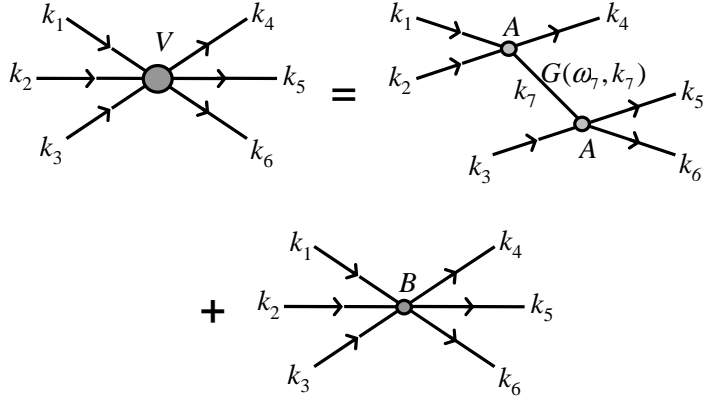}
\caption{The effective vertex of the three-kelvon scattering. The vertices $A$ and $B$ come from the terms $H_1$ (\ref{H1}) and $H_2$ (\ref{H2}) and are given by (\ref{A}) and (\ref{B}) respectively. $G(\omega,k)$ is the free kelvon propagator.}
\label{fig:diag}
\end{figure}

As already demonstrated, the leading elementary process in our case is the three-kelvon scattering and the processes involving four and more kelvons are much weaker due to the inequality (\ref{ALPHA}). The effective vertex,
$V_{1,2,3}^{4,5,6}$, for the three-kelvon scattering
[subscripts (superscripts) stand for the initial (final) momenta] consists of two different parts. The first part is due to the terms generated by the two-kelvon vertex, $A$ (corresponding to the Hamiltonian $H_1$) in the second order of perturbation theory. [All these terms are similar to each other; we explicitly specify just one of them: $A_{1,2}^{4,7} \,
G(\omega_7,k_7) \, A_{7,3}^{5,6}$. Here $G(\omega,k)=1/(\omega
-\omega_k)$ is the free-kelvon propagator,
$\omega_7=\omega_1+\omega_2-\omega_4$, $k_7=k_1+k_2-k_4$.] The
second part of the vertex $V$ is the bare three-kelvon vertex,
$B$, associated with the Hamiltonian $H_2$. The expression for the effective vertex is depicted diagrammatically in Fig. \ref{fig:diag}.

%%%%%%%%%%%%%%%%%%%%%%% FIGURE three-kelvon collisions %%%%%%%%%%%%%
%%%%%%%%%%%%%%%%%%%%%%%%%%%%%%%%%%%%%%%%%%%%%%%%%%%%%%%%%%%%%%%%

The explicit expressions for the bare vertices directly follow
from (\ref{H1}) and (\ref{H2}) after the Fourier transform $w(z) = L^{-1/2} \sum_k
w_k e^{ikz}$:
\begin{eqnarray}
A=(6D -E)/8\pi, \label{A} \\
D^{3,4}_{1,2}=\int_{a_*}^{L}(dx /x^5) \Bigl( \, 1 -
\bigl[ _{ 1 }\bigr]  - \bigl[ _{ 2 }\bigr] - \bigl[^{3}\bigr] \nonumber \\ -\bigl[ ^{4}\bigr]
 + \bigl[^{  3  }_{  2  }
\bigr] + \bigl[ ^{  43  }\bigr]  + \bigl[ ^{  4  }_{  2 }\bigr]
\Bigr) , \nonumber \\
E^{3,4}_{1,2}=\int_{a_*}^{L} (dx /x^3) \Bigl\{ k_4k_1 ( \bigl[
^{  4  }\bigr] +\bigl[ _{  1  } \bigr] -\bigl[ ^{ 43 }\bigr]
-\bigl[ ^{  4  }_{  2  }\bigr] )
+ \nonumber \\
k_3k_1
( \bigl[ ^{ 3   }\bigr]  +
 \bigl[ _{ 1  } \bigr] - \bigl[
^{ 43 } \bigr] - \bigl[ ^{  3  }_{  2  }\bigr]
 ) +k_3k_2 ( \bigl[ ^{  3  }\bigr] + \bigl[ _{  2  } \bigr]
  - \bigl[^{  43  } \bigr] -   \bigl[ ^{  3  }_{
1  } \bigr] ) \nonumber \\ + k_4k_2 ( \bigl[ ^{  4  } \bigr] +\bigl[ _{ 2
}\bigr] - \bigl[ ^{  43  }\bigr] -\bigl[ ^{  3  }_{ 2 }\bigr] )
\Bigr\}, \nonumber
\end{eqnarray}
\begin{eqnarray}
B=(3P-5Q)/4\pi, \label{B} \\
P^{4,5,6}_{1,2,3}=\int_{a_*}^{L} (dx
/x^5) \;k_6k_2 \Bigl\{ \bigl[ _{ 2   }\bigr]  -\bigl[ ^{  5 }_{ 2
  } \bigr] -\bigl[ _{  23   }\bigr]  +\bigl[ ^{  5   }_{ 23
  }\bigr] \nonumber \\
   -   \bigl[ ^{  4   }_{  2    }\bigr] +\bigl[ ^{  45
}_{  2   } \bigr] +\bigl[ ^{  4   }_{  23   }\bigr] -\bigl[ ^{ 6
  }_{ 1   }\bigr]  +\bigl[^{  6   } \bigr] -\bigl[ ^{ 56
}\bigr]  \nonumber \\
 -\bigl[ ^{  6   }_{ 3   } \bigr] +\bigl[ ^{  56   }_{ 3
  }\bigr]  -\bigl[ ^{   46   } \bigr] +\bigl[ ^{ 456   }
\bigr] +\bigl[ ^{ 46   }_{  3   }\bigr] -\bigl[ _{  12   }
\bigr] \Bigr\}, \nonumber \\
Q^{4,5,6}_{1,2,3}=\int_{a_*}^{L} (dx /x^7) \:
\Bigl\{1 -\bigl[ ^{ 4   }\bigr]  -\bigl[ _{  1   }\bigr]  +\bigl[ ^{
4 }_{  1   } \bigr] -\bigl[ ^{  6   }\bigr]  +\bigl[ ^{  46   }
\bigr] \nonumber \\
 +\bigl[ ^{  6   }_{  1   } \bigr] -\bigl[ ^{  46   }_{ 1
  }\bigr] -\bigl[ ^{  5   }\bigr]  + \bigl[ ^{45} \bigr]
 +\bigl[ ^{  5    }_{  1    }\bigr]
-\bigl[ ^{   45   }_{  1   } \bigr] +\bigl[ ^{  65    }\bigr]
-\bigl[ ^{  456   } \bigr] \nonumber \\
 -\bigl[ ^{  56    }_{  1   } \bigr]
+\bigl[ _{  23   } \bigr] -\bigl[ _{  3   } \bigr] +\bigl[ ^{ 4
  }_{  3    } \bigr] +\bigl[ _{  13   }\bigr] -\bigl[ ^{  4
  }_{ \! 13   }\bigr]  +\bigl[ ^{  6   }_{  3   }\bigr]
-\bigl[ ^{  46   }_{  3   } \bigr]
- \bigl[ ^{  6   }_{\! 13
}\bigr] \nonumber \\
 +\bigl[ ^{  5   }_{  2   } \bigr] +\bigl[ ^{  5   }_{ 3
  }\bigr] - \bigl[ ^{  45   }_{  3   }\bigr] -\bigl[ ^{5}_{\! 13} \bigr]
  +\bigl[ ^{  6   }_{2   }\bigr] -\bigl[ ^{ 65 }_{  3
}\bigr] +\bigl[ _{  12   } \bigr] +\bigl[ ^{  4 }_{  2 } \bigr]
-\bigl[ _{  2   }\bigr] \Bigr\}. \nonumber
\end{eqnarray}

Here $\bigl[ ^{\cdots}_{\cdots}
\bigr]$'s denote similar looking cosine functions, $\bigl[ _{  1
} \bigr] = \cos k_1 x$, $\bigl[ ^{  4
  }_{  1   }\bigr] = \cos (k_4 - k_1) x$, $\bigl[ ^{  4 5  }_{
1   } \bigr] = \cos (k_4 +k_5 - k_1) x$, $\bigl[ ^{  45   }_{ 12
  } \bigr] = \cos (k_4 +k_5 - k_1-k_2) x$, and so forth.

Since the microscopic physics at the scales $\sim a_0$ is quite complicated and usually not known precisely in strongly correlated systems like $^4$He, it is important to asses the systematic error of the theory in the case when $a_*$ can not be determined accurately. For this purpose, as well as for the convenience of the numerical analysis described below, we process the integrals (\ref{A}),
(\ref{B}) as follows. We introduce a characteristic wavelength $\lambda\equiv
\lambda(k_1,k_2,k_3,k_4,k_5,k_6)$, the particular functional form being not important. Then, from each of the integrals (\ref{A}), (\ref{B}), we subtract the corresponding LIA contributions, which are easily obtained by the same procedure that led to Eqs.~(\ref{A}), (\ref{B}) from the local Hamiltonian (\ref{ham_loc}) defined by
\begin{equation}
\beta_\lambda = (\kappa/ 2 \pi)\Lambda_\lambda, ~~~~~~~~~
\Lambda_\lambda=\ln(\lambda/a_*) .
\label{beta_lambda}
\end{equation}
As a result of the subtraction, up to system-specific
terms $\sim (a_*/\lambda)^2$ to be neglected below, we get  $a_*$-independent convergent integrals.
Thereby we arrive at a convenient decomposition: $A=\Lambda_\lambda
A^{(0)} + A^{(1)}$,  $B=\Lambda_\lambda B^{(0)} + B^{(1)}$, where $A^{(0)}$ and
$B^{(0)}$ are known analytically from LIA, while
$A^{(1)}$ and $B^{(1)}$ are easily calculated numerically (see below for some useful details).  On the technical side, the decomposition solves the problem of handling the logarithmic divergency, whereas physically, it clarifies the form of the dependence on the vortex-core details, which turn out to be entirely enclosed by LIA.

To proceed with estimating possible systematic errors, we introduce the expansion  of
the propagator $G$ in the powers of inverse
$\Lambda_\lambda$:
\begin{equation}
G=\Lambda_\lambda^{-1}[G^{(0)}+\Lambda_\lambda^{-1}G^{(1)} +
\Lambda_\lambda^{-2}G^{(2)} + \mathcal{O}(\Lambda_\lambda^{-3})] .
\label{G_expand}
\end{equation}
For the vertex $V=AGA+B$ (see Fig.~\ref{fig:diag}), we thus have
\begin{equation}
V=\Lambda_\lambda V^{(0)} +
V^{(1)}+ V^{(2)}/\Lambda_\lambda + \mathcal{O}(\Lambda_\lambda^{-2}),
\label{V_expand}
\end{equation}
with
\begin{eqnarray}
V^{(0)}=A^{(0)}G^{(0)}A^{(0)}+B^{(0)}, \nonumber \\
V^{(1)}= 2A^{(0)}G^{(0)}A^{(1)}+A^{(0)}G^{(1)}A^{(0)}+B^{(1)},\nonumber \\
 V^{(2)}=
A^{(0)}G^{(2)}A^{(0)}+2A^{(0)}G^{(1)}A^{(1)}+A^{(1)}G^{(0)}A^{(1)}. \nonumber
\end{eqnarray}
Note that by construction all the quantities
$A^{(i)}$, $B^{(i)}$, $G^{(i)}$, and $V^{(i)}$ are $a_*$-independent, since the
dependence on $a_*$---up to the neglected terms $\sim (a_*/\lambda)^2$---comes
exclusively through $\Lambda_\lambda$.
The term
$\Lambda_\lambda V^{(0)}$ is precisely the effective vertex that follows from
the LIA
Hamiltonian (\ref{ham_loc}), and thus it necessarily obeys $V^{(0)} \equiv 0$.
The leading contribution to $V$ is the $a_*$-independent $V^{(1)}$. Thus, the short-range physics enters the answer only as a
small in $1/\Lambda_\lambda$ correction.

Now we are in a position to specify  the
systematic error  due to  uncertainties in microscopic details. In the case
when it is possible to calibrate the pseudo-Hamiltonian,  that is to find
(analytically or experimentally, see
Sec.~\ref{sec:basic}) the accurate value of $a_*$, the systematic error
is of order $(a_*/\lambda)^2$. Otherwise, we are forced to set $a_* \sim a_0$,
meaning that the parameter $\Lambda_\lambda \gg 1$ is only known up to $\delta
\Lambda_\lambda
\sim 1$, which for the systematic uncertainty in $V$ gives $V^{(2)}\delta
\Lambda_\lambda/\Lambda_\lambda^2 \sim 1/\Lambda_\lambda^2$. We thus arrive at
an important conclusion
that even if the microscopic details, such as the vortex-core shape, are
completely unknown, by naively setting $\Lambda_\lambda=\ln(\lambda/a_0)$ we are
paying
by a relative error of only $1/\Lambda_\lambda^2$.

A comment is in order here on the technical issue of a convenient handling of the integrals, which we do numerically, upon subtracting the logarithmic singularity. The whole procedure is almost identical to the one that led us to the kelvon dispersion,
Eqs.~(\ref{disp_law})-(\ref{C_0}). The new aspect is the dependence on four/six momenta, rendering a direct tabulation of integrals computationally expensive. The trick is to split a four/six-parametric integral into a sum of single-parametric
integrals. A minor technical problem  comes from the power-law divergence at $x\to 0$ of each separate single-parametric integral in Eqs.~(\ref{A}), (\ref{B}). The problem is readily solved by introducing power-law counter-terms---the {\it net} contribution of which is identically equal to zero---rendering each individual single-parametric integral convergent. As an illustration, we present the final result for the regularized (by subtracting the LIA terms and introducing the counter-terms) integral $D$ of Eq.~(\ref{A}),
which we denote with $\tilde{D}$. With the re-scaled (by $x\to \lambda x$) integration variable, the expression reads
\begin{eqnarray}
\tilde{D}^{3,4}_{1,2}=\lambda^{-4}\int_{1}^{\infty}(dx /x^5) \Bigl( \, 1 -
\bigl[ _{ 1 }\bigr]  - \bigl[ _{ 2 }\bigr] - \bigl[^{3}\bigr] -\bigl[ ^{4}\bigr]
 + \bigl[^{  3  }_{  2  }
\bigr] + \bigl[ ^{  43  }\bigr]  + \bigl[ ^{  4  }_{  2 }\bigr] \Bigr)
\nonumber \\
+ \, \lambda^{-4}\int_{0}^{1}(dx /x^5) \Bigl( \, \bigl[\kern-0.15em\bigl[ ^{  3  }_{  2  }
\bigr]\kern-0.15em\bigr]+ \bigl[\kern-0.15em\bigl[  ^{  43  }\bigr]\kern-0.15em\bigr] + \bigl[\kern-0.15em\bigl[  ^{  4  }_{  2 }\bigr]\kern-0.15em\bigr] - \bigl[\kern-0.15em\bigl[  _{ 1 } \bigr]\kern-0.15em\bigr] - \bigl[\kern-0.15em\bigl[ _{ 2 }\bigr]\kern-0.15em\bigr]- \bigl[\kern-0.15em\bigl[ ^{3}\bigr]\kern-0.15em\bigr] -\bigl[\kern-0.15em\bigl[ ^{4}\bigr]\kern-0.15em\bigr]
  \Bigr) .
  \label{tilde_G}
\end{eqnarray}
Here $\bigl[ ^{\cdots}_{\cdots} \bigr]$'s denote the same cosines as previously, but with an extra factor $\lambda$ in the argument due to re-scaled $x$. The symbol $\bigl[\kern-0.15em\bigl[  ^{\cdots}_{\cdots} \bigr]\kern-0.15em\bigr] $ means that for the  cosine, we subtract first few---first three in the case of $\tilde{D}$---terms of its Taylor expansion to render corresponding single-parametric integral convergent (and thus individually tabulatable).

From Eqs. (\ref{A}), (\ref{B}) it is straightforward to obtain the scaling of the effective vertex $V$ with the momenta---at $k_1 \sim \ldots \sim k_6 \sim k$ we have $|V| \sim k^6$. Thus, $\nu=6$ in Eq. (\ref{spect_nu}) and the pure Kelvin-wave cascade spectrum (restoring all the dimensional coefficients) is
\begin{equation}
\langle \hat{w}^{\dagger}_k \hat{w}_k \rangle = \frac{2 \hbar\,
n_k}{\kappa \rho} = A \, k^{-17/5} \; . \label{spect}
\end{equation}

This result was corroborated in a direct numerical simulation by the authors \cite{KS_05}, where the spectrum (\ref{spect}) was resolved with a high accuracy allowing us to distinguish it from $n\propto k^{-3}$ suggested in an earlier simulation by Vinen \textit{et al.} \cite{Vinen_2003}. The high precision required to distinguish the close exponents, both in terms of the cascade inertial range extent and low noise, was achieved using a special numerical scheme developed to reduce the complexity of the non-local model (\ref{ham_pseudo}) to that of an effectively local one at an expense of a controllable systematic error. We refer to Ref.~\cite{KS_04} for more details.

Let us also express the spectrum in terms of the typical geometrical amplitude, $b_k$, of the KW
turbulence at the wavevector $\sim k$. By the definition of the
field $\hat{w}(z)$ we have: $b^2_k \sim L^{-1} \sum_{q \sim k}
\langle \hat{w}^{\dagger}_k \hat{w}_k \rangle \sim k \, \langle \hat{w}^{\dagger}_k \hat{w}_k \rangle$. Hence, using Eq.~(\ref{rel}), we obtain
\begin{equation}
b_k \sim (\Theta/\kappa^3 \rho)^{1/10} k^{-6/5}. \label{spect2}
\end{equation}
One can convert (\ref{spect}) into the curvature spectrum. For the
curvature ${\bf c}(\zeta)=
\partial^{2} {\bf s} /
\partial \zeta^{2}$ [where ${\bf s}(\zeta)$
is the radius-vector of the curve as a function of the arc length
$\zeta$], the spectrum is defined as Fourier decomposition of the
integral $I_c=\int |{\bf c}(\zeta)|^2 \, d \zeta $. The
smallness of $\alpha$, Eq.~(\ref{ALPHA}), allows one to write $I_c
\approx \int dz \langle \hat{w}''^{\dagger}(z) \hat{w}''(z)
\rangle = \sum_k k^4 \, n_k \propto \sum_k  k^{3/5}$, arriving
thus at the exponent $3/5$.

So far we were heavily relying on the assumption of locality of the kinetic processes in the wavenumber space. We checked \cite{KS_04} the validity of this assumption by a numerical analysis of the kinetic equation thereby (i) making sure that the collision term of the kinetic equation is local and
(ii) estimating the value of the dimensionless coefficient in
(\ref{rel}) (see, however, below). The analysis is based on the following idea \cite{Sv91}. Consider a power-law distribution of occupation
numbers, $n_k=A / k^{\beta}$, with the exponent $\beta$
arbitrarily close, but not equal, to the cascade exponent $\beta_0
=17/5$. Substitute this distribution in the collision term of the
kinetic equation---right-hand side of (\ref{KE}). Given the scale
invariance of the power-law distribution, the following
alternative takes place. Case (1): collision integral converges
for $\beta$'s close to $\beta_0$, and, in accordance with a
straightforward dimensional analysis, is equal to
\begin{equation}
{\rm Coll}([n_k=A/k^{\beta}],k) = C(\beta) A^5 \omega^{-1}_k /
k^{5\beta - 16} \; . \label{coll}
\end{equation}
Here $C(\beta)$ is a dimensionless function of $\beta$, such that
$C(\beta_0)=0$ since the cascade is a steady-state solution. Case
(2): collision integral diverges for $\beta$ close to $\beta_0$.
The case (2) means that the collision term is non-local and the
whole analysis in terms of the Kolmogorov-like cascade is irrelevant.
Fortunately, our numerics show that we are dealing with the case
(1). Substituting (\ref{coll}) for $\dot{n}_k$ in
(\ref{flux_def}), we obtain the expression $\theta_k(\beta)\, = \,
(A^5/2\pi) \,  k ^{17- 5\beta} \, C(\beta)/(17- 5\beta)$. Taking
the limit $\beta \to \beta_0$, we arrive at the $k$-independent
flux $\theta = - C'(\beta_0) A^5 /10 \pi $.

In Ref.~\cite{KS_04}, we used this formula to obtain the coefficient in (\ref{rel}) by calculating $C(\beta)$
and finding its derivative $C'(\beta_0)=-10 \pi \, C_{\theta}$. We simulated the collision integral by Monte Carlo method. The integrals (\ref{A}), (\ref{B}), were calculated numerically. However, a mistake was made at the level of combining these integrals into the effective vertex $V$. Due to the cancelation of the leading logarithmic terms one has to keep an accurate account of the sub-logarithmic corrections, including those in the kelvon propagator $G(\omega,k)=1/(\omega-\omega_k)$ as well, as we already demonstrated above.
Unfortunately, we failed to appreciate a simple fact that the constant $C_0$ appearing in the kelvon dispersion (\ref{disp_law}) results in an order-one contribution to the effective vertex $V$, and neglected this constant in the propagator. Since the coefficient $C_{\theta} \approx 10^{-5}$ was obtained in Ref.~\cite{KS_04} only as an order of magnitude estimate (with an error $\sim 75 \%$ due to a slowing down of the high-order numerical integration) the mistake might not significantly effect this result, but still makes it  questionable. A more accurate determination of $C_{\theta}$ is necessary.

%%%%%%%%%%%%%%%%%%%%%%%%%%%%%%
\section{Self-reconnection driven cascade}
\label{sec:self-rec}

%%%%%%%%%%%%%%%%%%%%%%%%%%%%
\subsection{Absence of Feynman's cascade}
For quite a long period of time it has been generally accepted that at $T=0$ the scenario of decay of superfluid turbulence is the one proposed by Feynman \cite{Feynman}. Namely, that vortex lines first decay into vortex rings, each of the rings then independently self-reconnects, producing smaller rings, each of the smaller rings self-reconnects to produce even smaller rings, and so forth. In this respect it is very characteristic that even after the absence of pure Feynman's cascade was directly shown,  and the scenario of Kelvin-wave cascade driven by local self-crossings was proposed \cite{Sv95},   simulations of the low-temperature decay within LIA performed in Ref.~\cite{Tsubota00} were still interpreted in terms of Feynman's cascade.

The problem with Feynman's cascade is that it is inconsistent with simultaneous conservation of energy and momentum \cite{Sv95}.  This is  clear from the mere fact that energy scales as the length of the ring (up to logarithmic corrections) while the momentum scales as the length squared, so that a decay into arbitrarily small rings with the net
line length conserved would result in vanishing total momentum.

%%%%%%%%%%%%%%%%%%%%%%%%%%%%%%%%%
\subsection{Generation of Kelvin waves in the process of line reconnection}
With a slight modification of the time dependence of the phase, the standard self-similar solution of the linear Schr\"odinger's equation---its Green's function---applies also to the non-linear equation (\ref{NLSE}). Indeed, the  absolute value of the solution is spatially homogeneous, so that  the non-linearity is equivalent to a spatially homogeneous time dependent external potential that can be immediately absorbed into the phase as an extra time-dependent term. The result is
\begin{equation}
\psi(\xi,t)={A\over \sqrt{t}}  \exp \left[ i\left( {\xi^2\over 4t}+ {A^2\over 2}\ln |t|\right) \right] .
 \label{self-sim}
\end{equation}
In accordance with (\ref{Hasimoto_repr}), this solution implies
\begin{equation}
\xi=A/\sqrt{t}  ,
 \label{xi}
\end{equation}
\begin{equation}
\tau=\xi /2t  .
 \label{tau}
\end{equation}
The physical meaning of the solution (\ref{xi})-(\ref{tau}), first revealed by Buttke \cite{Buttke}, is the relaxation of the
vortex angle, the value of which is controlled by the parameter $A$. This solution gives an accurate description (within LIA) of relaxation of two vortex lines after their reconnection, as long as the curvature  in the relaxation region  remains much larger than the curvatures
of the two lines away from the  region, so that the distant parts of the two lines can be approximately treated as
straight lines. By dimensional argument, the self-similarity regime should be achieved very rapidly upon the reconnection,
with the velocity of propagation of the fastest Kelvin waves (with wave vectors $\sim a_0$).

With the solution (\ref{xi})-(\ref{tau}) one can explicitly see that reconnections lift the integrability constraints. At $\xi\to \infty$ we have $\varphi_n \sim \xi^n$, meaning that a single reconnection renders all the integrals $I_n$ (\ref{constants}) divergent.

Let us look at the asymptotic form of the solution (\ref{xi})-(\ref{tau}) in Cartesian coordinates, taking the direction of the $z$-axis along one of the two lines, $z\to +\infty$ corresponding to the asymptotic limit:
\begin{equation}
x(z,t)+iy(z,t) = (4At^{3/2}/ z^2) {\rm e}^{iz^2/4t}, ~~~~z\gg \sqrt{t} .
 \label{asym}
\end{equation}
 Equation (\ref{asym}) reveals a helical Kelvin-wave structure moving away from the reconnection region. It is important that while arbitrarily small wavelengths are present in the solution (\ref{asym}), the integral for the total line lengths comes
 from the largest length scale $z\sim \sqrt{t}$. This provides  a support for two crucial points of the scenario of self-reconnectionsdriven Kelvin-wave cascade: (i) Reconnections push Kelvin waves to smaller lengthscales, (ii) apart from higher-order corrections, the line length associated with the curvature radius $R$ cannot go directly to the scales of curvature radius much smaller than $R$ (locality of the cascade in the wavenumber space).

When the angle between the two lines is close to $\pi$, the reconnection leads to a production of vortex rings \cite{Buttke}.  This process does not introduce a new  cascade channel. Qualitatively it is very similar to the pure production of Kelvin waves (\ref{asym}), because, up to higher-order corrections, the line length is being transferred to the
rings of the radii of the order of the curvature radii of reconnecting lines, i.e. to  adjacent scales in the kelvon wavenumber space.

%%%%%%%%%%%%%%%%%%%%%%%%%%%%%%%%%%
\subsection{Fractalization of lines. Kelvin-wave spectrum}
In this subsection we render the analysis of Ref.~\cite{Sv95} of the fractalization of a vortex line
by the self-crossings driven Kelvin-wave cascade. We start with introducing the crucial notion of a smoothed line length, ${\cal L}(\lambda_1)$, which is the length of a (fractalized) line upon smoothing out all the structures of the length scales smaller than  $\lambda_1$. Corresponding mathematical expression reads
\begin{equation}
\ln {\cal L}(\lambda_1) \sim \ln {\cal L}(\lambda_0) + \int_{\lambda_1}^{\lambda_0} (b_{\lambda}/\lambda)^2 d\lambda/\lambda ,
 \label{smooth}
\end{equation}
where $b_{\lambda}$ is the characteristic amplitude of the Kelvin-wave structure at the wavelength $\lambda$, while $\lambda_0$ is, generally speaking,  {\it any} fixed wavelength scale significantly larger than $\lambda_1$. In particular,
if $\lambda_0$  is  the largest wavelength of the problem, then ${\cal L}(\lambda_0)$ is the length of maximally smoothed vortex line. With  ${\cal L}(\lambda)$ we can estimate the number of local self-crossings at the scale $\lambda$ per unit time as
\begin{equation}
N_\lambda \sim {{\cal L}(\lambda) \over \lambda} \Omega (\lambda)\,  \omega_\lambda ,
 \label{N_self_cros}
\end{equation}
where the factor ${\cal L}(\lambda)/ \lambda$ gives the number of statistically independent pieces of the line (at the wavelength scale $\lambda$ typical correlation length is $\sim \lambda$), the factor  $\Omega (\lambda)$ is the probability to have a large enough amplitude $ \sim \lambda$ to produce a self-crossing within a given element of the line with the length  $\sim \lambda$, and $\omega_\lambda$ is the kelvon frequency playing the role of inverse correlation time.

To estimate $\Omega (\lambda)$, we can rely on the theoretical limit of $b_\lambda \ll \lambda$, in which kelvons are independent harmonic modes and, correspondingly, the statistics of  fluctuations of the amplitude is Gaussian. This readily yields
\begin{equation}
\Omega(\lambda) \sim (b_\lambda / \lambda) {\rm e}^{-(\lambda/b_\lambda)^2} .
 \label{prob_ampl}
\end{equation}
We do not introduce any dimensionless factor in the Gaussian exponent in view of the freedom of defining $b_\lambda$ up to a factor of order unity.

For any cascade the fundamental notion is the flux of corresponding conserved quantity. In our case, it is the flux of the vortex line length. The fractalization of the lines introduces certain subtleties. In contrast to a standard cascade in which the integral for the conserved quantity comes from a single length scale, in our case the line length is spread over all  scales of distance of the inertial range. Moreover, different wavelength scales  are not {\it entirely} independent in the sense that with the fractalized lines the short-range structures with their energy are slaved by the long-wave structures.
So that the line length coming from long waves to shorter ones is essentially carried by short-ranged structures slaved by the long-wave modes. The crucial observation now is that with respect to larger wavelengths the slaved shorter wavelengths play a rather passive role in energy balance, since short-wave contribution to line length carried by the longer waves is just proportional to the smoothed line length. Correspondingly, one can speak of the {\it smoothed}
line length flux, $Q(\lambda_1)$, where $\lambda_1$ is the smoothing parameter of  Eq.~(\ref{smooth}). By  definition of the cascade, the quantity $Q$ is one and the same for any wavelength scale $\lambda$, as long as  $\lambda \gg \lambda_1$.

Having fixed some wavelength scale $\lambda $, we note that the self-reconnections scenario implies
\begin{equation}
Q(\lambda_1) \sim N_\lambda {\cal R}_\lambda (\lambda_1),
 \label{Q_est}
\end{equation}
where $ {\cal R}_\lambda (\lambda_1)$ is the length of a $\lambda_1$-smoothed circle of the radius $\sim \lambda$.
In a direct analogy with (\ref{smooth}), we have
\begin{equation}
\ln {\cal R}_\lambda (\lambda_1) \sim \ln \lambda + \int_{\lambda_1}^{\lambda} (b_{\lambda'}/\lambda')^2 d\lambda'/\lambda' .
 \label{smooth_circ}
\end{equation}
From (\ref{smooth_circ}) and (\ref{smooth}) the follows a useful relation
\begin{equation}
{\cal L}(\lambda) {\cal R}_\lambda (\lambda_1) \sim  \lambda {\cal L}(\lambda_1)  .
 \label{useful}
\end{equation}
With $N_\lambda$ (\ref{N_self_cros}) , $\Omega (\lambda)$ (\ref{prob_ampl}),  the estimate $\omega_\lambda \sim \beta/\lambda^2$, and the relation (\ref{useful}), equation (\ref{Q_est}) yields
\begin{equation}
Q(\lambda_1) / {\cal L}(\lambda_1) \sim (\beta b_\lambda / \lambda^3) {\rm e}^{-(\lambda/b_\lambda)^2}  .
 \label{main_step}
\end{equation}
The left-hand side of this relation is a function of  $\lambda_1$, while the right-hand side is a function of $\lambda$, meaning that both sides are actually constants.  For the right-hand side this implies
\begin{equation}
  (b_\lambda / \lambda)^2 \sim {( b_{\lambda_0} / \lambda_0)^2 \over 1 + ( b_{\lambda_0} / \lambda_0)^2\ln (\lambda_0/\lambda )} .
 \label{amp_spec}
\end{equation}
(Since up to logarithmic corrections we have $b_\lambda \sim  \lambda$, we do not distinguish between $\ln b_\lambda$
and $\ln \lambda$.)
Expression (\ref{amp_spec}) yields the Kelvin-wave cascade spectrum in terms of the characteristic amplitude $b_\lambda$.
As far as the function ${\cal L}(\lambda)$ is concerned, from (\ref{amp_spec}) and (\ref{smooth}) we find
\begin{equation}
{\cal L}(\lambda) = {\cal L}(\lambda_0)  \left[ 1 + \left({b_{\lambda_0}\over \lambda_0} \right)^2 \ln {\lambda_0\over \lambda} \right]^{\nu},
 \label{length_spec}
\end{equation}
where $\nu$ is a constant of order unity the particular value of which cannot be established by our order-of-magnitude analysis.

It is clear from Eq.~(\ref{spect2}) that no matter how large the energy flux
(per unit vortex-line length) $\theta \propto Q$ transported by the
self-crossings regime is, at sufficiently high wavenumbers the pure Kelvin-wave
cascade will be capable of supporting it. As soon as the purely non-linear
kinetics become appreciable, the amplitudes of Kelvin waves must become smaller,
which in view of Eq.~(\ref{prob_ampl}), inhibits the reconnections. Thus, the
self-crossings-driven regime will inevitably be replaced by the pure Kelvin-wave
cascade at some scale $\lambda_*$. Note, however, that due to a small difference
between the Kelvin-wave spectra in the two regimes, Eqs.~(\ref{spect2}),
(\ref{amp_spec}), the crossover between them is likely to be extended in the
wavenumber space. A rough estimate of the scale $\lambda_*$ can be obtained by
setting $b_k \sim k^{-1} \sim \lambda_*$ in Eq.~(\ref{spect2}). The result
clearly depends on the cascade energy flux $\theta$, which is specific to the
physics at the energy-containing scale. In quasi-classical turbulence, $\theta$
is related to the Kolmogorov energy flux, in which case an estimate for
$\lambda_*$ will be obtained in Sec.~\ref{sec:crossover}. In the case of a
non-structured tangle, the energy flux is formed by reconnections at the scale
of interline separation $l_0$, $\theta_\mathrm{ns} \sim \kappa^3\rho
\Lambda^2/l_0^2$, with $\Lambda=\ln(l_0/a_*)$, which gives
\begin{equation}
\lambda_* \; \sim \; l_0\,/\,\Lambda \;\;\;\;\;\; \rm{(non-structured~
tangles).} \label{lambda_*_ns}
\end{equation}
According to this rough estimate, the inertial range for the regime driven by
local self-crossings is only about $\Lambda/10$ decades, which for realistic
values of $\Lambda$ could turn out to be an insignificant range without a
distinct spectral signature. Thus, it would be crucial to quantify the crossover
between the regimes by a direct numeric simulation.
%%%%%%%%%%%%%%%%%%%%%%%%%%%%%%%%%%
\subsection{Qualitative Hamiltonian model}
The Hamiltonian description (\ref{eq_motion}) implies single-valuedness of the function $w(z)$. Amazingly, the question of what happens to the mathematical solution of Eq.~(\ref{eq_motion}) when the
 physical function $w(z)$ becomes non-single valued turns out to be very relevant to the theory of self-crossings driven cascade. Clearly enough, the mathematical solution has to develop a certain
 singularity, and,  strictly speaking, become ill defined afterwards. However, if one introduces a discretized analog of the problem described by the Hamiltonian \cite{Sv95} (we confine ourselves to LIA),
 \begin{equation}
i\dot{w}_n={\partial H\over \partial w_n^*}, ~~~~~H=\sum_{n=0}^{N-1} \sqrt{1+|w_{n+1}-w_n|^2} ,
\label{discrete}
\end{equation}
the Hamiltonian dynamics remains well-defined at any time moment.
Numerical analysis of this model \cite{Sv95} shown
that that the above-mentioned singularity evolves into a finite jump between the values of $w_j$ and $w_{j+1}$, at a certain $j$. The jump exists for a certain time, and then relaxes, the relaxation process being qualitatively similar to the process of vortex angle evolution in the sense that Kelvin waves are being emitted, while the integrability constraint is naturally lifted by non-smothness of the function. With the precise geometric meaning of the Hamiltonian (\ref{discrete})---the length of the broken line defined by the points $\{ w_n\}$---we realize that the dynamics governed by it should be qualitatively equivalent to the self-reconnection induced cascade, leading to the fractalization of the line necessary to support jumps of arbitrarily small amplitude, the amplitude of the jump playing qualitatively the same role as the radius of the ring in the self-crossings driven cascade. And that is precisely what has been revealed by numeric simulation of the model (\ref{discrete}) in Ref.~\cite{Sv95}. The simulations also revealed the spectrum $b_\lambda \sim \lambda$,
consistent with Eq.~(\ref{amp_spec}),  the logarithmic factor going beyond numeric resolution.

\section{Crossover from   Richardson-Kolmogorov to
Kelvin-wave cascade}
\label{sec:crossover}

\subsection{Quasi-classical tangles at T=0: crossover to the quantized regime}

Even at absolute zero temperature the superfluid dynamics supports a turbulent regime, which under certain conditions is indistinguishable from classical turbulence \cite{Vinen06}. That may seem quite surprising since the only degrees of freedom in a superfluid at $T=0$ are quantized vortex lines, which are singular topological objects and thus are very different from the classical eddies responsible for turbulence in classical ideal incompressible fluids. Nonetheless, quantized vortex lines possess a mechanism that allows them to mimic classical vorticity---it is well known \cite{Donnelly} that macroscopic velocity profile of a rapidly rotated superfluid mimics solid-body rotation, which is accomplished
by formation of a dense array of vortex lines aligned along the
rotation axis. The basis of this mechanism is the strong coupling between the vortex lines in the dense array, which makes such a bundle behave as a single coherent classical object. Therefore, by essentially classical turbulence generation methods (i.e. ``stirring'') one can produce vorticity in the \textit{course-grained} up to length scales larger than the typical interline separation $l_0$ superfluid velocity field, indistinguishable from that of a normal fluid, the underlying vortex
tangle being organized in polarized ``bundles'' of vortex lines.

Over the last decade, experimental observations of the classical behavior exhibited by superfluids \cite{Maurer,Stalp, Oregon, Ladik, Bradley,Golov, Golov2} have largely led to a renaissance of general interest in superfluid turbulence. Perhaps the most attractive feature of this quasi-classical turbulence is that, unlike counterflow turbulence, it in principle allows generation and probing at temperatures close to absolute zero. [It is only very recently that a unique technique of non-structured tangle (\`{a} la counterflow turbulence) production at very low temperatures was developed by the Manchester group \cite{Golov2}.] With recent technological advances this opens an intriguing possibility of studying such essentially low-temperature phenomena as, e.g., the Kelvin-wave cascades.

However, it was recently realized \cite{Lvov} that, at $T=0$, the question of how the quasi-classical vortex tangle looks like when one zooms in down to scales of order $l_0$, where the vorticity is essentially discrete, is quite a puzzling problem. The only fact that is immediately clear is that near the scales $\lambda_\mathrm{ph} \ll l_0$ where the dissipation due to the sound radiation takes place, the energy flux must be transported by the pure Kelvin-wave cascade. What happens in the intermediate regime between the classical Kolmogorov cascade of eddies and the Kelvin-wave cascade on individual vortex lines is the subject of this section.

In their scenario,  L'vov, Nazarenko, and Rudenko \cite{Lvov}, noted that a simple picture in which the pure Kelvin-wave cascade supersedes the Kolmogorov regime at the scale $l_0$ is not possible. The difficulty is due to the fact that at this scale the pure Kelvin-wave cascade is unable to sustain the Kolmogorov energy flux (per unit mass of the fluid) $\varepsilon$. Correspondingly, L'vov, Nazarenko, and Rudenko put forward an idea of bottleneck accumulation of energy at the classical scales adjacent to $l_0$. The accumulation of energy in the form of a thermalized distribution of quasi-classical vorticity was suggested to be necessary to raise the level of turbulence to a value at which the pure Kelvin-wave cascade becomes efficient. Note, however, that the concept of bottleneck at a given scale fundamentally relies on the absence of \textit{any} efficient energy transport mechanism at this scale. Due to this fact, under the conditions of Ref.~\cite{Lvov}, vortex-line reconnections play a fatal role for the bottleneck scenario. Indeed, if the Kolmogorov cascade can reach the scale $l_0$ as assumed in Ref.~\cite{Lvov}, the typical vortex-line curvature at this scale is of the order $l_0$ meaning that the vortex-line reconnections must happen due to the tangle geometry, which makes them an alternative energy transport channel to the pure Kelvin-wave cascade. Estimating the energy flux $\varepsilon_{\rm rec}$ processed by the reconnections at this scale we see that $\varepsilon_{\rm rec}/\varepsilon \sim \Lambda^2 \gg 1$, i.e. we are actually dealing with an ``anti-bottleneck''---the reconnections transport an energy flux much larger than the one supplied from the larger scales. The anti-bottleneck is of course forbidden by the energy conservation, so we are forced to conclude that the transformation of the classical regime must happen already before the scale $l_0$ is reached, and the reconnections play a crucial role in this process.

Let us first analyze the tangle of vortex lines at $T=0$ at the length scales much larger than the typical interline separation under the condition of a developed Kolmogorov cascade. Although understood intuitively, the ability of the system of quantized vortex lines to mimic the classical Kolmogorov cascade is not trivial. We start by a rigorous demonstration of this fact.

At $T=0$, vortex lines are the only degrees of freedom and their dynamics are completely captured by the Biot-Savart equation (\ref{Biot-Savart}). The latter can be rewritten
in classical terms of vorticity $\mathbf{w}= \mathrm{curl} \,
\mathbf{v}$ in the momentum space, $\mathbf{w}_\mathbf{k}=\int
\mathbf{w}(\mathbf{r}) \exp[-i\mathbf{k}\cdot\mathbf{r}] \,
\mathrm{d}^3r = \kappa \int \exp[-i\mathbf{k}\cdot\mathbf{s}] \,
\mathrm{d} \mathbf{s} $. The result is \textit{identical} to the vorticity
equation for a normal ideal incompressible fluid:
\begin{equation}
\frac{\partial{\mathbf{w}}_\mathbf{k}}{\partial t} \, =\, \mathbf{k}
\times \int \frac{\mathrm{d}^3q}{(2\pi)^3} \, q^{-2}  \, \Bigl[ \,
\mathbf{w}_{\mathbf{k}-\mathbf{q}} \times [ \,
\mathbf{w}_{\mathbf{q}} \times \mathbf{q} \, ] \, \Bigr] \,  .
\label{Euler}
\end{equation}
In view of Eq.~(\ref{Euler}) we can formulate the conditions, under which the vortex tangle must be automatically equivalent to classical ideal-incompressible-fluid turbulence: (i) the energy must be concentrated at a sufficiently small wavenumber scale $k_{\mathrm{en}} \ll l_0^{-1}$, and (ii) the decay scenario must be local in the momentum space, so that the quantized nature of vorticity is irrelevant for the long-wavelength behavior. These conditions are not restrictive: (i) is automatically satisfied if turbulence is generated by classical means due to the large compared to $\kappa$ values of the velocity circulation, and (ii) is necessary for the existence of the Kolmogorov cascade in classical fluids as well.

Note that the circulation quantum $\kappa$ completely drops out of the vorticity equation. This is a manifestation of the known fact that the superfluid hydrodynamics is completely described by the classical Euler equation with respect to which the quantization of circulation is nothing but an imposed initial condition enforced by quantum mechanics---due to the Kelvin theorem, once preformed the velocity circulation is a constant of motion.
Since the dynamics of each individual vortex line are controlled by the circulation quantum $\kappa$, the independence of Eq.~(\ref{Euler}) on $\kappa$ leads to an important conclusion that any large-scale (classical) motion necessarily implies strong coupling of the underlying vortex lines and that the crossover to the quantized regime is due to the self-induced motion of the vortex lines starting to dominate over the inter-line coupling. To obtain the corresponding crossover scale $r_0$, let us formally decompose the integral (\ref{BS}) into the self-induced part,
$\mathbf{v}^\mathrm{SI}(\mathbf{s})$, for which the integration is
restricted to the vortex line containing the element $\mathbf{s}$,
and the remaining contribution induced by all the other lines
$\mathbf{v}^\mathrm{I}(\mathbf{s})$,
\begin{equation}
\mathbf{v}(\mathbf{s})=\mathbf{v}^\mathrm{SI}(\mathbf{s})+\mathbf{v}^\mathrm{I}(\mathbf{s}).
\label{v-decomposition}
\end{equation}

By the definition of $r_0$, at length scales $r \gg r_0$ the turbulence mimics classical
vorticity taking on the form of a dense coherently moving array
of vortex lines bent at a curvature radius of order $r$. The velocity
field of this configuration obeys the Kolmogorov law
\begin{equation}
v_r \sim (\varepsilon r)^{1/3}, ~~~~~~~~~ r \gg r_0\, ,
\label{Kolmogorov}
\end{equation}
where $\varepsilon$ is the energy flux per unit mass of the fluid formed at the energy-containing eddies and transferred by the cascade.
Here and below the subscript $r$ means typical variation of a
field over a distance $\sim r$. On the other hand, the value of
$v_r$ is fixed by the quantization of velocity circulation around
a contour of radius $r$, namely $v_r r \sim \kappa n_r r^2$, where
$n_r$ is the areal density of vortex lines responsible for the
vorticity at the scale $r$. Note that scale invariance requires
that on top of vorticity at the scale $r$ there be a fine
structure of vortex bundles of smaller sizes, so that,
mathematically, $n_r r^2$ is the difference between large numbers
of vortex lines crossing the area of the contour $r$ in opposite
directions. The quantity $n_r$ is related to the flux by
\begin{equation}
n_r \sim \left[ \frac{\varepsilon}{\, \kappa^3 \; r^2}
\right]^{1/3}\, , ~~~~~~ r \gg r_0\, . \label{n_r}
\end{equation}

The underlying dynamics of a single vortex line in the bundle is
governed by $v^\mathrm{I}_r$ and $v^\mathrm{SI}_r$. While by its
definition $v^{I}_r \sim v_r$, which is given by
Eq.~(\ref{Kolmogorov}), the self-induced part is determined by the
curvature radius $r$ of the vortex line according to the LIA,
Eq.~(\ref{LIA}),
\begin{equation}
v^\mathrm{SI}_r \sim \Lambda_r \frac{\kappa }{r} , \label{v-SI_r}
\end{equation}
where $\Lambda_r=\ln(r/a_*)$. Here and throughout this section, the logarithmic accuracy will be sufficient for the analysis, and, correspondingly, we replace $\Lambda_r$ with $\Lambda=\ln(l_0/a_0)$. At length scales where $v^\mathrm{I}_r \gg v^\mathrm{SI}_r$, the
vortex lines in the bundle move coherently with the same velocity
$\sim v^\mathrm{I}_r$. However, at the scale $r_0 \sim (\Lambda^3
\kappa^3/\varepsilon)^{1/4}$, the self-induced motion of the vortex
line becomes comparable to the collective motion, $v^\mathrm{SI}_r
\sim v^\mathrm{I}_r $. At this scale, individual vortex lines start
to behave independently from each other and thus $r_0$ gives the
lower cutoff of the inertial region of the Kolmogorov spectrum
(\ref{Kolmogorov}).

Since $r_0$ is the size of the smallest classical eddies, the
areal density of the vortex lines at this scale is given by the
typical interline separation, $n_{r_0} \sim 1/l_0^2$. In other words, vortex bundles at the scale $r_0$ consist of almost parallel vortex lines separated by the spacing $l_0$. With Eq.~(\ref{n_r}), we arrive at
\begin{equation}
r_0 \sim \Lambda^{1/2} l_0\, ,\label{r_0-l_0}
\end{equation}
\begin{equation}
l_0 \sim ( \Lambda \kappa^3/\varepsilon)^{1/4} \; . \label{l_0}
\end{equation}

The crossover can be also understood in slightly more visual terms. Let us introduce an effective number of vortex lines $N_r$ in a bundle of size $r$. This number is obtained as an algebraic sum of the number of lines going through the bundle cross-section in opposite directions and is given by $N_r=n_r r^2 \sim \varepsilon^{1/3} r^{4/3} / \kappa$. In view of Eqs.~(\ref{Kolmogorov}), (\ref{v-SI_r}), the number of lines in a bundle relative to $\Lambda$ determines whether it behaves as a classical eddy or a set of independent vortex lines: $N_r \gg \Lambda$ means that the coupling between the vortex lines dominates resulting in the crossover when $N_{r_0} \sim \Lambda$. Thus, in the theoretical limit of $\Lambda \gg 1$ the bundles still contain a large number of vortex lines at the scale where the classical regime breaks down.

In the next subsection we shall introduce the cascade mechanism that supersedes the Kolmogorov cascade of eddies at the scales immediately adjacent to $r_0$.

\subsection{Reconnections of  bundles}

The analysis presented in this and the following section relies on the fact that the number of vortex lines in a bundle at the crossover scale is large, $N_{r_0} \gg 1$, which, in view of Eq.~(\ref{r_0-l_0}), is guaranteed in the limit of large $\Lambda$.

At the scale $r_0$, turbulence consists of randomly oriented
vortex-line bundles of size $r_0$ formed by the classical regime. The length $r_0$ plays the role of a
correlation radius in the sense that relative orientation of two
vortex lines (with the short-wavelength structure smoothed out) becomes uncorrelated only if they are a distance $\gtrsim r_0$ apart. On the other hand, the crossover to the quantized regime means that each line starts moving according to
its geometric shape, as prescribed by Eq.~(\ref{LIA}). Therefore,
reconnections, at least between separate bundles, are inevitable
and, as we show below, capable of sustaining the flux
$\varepsilon$.

The quantity that will play an important role in the analysis is the energy
transferred to a lower scale after one reconnection of vortex lines
at the scale $k^{-1}$, which, following Ref.~\cite{Sv95}, can be
written as
\begin{equation}
\epsilon_k \sim f(\gamma) \, \Lambda \; \rho \; \kappa^2 k^{-1} \,
. \label{energy-reconn}
\end{equation}
Here, $f(\gamma)$ is a dimensionless function of the angle
$\gamma$ at which the vortex lines cross ($\gamma=0$
corresponds to parallel lines). Its asymptotic form is
\begin{equation}
f (\gamma) \sim \gamma^{2}, ~~~~~~~~ \gamma \ll 1 \, .
\label{f_gamma}
\end{equation}

Although at the scale $r_0$ there is already no coupling between
vortex lines to stabilize the bundles, they should still move
coherently on the time scale of their turnover time since, by the definition of the bundle size $r_0$, geometry of neighboring lines at this scale is essentially the same over distances $\lesssim r_0$. On the other hand, during about one turnover the bundle must cross a neighboring bundle and reconnect providing a mechanism of energy transfer to the lower scales. It is possible, however, that vortex lines \textit{within} the bundle reconnect. One can show that such processes can not lead to any significant redistribution of
energy at the scale $r_0$ (but they will play an important role at smaller scales) and thus to a deformation of the bundle at this scale
because they happen at small angles so that the energy
(\ref{energy-reconn}) is too small. Indeed, the dimensional upper
bound on the rate at which two lines at a distance $l \ll r_0$ can
cross each other is, from Eq.~(\ref{LIA}), $\Lambda \kappa/r_0 l$,
while the actual value should be much smaller due to the strong
correlations between line geometries. Taking into account that the
number of lines in the bundle is $(r_0/l_0)^2$ and that $\gamma
\sim l/r_0$, the contribution to the energy flux from these
processes is bounded by $(l/r_0)\varepsilon $. Only when $l \sim r_0$ the reconnections become important, which are the reconnections between the whole bundles of size $r_0$.

Crossing of the bundles results in reconnections between all their
vortex lines and Kelvin waves with a smaller but adjacent wavelength
$\lambda$ are generated. This picture of bundle crossing was recently corroborated by direct numerical simulations \cite{Barenghi}. The coherence of the initial bundles
implies that the waves on different vortex lines of the same bundle must be generated
coherently. Thus, at the scale $k^{-1} \lesssim r_0$, adjacent vortex
lines should still be almost parallel with Kelvin waves on them of the wavenumber $k$ and amplitudes $b_k$---vortex lines at the scale $k^{-1}$ also form bundles. Similarly, these bundles can reconnect transporting the energy to a lower scale, where similar bundle reconnections happen, and so on down the scales until the bundle size is comparable to the interline distance and the self-similar regime is cut off.

An important ingredient of this scenario characterizing the bundles at a scale $k^{-1}$ is the correlation radius $r^{(c)}_{k}$ of the vortex-line geometry (with the short-wavelength structure smoothed out up to $k^{-1}$) in the transverse to the bundle direction. The value of $r^{(c)}_{k}$ determines the distance over which neighboring vortex lines can be considered as parallel, i.e. it gives the transverse size of the bundle. This size is due to a finite time required for the Kelvin-wave amplitude at the scale $k_2^{-1}$ to build up (after a reconnection at a larger scale $k_1^{-1}>k_2^{-1}$), which is of order of the wave turnover time, $\tau_{k_2} \sim 1/\kappa\Lambda k_2^{2}$. The Kelvin waves that were generated within the time $\sim \tau_{k_2}$ are coherent. In other words, as a reconnection of two lines happens at time $t=\tau_{k_2}$ at the scale $k_1^{-1}>k_2^{-1}$, the lines that reconnected in the same bundle at $t=0$ have already-developed waves, which can not be coherent with the waves about to be generated at $t \gtrsim \tau_{k_2}$. Thus, the distance traveled by the bundle at the scale $k_1^{-1}$ over the time $\sim \tau_{k_2}$ determines the orientational correlation radius at the scale $k_2$, $r^{(c)}_{k_2}=r^{(c)}_{k_2}(k_1)= b_{k_1} k_1^2/k_2^2$. Since the scales $k_1$ and $k_2$ are actually adjacent (i.e. different only by a factor of order unity), we finally get $r^{(c)}_{k} \sim b_k$.

The spectrum of Kelvin waves $b_k$ in this regime can be obtained from
the condition $\tilde{\varepsilon}_k \equiv \varepsilon$, where
$\tilde{\varepsilon}_k$ is the energy flux per unit mass
transported by the reconnections at the scale $k^{-1}$,
\begin{equation}
\tilde{\varepsilon}_{k} \sim \bigl(k/\rho \, [r^{(c)}_k]^2 \bigr) \; N_{k} \; \epsilon_{k} \;
\tau_{k}^{-1}. \label{flux-k}
\end{equation}
Here, we take into account that the correlation volume of the reconnection is
$[r^{(c)}_{k}]^2/{k}$, $N_{k} \sim (b_{k}/l_0)^2$ is the number of vortex
lines participating in the reconnection, and $\tau_{k}^{-1} \sim \kappa \Lambda  k^2$ is
the rate at which the bundles cross. Physically, $b_{k}$ determines
the typical crossing angle, $\gamma \sim b_{k} k$, thereby
controlling the energy lost in one reconnection. From Eq.~(\ref{flux-k}), the spectrum of Kelvin waves in the bundle-crossing regime has the form
\begin{equation}
b_k \sim r_0^{-1} k^{-2}. \label{spectrum-bundles}
\end{equation}

At the wavelength $\sim \lambda_\mathrm{b}=\Lambda^{1/4} l_0$, the
amplitudes become of order of the interline separation $b_k \sim
l_0$ and the notion of bundles looses meaning---the cascade of bundles is cut off.
The scenario at the scales $\lambda \lesssim \lambda_\mathrm{b}$ is rather peculiar and the next subsection is devoted to its description.

\subsection{Reconnections of adjacent lines}

In the regime of bundle crossings, in view of Eq.~(\ref{spectrum-bundles}), the Kelvin-wave amplitudes are steeply decreasing with the wavenumber, so that at the wavelength $\lambda_\mathrm{b}=\Lambda^{1/4} l_0$, where the amplitudes become of order of the interline spacing $l_0$, the vortex lines are only slightly bent, $b_k k \ll 1$. This poses an interesting question of what is driving the cascade at the wavelengths below $\lambda_\mathrm{b}$. Since the amplitude spectrum can not change discontinuously, at the scales adjacent to $\lambda_\mathrm{b}$ we must still have $b_k k \ll 1$, so that the mechanism of self-reconnections is strongly
suppressed. On the other hand, the kinetic times of the purely non-linear regime are still too long to carry the flux $\varepsilon$ \cite{KS_04}. We thus conclude that in some range of length scales $\lambda_\mathrm{c} \lesssim \lambda\lesssim \lambda_b$ there should take place a build up of the relative Kelvin-wave amplitude supported by the energy flux from the scale $\lambda_\mathrm{b}$ until $b_k k \sim 1$ and the self-reconnections can take over the cascade, the amplitudes $b_k$ being defined by the condition of constant energy
flux per unit length and the crossover scale $\lambda_\mathrm{c}$ being
associated with the condition $b_{k\sim 1/\lambda_\mathrm{c}} \sim
\lambda_\mathrm{c}$. The observation crucial for
understanding the particular mechanism of this regime and thus
finding $b_k$ is that each nearest-neighbor reconnection
(happening at the rate $\propto \Lambda/\lambda_b^2$ per each line
element of the length $\sim \lambda_b$) performs a sort of {\it
parallel processing} of the energy distribution for {\it each} of the
wavelength scales in the range $[\lambda_\mathrm{c},\, \lambda_b]$. For the
given wavelength scale $\lambda \sim k^{-1}$, the energy transferred
by a single collision is $\propto \Lambda (b_k k)^2 \lambda$, and
with the above estimate of the collision rate per the length
$\lambda_b$, this readily yields the spectrum
\begin{equation}
b_k\sim l_0 (\lambda_b k)^{-1/2}, \label{spectrum-nncross}
\end{equation}
and, correspondingly, $\lambda_\mathrm{c} \sim l_0/\Lambda^{1/4}$. Note that the rise of the relative amplitude $b_k k \propto k^{1/2}$ implied by Eq.~(\ref{spectrum-nncross}) describes the process of fractalization of the vortex lines with its culmination around the scale $\lambda_\mathrm{c}$. Thus, in this regime, the energy is actually contained at the low-end of the inertial range. We refer to this circumstance as a \textit{deposition} of energy and discuss the relevance of this term in Sec.~\ref{sec:conclusions}. Although the deposition of energy makes this regime formally different from a standard cascade setup, its presence has no effect on the rest of the inertial range and thus on the overall cascade efficiency.

\subsection{To the Kelvin-wave cascade}

The stages of vortex-bundle reconnections and the reconnections between adjacent vortex lines essentially complete the transformation of the quasi-classical tangle that is necessary to connect the classical regime to the Kelvin-wave cascades on individual vortex lines. At the wavelengths  $\sim \lambda_\mathrm{c}$ the self-reconnections on the vortex lines take over the cascade and the picture at the shorter scales is qualitatively not different from that of the non-structured tangles. The self-reconnection regime continues in the range $\lambda_* \ll \lambda \ll \lambda_c$ with the spectrum $b_k \sim k^{-1}$. At a sufficiently small wavelength $\lambda_*$, the strongly
turbulent cascade of Kelvin waves is replaced by the purely
non-linear cascade. The spectrum of Kelvin-wave amplitudes in the non-linear cascade is given by Eq.~(\ref{spect2}). The value of $\lambda_*$ can
be determined by matching the energy flux $\varepsilon$ with
$\Theta/\rho l_0^2$, where $b_{k \sim 1/\lambda_*} \sim k^{-1} \sim \lambda_*$. With
Eq.~(\ref{l_0}), we then obtain
\begin{equation}
\lambda_*=l_0/\Lambda^{1/2}. \label{lambda_*}
\end{equation}

At $T=0$ Kelvin waves decay emitting phonons.
This dissipation mechanism is negligibly weak all the way down to
wavelengths of order $\lambda_\mathbf{ph}$, where the rate of energy radiation by the vortex lines becomes comparable to $\varepsilon$. As we show in Sec.~\ref{sec:kelvon_phonon} this condition with the estimate for the power of sound radiation by kelvons yields
\begin{equation}
\lambda_\mathrm{ph} \, \sim \, \Lambda^{27/31} \left[\kappa/c\,l_0\right]^{25/31}l_0 . \label{lambda_ph}
\end{equation}
The scale $\lambda_\mathbf{ph} \lll \lambda_*$
gives the lower dissipative cutoff of the Kelvin-wave cascade.

\begin{figure}[htb]
\includegraphics[width = 0.8\columnwidth,keepaspectratio=true]{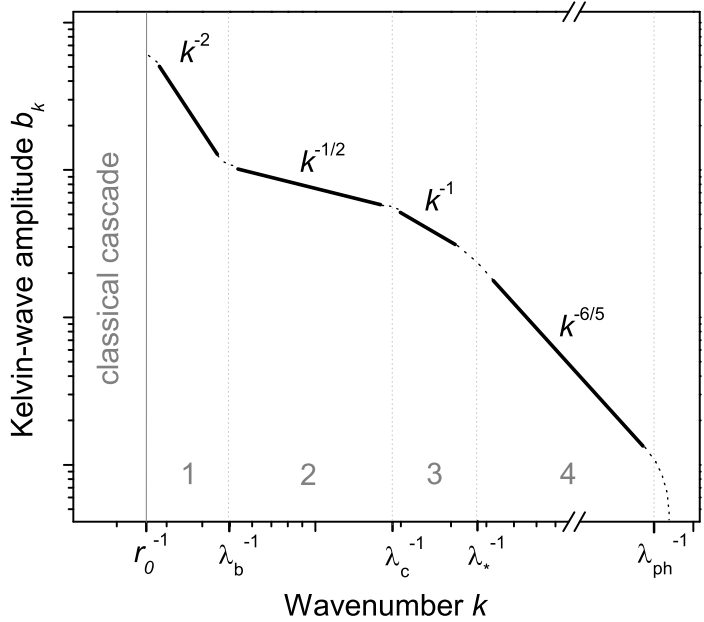}
\caption{Spectrum of Kelvin waves in the quantized regime of quasi-classical tangles. The
inertial range consists of a chain of cascades driven by different
mechanisms: (1) reconnections of vortex-line bundles, (2)
reconnections between nearest-neighbor vortex lines in a bundle,
(3) self-reconnections on single vortex lines, (4) non-linear
dynamics of single vortex lines without reconnections.}
\label{fig:spec}
\end{figure}

The overall decay scenario is summarized in Fig.~\ref{fig:spec}, where we plot the resulting Kelvin-wave spectrum in the whole quantized regime.

%%%%%%%%%%%%%%%%%%%%%%%%%%%%%%%% FIGURE spectrum %%%%%%%%%%%%%
%%%%%%%%%%%%%%%%%%%%%%%%%%%%%%%%%%%%%%%%%%%%%%%%%%%%%%%%%%%%%%%%

\subsection{Scanning by finite temperature}

In this subsection, on the basis of the presented decay scenario at $T=0$, we describe a theory of the low-temperature dissipative cutoff of the cascade \cite{KS_05_vortex_phonon}. We shall demonstrate that the temperature dependence of the total vortex-line length density $L$ comes from the dependence of the wavelength scale $\lambda_\mathrm{cuttoff}$, at which the cascade
ceases due to the mutual friction of vortex lines with the normal
component, on the dimensionless friction coefficient $\alpha
\propto T^5~~(T\to 0)$. The main result of the theory is the prediction for the function $\ln L(\ln \alpha)$, which turns out to have a very characteristic form reflecting the four qualitatively distinct wavenumber regions of the cascade in the quantized regime. The results of the theory are in an excellent agreement with the recent experiments by Walmsley {\it et al.} \cite{Golov, Golov2} giving a strong evidence for the highly nontrivial scenario of low-temperature turbulence decay.

If the vortex lines were smooth, the vortex-line density $L$ would
be simply related to the interline separation as $L = l_0^{-2}$. However, the presence of a fine wave structure on the lines can make the total length many times as large as $l_0^{-2}$ (or even infinite in the limit of fractal lines).
The increase of $L$ due to the presence of KWs is related to their spectrum by \cite{Sv95} (cf. Eq.~(\ref{smooth}))
\begin{equation}
\ln \left[L(\alpha)/L_0\right] \; = \;
\int_{\tilde{k}}^{k_\mathrm{cutoff}(\alpha)} (b_k k)^2 \; dk/k \;
. \label{line_density}
\end{equation}
Here $k_\mathrm{cutoff}\sim 1/\lambda_\mathrm{cutoff}$,
$\tilde{k}$ is the smallest wavenumber of the KW cascade (not to
be confused with the smallest wavenumber of the Kolmogorov
cascade) at which the concept of a definite cutoff scale is
meaningful, and $L_0$ is the ``background" line density
corresponding to $k_\mathrm{cutoff} \sim \tilde{k}$. There is an
ambiguity in the definition of $b_k$ associated with the choice of
the spectral width of the scale $k$, which is fixed in
Eq.~(\ref{line_density}) by setting the proportionality constant
on each side to unity.

At $T=0$, the cascade is cut off by the radiation of sound (at
least in $^4$He) at the length scale
$\lambda_\mathrm{cutoff} = \lambda_\mathrm{ph}$ given by Eq.~(\ref{lambda_ph}). Changing the temperature one controls $\lambda_\mathrm{cutoff}(T)
> \lambda_\mathrm{ph}$ in Eq.~(\ref{line_density}), which allows one
to \textit{scan} the KW cascade observing qualitative changes in
$L(T)$ as $\lambda_\mathrm{cutoff}$ traverses different cascade
regimes. The existence of a well-defined cutoff is due to the fact
that the cascade is supported by rare kinetic events in the sense
that the collision time $\tau_{\rm coll}\equiv \tau_{\rm coll}(\varepsilon, k)$ is
much larger than the KW oscillation period, $\tau_{\rm per}\equiv
\tau_{\rm per}(k)$. The dissipative time $\tau_{\rm dis}\equiv
\tau_{\rm dis}(\alpha, k)\sim \tau_{\rm per} /\alpha$, as we show
below, is the typical time of the frictional decay of a KW at the
scale $k$. Thus, the cutoff condition is
$\tau_{\rm dis}(\alpha, k)\, \sim \tau_{\rm coll}(\varepsilon, k)$,
which implies that the energy dissipation rate at a
given wavenumber scale becomes comparable to the energy being
transferred to higher wavenumber scales per unit time by the
cascade. It is this condition that defines the cutoff wavenumber
$k_{\rm cutoff}\equiv k_{\rm cutoff}(\varepsilon, \alpha)$. Decreasing
$T$ and thus $\alpha(T)$, one gradually increases $k_{\rm cutoff}$,
thereby scanning the cascade. In view of
Eq.~(\ref{line_density}), this in principle allows one to extract
the KW spectrum.

At finite $T$, dissipative dynamics of a vortex line element are
described by the equation (omitting the third term in the r.h.s.,
which is irrelevant for dissipation) \cite{Donnelly,Cambridge_workshop}
\begin{equation}
\dot{\mathbf{s}}=\mathbf{v}(\mathbf{s})+\alpha\mathbf{s}'\times\left[\mathbf{v}_\mathrm{n}(\mathbf{s})-\mathbf{v}(\mathbf{s})\right]\;
. \label{BS+friction}
\end{equation}
Here $\mathbf{v}(\mathbf{r})$ is the superfluid velocity field,
$\mathbf{v}_\mathrm{n}(\mathbf{r})$ is the normal velocity field,
$\textbf{s}=\textbf{s}(\xi, t)$ is the time-evolving radius-vector
of the vortex line element parameterized by the arc length, the
dot and the prime denote differentiation with respect to time and
the arc length, respectively.

At $\alpha \sim 1$ the superfluid and normal components are
strongly coupled and the KWs are suppressed. In this
case, the cascade must cease before it enters the quantized
regime, i.e.
$\lambda_\mathrm{cuttoff} \gtrsim r_0$. If, however, the mutual friction is small, then $\alpha^{-1} \gg
1$ gives the characteristic number of KW oscillations required for
the wave to decay. Indeed, to the first approximation in
$1/\Lambda$, $\mathbf{v}(\mathbf{s})$ in Eq.~(\ref{BS+friction})
is given by the local induction approximation (\ref{LIA}). In the range $\lambda \ll r_0$, which we will be interested in, the normal component is already laminar and thus the field $\mathbf{v}_\mathrm{n}$ has no structure at these small scales. In addition, as long as $\alpha \ll 1$, the disturbance of the normal component caused by vortex line motion as these scales
can be neglected. Therefore, $\mathbf{v}_\mathrm{n}(\mathbf{s})$ can be treated as a constant in Eq.~(\ref{BS+friction}), which makes it irrelevant for KW
dissipation. In this case, Eqs.~(\ref{BS+friction}), (\ref{LIA})
give the rate at which the amplitude $b_k$ of a KW with the
wavenumber $k$ decays due to the mutual friction,
\begin{equation}
\dot{b}_k \, \sim \, - \, \alpha \, \omega_k \, b_k, \label{KW_decay}
\end{equation}
while the KW dispersion is $\omega_k=(\kappa/4\pi)\Lambda
k^2$. Here and below we omit factors of order unity, which are
subject to the definition of the spectral width of the wave. These
factors can not be found within our theory, but can be extracted
from experimental data (as we do it below) and from numerical
simulations. Since the energy per unit line length associated with
the wave is $E_k \sim \kappa \rho \omega_k b_k^2$, the power
dissipated (per unit line length) at the scale $\sim k^{-1}$ is
given by
\begin{equation}
\Pi(k) \, \sim \, \alpha \, \kappa \, \rho \, \omega_k^2 \, b_k^2, \label{power}
\end{equation}
where $\rho$ is the fluid density. In the following, we analyze ST
as $\alpha(T)$ scans through the regimes summarized in Fig.~\ref{fig:spec}.

\textit{Regime (1).}---At this stage
the vortex lines are organized in bundles; the amplitudes of waves
on these lines are given by Eq.~(\ref{spectrum-bundles}). One can estimate the total power lost due to the friction
of these bundles at the wavenumber scale $r_0^{-1} \ll k \ll
k_\mathrm{b}\sim \lambda_\mathrm{b}^{-1}$ per unit mass of the fluid as
\begin{equation}
\varepsilon_\mathrm{dis}(k)\, =\, \frac{1}{\rho b_k^2} \;
\frac{b_k^2}{l_0^2} \; \Pi(k) \; \sim \; \alpha \, \kappa^3 \,
\Lambda \, / \, l_0^4 \; .\label{reg1}
\end{equation}
Here, the first factor in the r.h.s. is associated with the
correlation volume at this scale $\sim b_k^2 k^{-1}$
and the second one stands for the number of vortex lines in the
volume. Note that the dissipated power is constant at all the
length scales within this regime and, since the interline separation is related to the Kolmogorov flux by Eq.~(\ref{l_0}), it
is simply given by $\alpha \varepsilon$. Thus, when $\alpha \ll 1$
the kinetic channel in the whole regime (1) becomes efficient and
the cascade reaches the scale $\lambda_\mathrm{b}$, where the
notion of bundles becomes meaningless. That is, the regime (1), as
opposed to the regimes (2)-(4), is \textit{not} actually scanned
by $\alpha$---its inertial range develops as a whole while $\alpha$ evolves from $\alpha \sim 1$ to $\alpha \ll 1$. [In the purely theoretical
limit of exponentially large $\Lambda$, when the inertial range
of the regime (1) occupies many decades, a (loose) dissipative
cutoff appears in the regime (1) as well; the cutoff wavelength in
this case can be roughly estimated as $\ln (r_0/\lambda_{\rm
cutoff})\sim 1/\alpha$.]

\textit{Regime (2).}---In the range $k_\mathrm{b} \ll k \ll k_\mathrm{c}\sim \lambda_\mathrm{b}^{-1}$ ,
the spectrum of KWs is given by $b_k \sim
l_0 (k_\mathrm{b}/k)^{1/2}$ (Eq.~(\ref{spectrum-bundles})) which for the dissipated power yields
\begin{equation}
\varepsilon_\mathrm{dis}(k)  \sim \frac{1}{\rho l_0^2} \, \Pi(k)
\sim \alpha \kappa^3 \Lambda^{7/4}k^{3}/l_0\; . \label{reg2}
\end{equation}
The condition $\varepsilon_\mathrm{dis}(k) \sim \varepsilon$ gives
the cutoff wavenumber
\begin{equation}
k_\mathrm{cutoff} \; = \; \xi_2 \, \Lambda^{-1/4} \alpha^{-1/3} \,
(2\pi/l_0)\; , \;\;\; k_\mathrm{b} \ll k_\mathrm{cutoff} \ll k_\mathrm{c}\;
, \label{cutoff_reg2}
\end{equation}
where $\xi_2$ is some constant of order unity.  Then, from
Eq.~(\ref{line_density}) we obtain
\begin{equation}
\ln\frac{L(k_\mathrm{cutoff})}{L_0} = C^2 (k_\mathrm{b} l_0)^2
[k_\mathrm{cutoff}/k_\mathrm{b} - 1]\; .\label{L_reg2}
\end{equation}
Here, we set $b_k=C \, l_0 (k_\mathrm{b}/k)^{1/2}$, where C is a constant of order unity. The overall
magnitude of $b_k$ in the other regimes follows then from the
continuity.

\textit{Regime (3).}---In this regime, supported by
self-reconnections, the spectrum is given by $b_k \sim k^{-1}$ (up
to a logarithmic prefactor). The corresponding energy balance
condition yields the cutoff scale:
\begin{equation}
k_\mathrm{cutoff} \; = \; \xi_3 \,(\Lambda \alpha)^{-1/2}\,(2
\pi/l_0), \;\;\; k_\mathrm{c} \ll k_\mathrm{cutoff} \ll k_* \; .
\label{cutoff_reg3}
\end{equation}
With the logarithmic prefactor taken into account, Eq.~(\ref{amp_spec}),
the spectrum in this regime reads $b_k=C
[1+c_3^2\ln(k/k_\mathrm{c})]^{-1/2}( \sqrt{ k_\mathrm{c} k_\mathrm{b} }/k)l_0$,
where $c_3$ is a constant of order unity. Then the relative
increase of vortex line density through this regime is given by
\begin{equation}
\frac{L(k_\mathrm{cutoff})}{L(k_\mathrm{c})} \; =  \; \left[ \, 1 \; + \;
c_3^2 \; \ln\frac{k_\mathrm{cutoff}}{k_\mathrm{c}}\, \right]^{\nu}\; ,
\label{L_reg3}
\end{equation}
where $\nu = C^2 k_\mathrm{c} k_\mathrm{b} l_0^2/c_3^2$.

\textit{Regime (4).}---Since the spectrum of the purely nonlinear
regime, $b_k \sim k^{-6/5}$, (Eq.~(\ref{spect2})) is steeper than the marginal $b_k
\sim k^{-1}$ meaning that the integral in Eq.~(\ref{line_density})
builds up at the lower limit, as soon as $k_\mathrm{cutoff}
\gtrsim k_*$ the line density $L(k_\mathrm{cutoff})$ starts to
saturate and becomes independent of $k_\mathrm{cutoff}$ at
$k_\mathrm{cutoff} \gg k_*$. The energy balance gives the
dependence $k_\mathrm{cutoff}(\alpha)$,
\begin{equation}
k_\mathrm{cutoff} = \xi_4 \, \Lambda^{-3/4} \, \alpha^{-5/8} \,
(2\pi/l_0), \;\;\;  k_\mathrm{cutoff} \gg k_*  \; ,
\label{cutoff_reg4}
\end{equation}
where $\xi_4$ is an unknown constant. The coefficient in the
KW spectrum is fixed by continuity with the previous
regime, yielding $ \\b_k =C [1+c_3^2 \ln(k_*/k_\mathrm{c})]^{-1/2} \sqrt{k_\mathrm{c}
k_\mathrm{b} } \, k_*^{1/5} k^{-6/5} l_0$.
Eq.~(\ref{line_density}) thus yields
\begin{equation}
\frac{L(k_\mathrm{cutoff})}{L(k_*)} \approx 1+  \frac{(5C^2\,/2) k_\mathrm{c} k_\mathrm{b}l_0^2}{1+c_3^2
\ln(k_*/k_\mathrm{c})} \left[ 1-
\left(\frac{k_*}{k_\mathrm{cutoff}}\right)^{2/5}\right]\; .
\label{L_reg4}
\end{equation}

The continuity of $k_\mathrm{cutoff}$ leads to the following
constraints on the free coefficients in Eqs.~(\ref{cutoff_reg2}),
(\ref{cutoff_reg3}), and (\ref{cutoff_reg4}), $\xi_3\;=\;\xi_2\;
\,\Lambda^{1/4} \, \alpha_\mathrm{c}^{1/6}$, $\xi_4\;=\;\xi_2\;
\Lambda^{1/2}\,\alpha_\mathrm{c}^{1/6}\alpha_*^{1/8}$, where
$\alpha_\mathrm{c} \sim \Lambda^{-3/2}$ is the value of the
friction coefficient at which the cascade is cut off at the
crossover between the regimes (2) and (3) and $\alpha_* \sim
1/\Lambda^2$ corresponds to the one between (3) and (4).
Introducing $\alpha_b \lesssim 1$, corresponding to the crossover
from the regime (1) to (2), we rewrite Eqs.~(\ref{L_reg2}),
(\ref{L_reg3}), and (\ref{L_reg4}) in terms of $\alpha$:
\begin{eqnarray}
\ln \frac{L(\alpha)}{L_0} = A_2 \left[ (\alpha_\mathrm{b}/\alpha)^{1/3} -1 \right], \;\;\; \alpha_\mathrm{c} \ll \alpha \ll \alpha_b, \nonumber \\
\frac{L(\alpha)}{L(\alpha_\mathrm{c})}=\left[ \, 1 \; + \; (c_3^2/2) \; \ln\frac{\alpha_\mathrm{c}}{\alpha}\, \right]^{\nu}, \; \; \; \alpha_* \ll \alpha \ll \alpha_\mathrm{c}, \nonumber \\
\frac{L(\alpha)}{L(\alpha_*)} \approx 1+ A_4 \left[ 1-
\left(\frac{\alpha}{\alpha_*}\right)^{1/4}\right],
\;\;\;\;\; \alpha \ll \alpha_*, \label{fit}
\end{eqnarray}
where
\begin{eqnarray}
A_2= (2 \pi)^2 \xi_2^2 C^2/\Lambda^{1/2}
\alpha_\mathrm{b}^{2/3}, \nonumber \\
\nu = (2 \pi)^2 \xi_2^2
C^2/\Lambda^{1/2} (\alpha_\mathrm{c} \alpha_\mathrm{b})^{1/3}
c_3^2, \label{fitting_params} \\
A_4=5(2 \pi)^2 \xi_2^2 C^2/2 \Lambda^{1/2}
[1+(c_3^2/2) \ln(\alpha_\mathrm{c}/\alpha_*)] (\alpha_\mathrm{c}
\alpha_\mathrm{b})^{1/3}. \nonumber
\end{eqnarray}

%%%%%%%%%%%%%%%%%%%%%%%%%%%%%%%% FIGURE fit %%%%%%%%%%%%%
\begin{figure}[htb]
\includegraphics[width = 0.8\columnwidth,keepaspectratio=true]{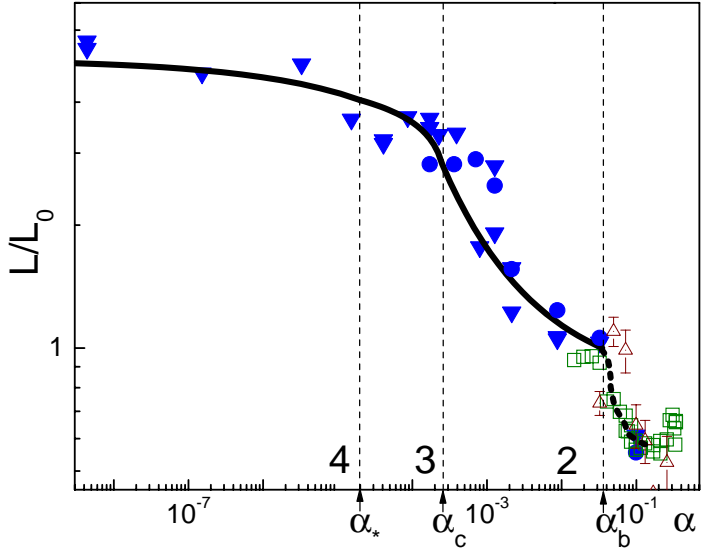}
\caption{(Color online.) Fit of the experimental data (closed circles and triangles) adapted from Ref.~\cite{Golov2} ($l_0 \approx
5\times10^{-3} cm$ giving $\Lambda \approx 13$). The high
temperature measurements of Refs.~\cite{Oregon}, \cite{Ladik} are
represented by open squares and open triangles respectively.  The
form of $\alpha(T)$ is taken according to
Ref.~\cite{Samuels_Donnelly} at $T \gtrsim 0.5K$ (roton
scattering) and according to Ref.~\cite{Iordanskii} at $T \lesssim
0.5K$ (phonon scattering). The fitting parameters are
$A_2\approx0.25$, $A_4\approx0.26$, $c_3\approx 3.0$,
$\nu\approx0.14$ with $\alpha_\mathrm{b} \approx 3.5 \times
10^{-2}$, $\alpha_\mathrm{c} \approx 2.5 \times 10^{-4}$,
$\alpha_* \sim 2.0\times10^{-5}$.}
\label{fig:fit}
\end{figure}
%%%%%%%%%%%%%%%%%%%%%%%%%%%%%%%%%%%%%%%%%%%%%%%%%%%%%%%%%%%%%%%%

%%%%%%%%%%%%%%%%%%%%%%%%%%%%%%%% FIGURE bk %%%%%%%%%%%%%
\begin{figure}[htb]
\includegraphics[width = 0.8\columnwidth,keepaspectratio=true]{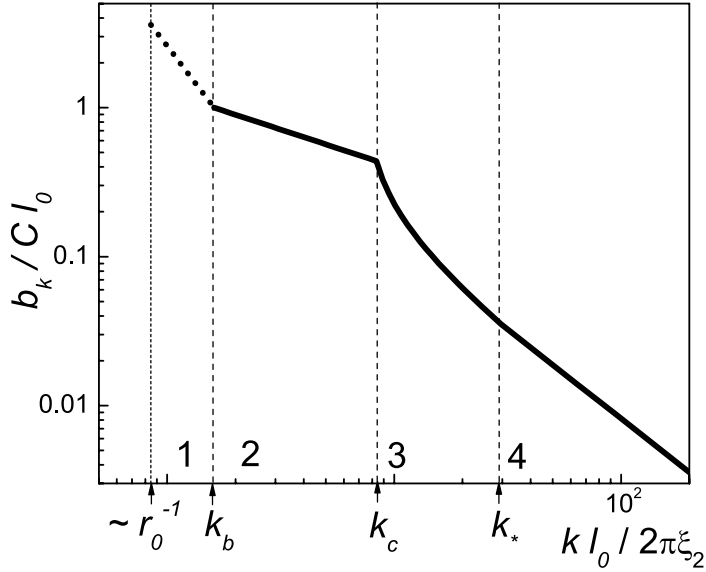}
\caption{Spectrum of Kelvin waves in the quantized regime quantified, apart from the
regime (1), by the fit to experimental data, Fig.~\ref{fig:fit}. In
view of Eq.~(\ref{line_density}), the constants $C$ and $\xi_2$
can be found only in the combination $\xi_2 C \approx 0.049$. The
regimes (1)-(4) correspond to those in Fig.~\ref{fig:spec}. }
\label{fig:bk}
\end{figure}
%%%%%%%%%%%%%%%%%%%%%%%%%%%%%%%%%%%%%%%%%%%%%%%%%%%%%%%%%%%%%%%%

Although the excellent quantitative agreement with the experiments can be sceptically attributed to the large number of fitting parameters, it is remarkable that the nontrivial qualitative behavior of $L(\alpha)$ exhibits the form peculiar to our decay scenario. As $\alpha$ decreases from $\alpha
\sim 1$ to the values significantly smaller than unity, the line
density $L$ increases only by some factor close to unity ($\sim
1.5$ in the experiment), which reflects the formation of the
regime (1) driven by the reconnections of vortex bundles. During
the crossover from the region (1) to (2), the increase of $L$ is
minimal---a \textit{shoulder} in the curve $L(\alpha)$ arises.  It is only well inside the region (2) that the increase
of $L$ becomes progressively pronounced, and, at the crossover to
the region (3), the function $L(\alpha)$ achieves its maximal
slope, determined by the developed fractalization of the vortex lines
necessary to support the cascade within the interval (3). As the
cutoff moves along the interval (3) towards higher wavenumbers,
the slope of $L(\alpha)$ becomes less steep due to the logarithmic decrease of
the characteristic amplitude of KW turbulence. When the cutoff
passes the crossover to the regime (4), the curve $L(\alpha)$
gradually levels.
[Note that because of the closeness of the KW spectra in the regimes (3) and (4) the crossover between them, as expected, is not distinct, which results in a significant uncertainty in the choice of $\alpha_*$.]
Fitting the experimental data fixes the values of the
dimensionless parameters, thus revealing the {\it quantitative}
form of the KW spectrum shown in Fig.~\ref{fig:bk}. Note, however, that in
view of the fact that the experimental $\Lambda \sim 10$ is not so
large we would expect certain systematic deviations, but a
significant scatter of the experimental data does not allow us to
assess them.

%%%%%%%%%%%%%%%%%%%%%%%%%%%%%%%%%%%%%%%%%%%%%%%%%%%%%%%%%%%%%%%%%%%%%%%%%%%%%%
\section{Kelvon-Phonon interaction}
\label{sec:kelvon_phonon}

\subsection{Hydrodynamic action. Hamiltonian of Kelvon-Phonon interaction}
\label{subsec:hydro}

The general problem of kelvon interaction with other modes is relevant far beyond the scope of superfluid turbulence. Since the early study of phonon scattering by vortex lines
\cite{Fetter_scattering}, much interest has been attracted by the
problems of the interaction of Kelvin waves with density-distortion
modes \cite{Vinen2000, Vinen2001, Epstein_Baym, Bretin, Mizushima,
Stoof}. In neutron stars, the excitation of Kelvin waves due to the interaction with the nuclei in the solid crust is
suggested to be the main mechanism of pulsar glitches
\cite{Epstein_Baym}. Nonlinear kelvon dynamics is also becoming an attractive topic in the field of ultra-cold gases, where one can study kelvon coupling to the modes of the Bose-Einstein condensate \textit{in situ}, as, e.g., in the problem of kelvon excitation by the quadrupole mode \cite{Bretin, Mizushima, Stoof}

In this section we describe a systematic approach (developed by the authors in Ref.~\cite{KS_05_vortex_phonon}) to the problem of interaction of phonons with vortices in the hydrodynamic regime, i.e. when any
physical length scale is much larger than the vortex core size
$a_0$, which allows one to describe vortices as geometrical lines
\cite{Donnelly}. We derive the interaction Hamiltonian basing the
analysis on the small parameter
\begin{equation}
\beta = a_0 \tilde{k} \ll 1, \label{beta:kelvon-phonon}
\end{equation}
where $\tilde{k}$ is the largest wave number among kelvons and phonons. To
employ the transparent description in terms of the normal modes, we
confine ourselves to the case of weak nonlinearity, which for kelvons implies the small parameter $\alpha$, Eq.~(\ref{ALPHA}), meaning that the amplitudes $b_k$ of the Kelvin waves of the typical wavelength $\lambda \sim k^{-1}$ are much smaller than $\lambda$, $\alpha_k = b_k k \ll 1$.
For phonons this requires that $\eta \ll n$, where $\eta$ is the
number density fluctuation in a sound wave and $n$ is the average
number density. The obtained result allows us to rigorously describe
the radiation of sound by kelvons, which we apply to the problem of
superfluid turbulence decay at zero
temperature in the next subsection.

Let us first describe the qualitative ideas behind the derivation. The condition (\ref{beta:kelvon-phonon}) implies that the typical vortex-line velocities are much smaller than the speed of sound.
Along with $\eta \ll n$ it leads to the fact that the vortex-phonon
coupling contributes only small corrections to the dynamics of the
non-interacting vortex and phonon subsystems. Therefore, a
perturbative approach is applicable, provided the interaction energy
is written in terms of the canonical variables. Normally, the form
of the canonical variables comes from the solution of the
interaction-free dynamics. However, when studying vortices separate
from phonons, one naturally neglects the compressibility of the
fluid, which we did in Sec.~\ref{sec:basic}, since finite compressibility leads only to higher-order ``relativistic'' corrections to vortex dynamics. As a result, the dynamics
of vortices are described by the Hamiltonian (\ref{ham_pseudo}), written in terms of
the geometrical configuration of the vortex lines. When
the vortices are absent, the phonon modes come from the bilinear
Hamiltonian for the density fluctuation $\eta(\mathbf{r}, t)$ and
the phase field $\varphi(\mathbf{r}, t)$, which determines the
velocity in the density wave (see, e.g., \cite{LLStatMech2}). If
finite compressibility of a superfluid is taken into account in
order to join the subsystems, the positions of vortices and the
fields $\eta(\mathbf{r}, t), \varphi(\mathbf{r}, t)$ are no longer
the sets of canonical variables because of the variable-mixing term
in the Lagrangian. This makes this perturbative problem quite peculiar since introducing the interaction, one has to simultaneously reconsider the canonical variables.

One can easily see that the standard vortex parametrization used in Sec.~\ref{sec:basic} fails to capture the physics of vortices when phonons are present. In Eq.~(\ref{BS}), the geometrical position of the vortex lines unambiguously determines the instantaneous velocity field configuration at arbitrary distances, which is inconsistent with the finite velocity of propagation of excitations in a compressible medium. From this simple physical argument, we can already guess the form of the canonical variables: the long-distance part of the velocity field produced by vortices in Eq.~(\ref{BS}) should actually belong to phonons.

The small parameters allow us to obtain an asymptotic expansion of
the canonical variables by means of a systematic iterative
procedure. Physically, the procedure restores the retardation in the
adjustment of the superfluid velocity field to the evolving vortex
configuration. It qualitatively changes the structure of the
Hamiltonian with respect to the terms responsible for the radiation
of sound and the relativistic corrections to the vortex dynamics.

\textit{Hydrodynamic Lagrangian: standard parametrization.} Long-wave superfluid dynamics at zero temperature are described by
the Popov's hydrodynamic action \cite{Popov}:
\begin{eqnarray}
S=\int \! \! \! dt \, d^3 r \left[-(n+\eta)\dot{\Phi}
 - \frac{(n + \eta)}{2m_0}
|\nabla \Phi |^2 - \frac{1}{2\varkappa} \eta^2 \right].
\label{Action}
\end{eqnarray}
Here the spatial integral is taken over the macroscopic fluid
volume, $\Phi(\mathbf{r}, t)$ is the phase field, which determines
the velocity according to $\mathbf{v}=(1/m_0)\nabla \Phi$ ($\hbar =
1$), $m_0$ is the mass of a particle, $\varkappa$ is the
compressibility, the dot denotes the derivative with respect to
time, and the microscopic length scale is defined by $a_0 = \sqrt{\varkappa/n m_0}$. The vortex core radius is of order $a_0$.

The phase $\Phi$ is non-single-valued and contains topological
defects, the vortex lines. The velocity circulation around each
vortex line is quantized:
\begin{equation}
\oint \nabla \Phi(\mathbf{r}) \cdot d \mathbf{r} = 2 \pi.
\label{circulation}
\end{equation}
The defects can be separated from the regular contribution:
\begin{equation}
\Phi=\Phi_0 + \varphi, \label{Phi_0_plus_phi}
\end{equation}
where $\Phi_0$ is non-single-valued and satisfies
(\ref{circulation}), while $\varphi$ is regular and $\nabla \varphi$
is circulation-free. The standard decomposition into vortices and
phonons is done by introducing an additional constraint that to the
zeroth approximation eliminates the coupling between $\Phi_0$ and
$\varphi$ in the Hamiltonian, namely
\begin{equation}
\Delta \Phi_0(\mathbf{r})=0. \label{Laplacian}
\end{equation}
(Physically, this parametrization is suggested by the velocity
potential of an incompressible fluid.) With
Eqs.~(\ref{Phi_0_plus_phi}),(\ref{Laplacian}) the Lagrangian becomes
\begin{equation}
L= \int \! \! \! d^3 r \left[- n \dot{\Phi}_0 - \eta\dot{\varphi} -
\eta \dot{\Phi}_0 \right] - H, \label{Lagrangian}
\end{equation}
where $H=H_{\mathrm{vor}} + H_{\mathrm{ph}} + H'_{\mathrm{int}}$,
%\begin{equation}
%H=H_{\mathrm{vor}} + H_{\mathrm{ph}} + H'_{\mathrm{int}},
%\label{Ham.total}
%\end{equation}
\begin{eqnarray}
H_{\mathrm{vor}} = \frac{n}{2m_0} \int \! \! \! d^3 r \;
\bigl| \nabla \Phi_0 \bigr|^2, \label{H_vor} \\
H_{\mathrm{ph}} = \int \! \! \! d^3 r \left[ \frac{n}{2m_0} \bigl|
\nabla \varphi \bigr|^2 + \frac{1}{2 \varkappa}\eta^2
\right], \label{H_ph} \\
H'_{\mathrm{int}} = \frac{1}{2m_0} \int \! \! \! d^3 r \Bigl[ \;
\eta \bigl|\nabla \Phi_0\bigr|^2 + 2 \eta \nabla \varphi \cdot
\nabla \Phi_0 \Bigr]. \label{H_int}
\end{eqnarray}
The coupling between the vortex variable, $\Phi_0$, and the density
waves, $\{\eta, \varphi\}$, is determined by $H'_{\mathrm{int}}$ and
the time derivative term $\int \! d^3 r \, \eta \dot{\Phi}_0$, both
being first-order corrections to the non-interacting parts.

%%%%%%%%%%%%%%%%%%%%%%%%%%%%%%%%%%%%%%%%%%%%%%%%%%%%%%%%%%%%%%%%%%%%%%%%%%%
\textit{Noninteracting case.} Following standard perturbative procedure, we first neglect the
coupling between the vortex and phonons to find the non-interacting
normal modes. The vortex part of the Lagrangian is then given by
$L_{\mathrm{vor}}= - n \int \! d^3 r \; \dot{\Phi}_0 -
H_{\mathrm{vor}}$. For the sake of simplicity, from now on we
consider a solitary vortex line; the generalization is
straightforward. Let the two-dimensional vector
$\boldsymbol{\rho}_0(z_0) = \bigl(x_0(z_0) , y_0(z_0), 0\bigr)$
describe the position of the vortex line in the plane $z=z_0$ of a
Cartesian coordinate system, where the $z$-direction is chosen along
the vortex line. The field $\Phi_0$ is a functional of
$\boldsymbol{\rho}_0(z)$, hence
\begin{equation}
\int \! \! \! d^3 r \; \dot{\Phi}_0 = \int \! \! \! d z \;
\dot{\boldsymbol{\rho}}_0 \cdot \frac{\delta}{\delta
\boldsymbol{\rho}_0} \int \! \! \! d^3 r \; \Phi_0. \label{int1}
\end{equation}
To obtain $\int \! d^3 r \; \dot{\Phi}_0$, it is sufficient to
calculate the integral $\int \!  d^3 r \; \delta \Phi_0$, where
$\delta \Phi_0$ is the variation of the phase field due to the
distortion of the vortex line by $\delta \boldsymbol{\rho}_0(z)$.
Using the identity $\Delta(\boldsymbol{\rho}^2)=4$, with
$\boldsymbol{\rho}=(x, y, 0)$, and Eq.~(\ref{Laplacian}), obtain
\begin{equation}
\int \! \! \! d^3 r \; \delta \Phi_0 = \frac{1}{4} \int \! \! \!
d^3r \; \nabla \cdot \left[ \; \delta \Phi_0 \nabla
(\boldsymbol{\rho}^2) \; \right]. \label{int2}
\end{equation}
The variation $\delta \Phi_0 (\mathbf{r})$ can be viewed as being
produced by two vortex lines with opposite circulation quanta and
separated by $\delta \boldsymbol{\rho}_0(z)$. In view of
Eq.~(\ref{circulation}) the field $\delta \Phi_0(\mathbf{r})$
experiences a jump of $2\pi$ across the surface $\mathcal{S}$ that
extends between these vortex lines along the vector field $\delta
\boldsymbol{\rho}_0(z)$, therefore the integration volume must have
a cut along $\mathcal{S}$. Applying the Gauss theorem, we rewrite
Eq.~(\ref{int2}) as the surface integral $(1/2) \oint_{\mathcal{S}}
\delta \Phi_0 \; (\boldsymbol{\rho} \cdot d \mathbf{S} )$ yielding
\begin{equation}
\int \!\!\! d^3r \; \dot{\Phi}_0 = \pi \int \!\!\! d z \;
 [ \mathbf{\hat{z}} \times \boldsymbol{\rho}_0(z) ] \cdot
\dot{\boldsymbol{\rho}}_0(z) . \label{int4}
\end{equation}
Introducing the complex variable $w(z)= \sqrt{n m_0 \kappa/2} \;
\bigl[ x(z)+iy(z) \bigr]$, where $\kappa = 2\pi/m_0$ is the velocity
circulation quantum, obtain
\begin{equation}
L_{\mathrm{vor}}= \frac{1}{2} \int \!\!\! dz \; \bigl[ i w^* \dot{w}
- i \dot{w}^* w \bigr] - H_{\mathrm{vor}}[w, w^*], \label{L_vor}
\end{equation}
which implies that $w(z)$ and $w^*(z)$ are the canonical variables
with respect to $L_{\mathrm{vor}}$. The energy (\ref{H_vor}),
rewritten in terms of $w(z)$ and $w^*(z)$ gives the vortex
Hamiltonian (\ref{ham}), or the pseudo-Hamiltonian (\ref{ham_pseudo}) with an appropriate regularization. We need only the bilinear
term of the expanded with respect to $\alpha_k \ll 1$ Hamiltonian since the kelvon interactions are irrelevant for our problem. We change the notation for the kelvon spectrum reserving $\omega_q$ for phonons:
\begin{equation}
 H_{\mathrm{vor}} \approx \sum_k \varepsilon_k a^{\dagger}_k a_k \;,
\;\; \varepsilon_k = \frac{\kappa}{4\pi} \left[\ln (1/k a_0)+C_0\right] k^2,
\label{kelvons}
\end{equation}
with $a_k$ and $a^{\dagger}_k$ defined in Sec.~\ref{sec:basic}.
The sound waves are described by the Lagrangian $L_{\mathrm{ph}}=
\int \! d^3 r \left(- \eta\dot{\varphi}\right) -
H_{\mathrm{ph}}[\eta, \varphi] $, with $H_{\mathrm{ph}}$ given by
(\ref{H_ph}). We assume that the system is contained in a cylinder
of radius $R$ with the symmetry axis along the $z$-direction and
that the system is periodic along $z$ with the period $L$. In the
cylindrical geometry, the phonon fields $\eta(r, \theta, z)$,
$\varphi(r, \theta, z)$ are parametrized by phonon creation and
annihilation operators $c_s$, $c^{\dagger}_s$ as
\begin{eqnarray}
\eta = \sum_s \sqrt{\,\omega_s \, \varkappa\,/\,2} \; \left[ \;
\chi_s \, c_s + \chi^*_s\, c^{\dagger}_s \; \right], \nonumber
\\ \varphi = - i \sum_s \sqrt{1\, /\, 2\,\omega_s \, \varkappa} \; \left[ \;
\chi_s \, c_s - \chi^*_s\, c^{\dagger}_s \; \right],
\label{phonon_fields} \\
\chi_s=\chi_s (r, \theta, z) = \mathcal{R}_{m q_r} \!(r) \; Y_m
\!(\theta) \; \mathcal{Z}_{q_z}\!(z), \nonumber
\end{eqnarray}
where $\mathcal{R}_{m q_r}(r)= (\pi q_r/R)^{1/2} J_m(q_r r)$, $Y_m
\! (\theta) = (2\pi)^{-1/2}  \exp(i m \theta)$,
$\mathcal{Z}_{q_z}\!(z) = L^{-1/2}  \exp ( i q_z  z )$, $s$ stands
for $\{ q_r, m, q_z \}$, and $J_m(x)$ are the Bessel functions of
the first kind. The phonon Hamiltonian then reads
\begin{equation}
H_{\mathrm{ph}}=\sum_s \omega_s c^{\dagger}_s c_s, \;\;\; \omega_s=
c \, q,\label{phonons}
\end{equation}
with $q= \sqrt{q_r^2 + q_z^2}$ and $c=\sqrt{n/\varkappa m_0}=1/m_0a_0$.

\textit{The effect of coupling: change of canonical variables.} Now we address the coupling between phonons and vortices. In terms
of the obtained variables, the Lagrangian (\ref{Lagrangian}) takes
on the form
\begin{equation}
L = \sum_k i \, \dot{a}_k a^{\dagger}_k + \sum_s i \, \dot{c}_s
c^{\dagger}_s - T  - H , \label{Lag_old_variables}
\end{equation}
where $T = \int \! d^3r \; \eta\dot{\Phi}_0 \equiv T\{a_k,\dot{a}_k,
a^{\dagger}_k,\dot{a}^{\dagger}_k, \; c_s,\dot{c}_s, c^{\dagger}_s,
\dot{c}^{\dagger}_s\}$, and
$H=H_{\mathrm{vor}}+H_{\mathrm{ph}}+H'_{\mathrm{int}}$. The coupling
term $T$ plays a special role in the Lagrangian
(\ref{Lag_old_variables}). This term is linear in time derivatives
of the variables and thus can not contribute to the energy in
accordance with the Lagrangian formalism. Moreover, because of the
time derivatives in $T$, the equations of motion in terms of $\{a_k,
a^{\dagger}_k\}, \{c_s, c^{\dagger}_s\}$ take on a non-Hamiltonian
form. This implies that the chosen variables become non-canonical in
the presence of the interaction, and therefore the total energy,
$H$, \textit{in these variables} can not be identified with the
Hamiltonian.

There must exist such a variable transformation $\{a_k,
a^{\dagger}_k\}, \{c_s, c^{\dagger}_s\} \rightarrow \{\tilde{a}_k,
\tilde{a}^{\dagger}_k\}, \{\tilde{c}_s, \tilde{c}^{\dagger}_s\}$
that restores the canonical form of the Lagrangian, $L = \sum_k i \,
\dot{\tilde{a}}_k \tilde{a}^{\dagger}_k + \sum_s i \,
\dot{\tilde{c}}_s \tilde{c}^{\dagger}_s  - H \{\tilde{a}_k,
\tilde{a}^{\dagger}_k,\tilde{c}_s, \tilde{c}^{\dagger}_s\}$. The
canonical variables are obtained by the following iterative
procedure. The term $T$ is expanded with respect to $\alpha_k \ll
1$, $\beta \ll 1$ and $\eta \ll n$ yielding
$T=T^{(1)}+T^{(2)}+\cdots$. Then the variables are adjusted by $a_k
\rightarrow a_k + a^{(1)}_k$, $c_s \rightarrow c_s + c^{(1)}_s$,
where $a^{(1)}_k ( \{a_k, a^{\dagger}_k, c_s, c^{\dagger}_s\})$ and
$c^{(1)}_s (\{a_k, a^{\dagger}_k, c_s, c^{\dagger}_s\})$ are chosen
to eliminate the term $T^{(1)}$ in (\ref{Lag_old_variables}). As a
result, $T \rightarrow 0 + T'^{(2)} + \cdots$, where the prime means
that the structure of the remaining terms has changed. At the next
step, $T'^{(2)}$ is eliminated by $a_k \rightarrow a_k + a^{(2)}_k$,
$c_s \rightarrow c_s + c^{(2)}_s$ and so on. By construction, the
canonical variables are given by $\tilde{a}_k = a_k + a^{(1)}_k +
a^{(2)}_k + \cdots$, $\tilde{c}_s = c_s + c^{(1)}_s + c^{(2)}_s +
\cdots$, and likewise for the conjugates. In practice, only the
first few terms are enough, as the rest ones give just higher-order
corrections.

The explicit expression for $T$ is obtained following the steps of
the derivation of Eq.~(\ref{int4}). The only difference here is that
the role of the auxiliary function $\boldsymbol{\rho}^2$ is played
by $Q$, defined by $\Delta Q (\mathbf{r}) \equiv \eta(\mathbf{r})$:
\begin{equation}
T = 2 \pi  \int \!\!\! d z \;  [ \mathbf{\hat{z}} \times \nabla
Q(\boldsymbol{\rho}_0(z), z) ] \cdot \dot{\boldsymbol{\rho}}_0(z) .
\label{T1}
\end{equation}

In accordance with $q_r \rho_0 \sim \beta k \rho_0 \ll 1$, we expand
the radial functions, retaining only the greatest term for each
particular angular momentum $m$. Switching to the cylindrical
coordinates, $\boldsymbol{\rho}_0 = (\rho_0 \, \cos \gamma, \;
\rho_0 \, \sin \gamma, \; 0 )$, and noticing that
\begin{eqnarray}
\left[ \dot{\gamma} \, \rho_0 \, \frac{\partial}{\partial r} -
\frac{\dot{\rho}_0}{\rho_0}\, \frac{\partial }{\partial \theta}
\right]\mathcal{R}_{m q_r}\!(r)Y_m \! (\theta) \Bigl|_{r=\rho_0,
\theta=\gamma} \nonumber \\
\propto \left\{ \begin{array}{cc}   d( w^{m})/dt ,& m\geq0
\\ d( w^{*|m|})/dt, & m<0
\end{array} \right. , \label{w.z}
\end{eqnarray}
obtain
\begin{eqnarray}
T \approx \sum_{s, k_1 \ldots k_m} \Bigl[ \;
 - iA_{s,k_1\ldots k_m}\, c_s \; \frac{d}{dt}(a_{k_1}\ldots
a_{k_m}) \nonumber \\ - iB_{s,k_1\ldots k_m}\, c_s \;
\frac{d}{dt}(a^{\dagger}_{k_1}\ldots a^{\dagger}_{k_m})\; \Bigr] +
\mathrm{H.c.}, \label{T_final}
\end{eqnarray}
where the sum is over all $s$ with $m\neq0$ and
\begin{eqnarray}
A_{s,k_1\ldots k_m}= -\Theta(m) \, A_s \, \delta_{k_1+\ldots +k_m, -q_z}, \nonumber \\
B_{s,k_1\ldots k_m}= (-1)^{|m|} \Theta(-m -1) \, A_s \,
\delta_{k_1+\ldots +k_m, q_z}, \nonumber \\ A_s=\frac{\sqrt{q/m_0\,c}}{2^{|m|/2+1}\; |m|!} \, \frac{n^{\frac{1-|m|}{2}} (m_0
\kappa)^{\frac{2-|m|}{2}} \, q_r^{|m| + \frac{1}{2}} \,
q^{-2}}{L^{(|m|-1)/2} \; R^{1/2}}, \label{vertex}
\end{eqnarray}
where $\Theta(m) = \left\{
\begin{array}{cc}   1,& m\geq0
\\ 0, & m<0
\end{array} \right.$. Thus,
\begin{equation}
a_k=\tilde{a}_k \label{transf.kelvon}
\end{equation}
(we omitted the terms that do not contain phonon operators and thus
result only in relativistic corrections to the kelvon spectrum and
kelvon-kelvon interactions),
\begin{eqnarray}
c_s=\tilde{c}_s + (1-\delta_{m,0}) \sum_{k_1 \ldots k_m}
\Bigl[A_{s,k_1\ldots k_m}\tilde{a}^{\dagger}_{k_1}\ldots
\tilde{a}^{\dagger}_{k_m} \nonumber \\
+B_{s,k_1\ldots k_m} \tilde{a}_{k_1}\ldots \tilde{a}_{k_m} \Bigr],
\;\;\; s=\{ q_r, m, q_z \}. \label{transf.phonon}
\end{eqnarray}

\textit{The interaction Hamiltonian.} The Hamiltonian $H$ is given by the energy
(\ref{H_vor})-(\ref{H_int}) in terms of the variables
$\{\tilde{a}_k, \tilde{a}^{\dagger}_k\}, \{\tilde{c}_s,
\tilde{c}^{\dagger}_s\}$. Up to neglected relativistic corrections,
the variable transformation does not change the spectrum of the
elementary modes: the zero-order Hamiltonians are given by
(\ref{kelvons}), (\ref{phonons}) in terms of $\{\tilde{a}_k,
\tilde{a}^{\dagger}_k\}, \{\tilde{c}_s, \tilde{c}^{\dagger}_s\}$.
The transform (\ref{transf.phonon}) applied to (\ref{phonons})
generates the interaction term
\begin{eqnarray}
H^{(\mathrm{rad})}_{\mathrm{int}} = \sum_{s,
\{k_i\}}(1-\delta_{m,0})\Bigl[\;\omega_s A_{s,k_1\ldots k_m} \;
\tilde{a}^{\dagger}_{k_1}\ldots
\tilde{a}^{\dagger}_{k_m}\tilde{c}^{\dagger}_s \nonumber \\
+ \omega_s B_{s,k_1\ldots k_m} \; \tilde{a}_{k_1}\ldots
\tilde{a}_{k_m}\tilde{c}^{\dagger}_s\; \Bigr] + \mathrm{H.c}.
\label{H_radiation}
\end{eqnarray}
Remarkably, the energy term $ \propto \int \! d^3r \, \eta
\bigl|\nabla \Phi_0\bigr|^2$ in (\ref{H_int}), which results in the
same as (\ref{H_radiation}) operator structure, is irrelevant, being
smaller in $\beta \ll 1$. It can be checked straightforwardly, that
the term $\propto \int \! d^3 r \; \eta \nabla \varphi \cdot \nabla
\Phi_0$ in Eq.~(\ref{H_int}) gives Fetter's amplitudes of the
elastic and inelastic scattering of phonons \cite{Fetter_scattering}
(see, however, Sec.~\ref{subsec:scatter}). In addition, this term leads to a
macroscopically small splitting of the phonon spectrum due to the
superimposed fluid circulation.

A kelvon carries a quantum of (negative) angular momentum
projection. The interaction (\ref{H_radiation})
explicitly conserves the angular momentum: a real process of the
emission of a phonon with the angular momentum $(-m)$ requires an
annihilation of $m$ kelvons.

\subsection{Sound emission by Kelvin-wave cascade}

Since $\varepsilon_k \sim (a_0 k)
\omega_k$, the total momentum transferred to phonons in a radiation
event should be small in order to satisfy the energy conservation.
Thus, the radiation by one kelvon on an infinite line is kinematically suppressed. The
leading radiation process is the emission of the $m=-2$ (quadrupole)
phonon mode, the events involving more than two kelvons being
suppressed by $\alpha_k \ll 1$. First-order processes of the
two-phonon emission come from the term $\propto \int \! d^3 r \;
\eta \nabla \varphi \cdot \nabla \Phi_0$ of (\ref{H_int}). The
amplitude of these processes is suppressed by the relativistic
parameter $\beta \ll 1$.

The qualitative observation that the sound emission is due to the quadrupole radiation is already sufficient to restore the formula for the power $\Pi_k$ radiated by the Kelvin waves at the wavenumber scale $k$ per unit vortex-line length. From general hydrodynamics \cite{LLHydrodynamics}, $\Pi_k$ must be proportional to the square of the third-order time derivative of the quadrupole moment, $ \propto \varepsilon_k^6 b_k^2 b_{-k}^2$, and inversely proportional to the fifth power of the sound velocity $c$. The rest of the dimensional coefficients are restored unambiguously, since the only remaining time and length scales are set by $\kappa$ and $k$ and the dimensions of mass come from the fluid density $\rho=nm_0$, which yields
\begin{equation}
\Pi_k \; \sim \; \frac{\kappa^2 \rho}{c^5k} \; \varepsilon_k^6 \; b_k^2 \, b_{-k}^2 \; \sim \;
\frac{\varepsilon_k^6 k}{c^5 \rho} \; n_k \, n_{-k}. \label{rad_power}
\end{equation}
This result is valid up to dimensionless prefactors of order unity.

We can use the Hamiltonian (\ref{H_radiation}) to obtain an accurate formula for the turbulence decay rate due to the sound radiation. The kelvon occupation
number decay rate is given by $\dot{n}_k = - \sum_{s,k_1} W_{s,k,k_1}$, where
$W_{s,k,k_1}$ is the probability of the event $|0_s,\, n_k,\,
n_{k_1}\rangle \rightarrow |1_s,\, n_k\!-\!1, \, n_{k_1}\!-\!1 \,
\rangle$ per unit time. Applying the Fermi Golden Rule to
$W_{s,k,k_1}$ with the interaction (\ref{H_radiation}) and replacing
the sums by integrals, obtain
\begin{equation}
\dot{n}_k = - \frac{ (\kappa/2 \pi)^5 }{15 \pi  \rho} \,
\bigl[ \ln(1/a_* k) + C_0 \bigr]^5 (k/c)^5 \, k^5\, n_k^2. \label{kelvon_decay}
\end{equation}
The total power radiated by kelvons per unit vortex-line length is then obtained from $\Pi_k= -\sum_{k' \sim k} \varepsilon_{k'} \dot{n}_{k'}/L$, which allows to restore the dimensionless prefactors in Eq.~(\ref{rad_power}). However, since the theories of non-structured and quasi-classical tangles can not trace coefficients of order unity we shall confine ourselves to using Eq.~(\ref{rad_power}) for determining the cascade cutoff scale $\lambda_\mathrm{ph}$.

In both non-structured and quasi-classical tangles, at the high wavenumbers where sound radiation becomes appreciable, turbulence decay is due to the purely nonlinear Kelvin-wave cascade with the spectrum given by Eq.~(\ref{spect2}). The condition for finding the cutoff scale $\lambda_\mathrm{ph}$ is $\Pi_{k_\mathrm{ph}} \sim \theta$. Clearly, the value of $\theta$ depends on the structure of the tangle. In non-structured tangles $\theta_\mathrm{ns} \sim \kappa^3\rho \Lambda^2/l_0^2$, where $l_0$ is the interline separation and $\Lambda=\ln(l_0/a_*)$, which gives
\begin{equation}
\lambda_\mathrm{ph} \,\sim\, \Lambda^{24/31} \left[\kappa/c\,l_0\right]^{25/31}l_0  \;\;\;\; {\rm (non-structured~tangles).} \label{lambda_ph_ns}
\end{equation}

For the quasi-classical tangles $\theta_\mathrm{qc} \sim \varepsilon \rho l_0^2$, which with the Eq.~\ref{l_0} gives $\theta_\mathrm{qc} \sim \kappa^3\rho \Lambda/l_0^2 \sim \theta_\mathrm{ns}/\Lambda$. Thus, for the cascade cutoff we obtain
\begin{equation}
\lambda_\mathrm{ph} \, \sim \, \Lambda^{27/31} \left[\kappa/c\,l_0\right]^{25/31}l_0 \;\;\;\; {\rm (quasi-classical~tangles).} \label{lambda_ph_qc}
\end{equation}

A comment is in order here. In superfluid turbulence, kelvons do not live on infinite vortex lines, as we assumed throughout this section, but are superimposed on vortex kinks with a typical curvature radii $R_0$ much larger than the kelvon wavelength, $R_0 \gg k^{-1}$. This circumstance led Vinen \cite{Vinen2001} to obtain a dimensional estimate $\Pi'_k \propto b^2_k \propto n_k$ for the power radiated per unit length of the vortex line, which is qualitatively different from our Eq.~(\ref{rad_power}). In the quasi-particle language, $\Pi'_k$ implies that the radiation is governed by the conversion of \textit{one}
kelvon into a phonon. Vinen argues that this process becomes allowed
in due to the finite size of the kinks, or,
equivalently, the kelvon coupling to a kink lifts the ban on
single-kelvon radiative processes by effectively removing the
momentum conservation constraint. We note that the probabilities of
such elementary events are likely to be suppressed
\textit{exponentially}, as it is generically the case, say, for
soliton-phonon interactions (see, e.g., Ref.~\cite{soliton1} and
references therein.) The processes involving kelvon-kink coupling
should contain an exponentially small factor $\sim \exp(-R_0 k)$,
which arises from the convolution of the smooth kink profile with
the oscillating kelvon mode.

\subsection{Elastic and inelastic phonon scattering}
\label{subsec:scatter}

The main advantage of the Hamiltonian formalism developed here is that it allows to reduce dynamics to elementary
scattering events tractable by the standard scattering theory,
thereby providing a general systematic and physically transparent
theoretical tool. So far, the Hamiltonian framework allowed us to address the
most important remaining problems of vortex dynamics---reconnection-free kinetics of Kelvin waves and phonon radiation by Kelvin waves. The purpose of this section is to use the developed formalism to revisit the longstanding problem of phonon scattering on vortex lines. The interest to this problem has been maintained by the controversy on the value of the elastic
scattering cross-section and the corresponding force between a
vortex and the normal component (see Ref.~\cite{Sonin} and references
therein). The elastic scattering cross-section was originally obtained by Pitaevskii \cite{Pitaevskii} and later confirmed by Sonin (Ref.~\cite{Sonin} and references therein) basing on hydrodynamic equations of motion. However, the Hamiltonian approach to this problem suggested by Fetter \cite{Fetter_scattering} (and rederived later in Ref.~\cite{Demircan}) gave a different result. Within the formalism developed in this section it is straightforward to see, as we do here by means of a direct calculation, that the discrepancy is due to the non-canonicity of the standard vortex and phonon parametrization used in the Hamiltonian of Ref.~\cite{Fetter_scattering}. The result of Ref.~\cite{Fetter_scattering} actually corresponds to the case of a pinned vortex, which was first noted by Sonin \cite{Sonin}, and the effect of vortex motion during the scattering is properly accounted for by the transformation to the canonical variables.

\begin{figure}
\includegraphics[width = 0.7\columnwidth,keepaspectratio=true]{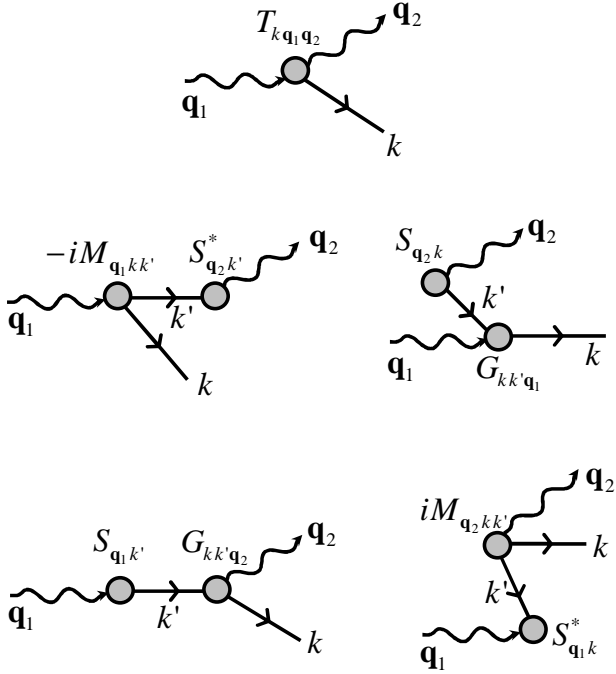}
\caption{Inelastic phonon scattering. The diagrams give the total
transition amplitude. The wavy and straight lines represent phonons
and kelvons respectively. The virtual kelvons are described by
vacuum propagators.}
\label{fig:inel_sc}
\end{figure}

We also discuss inelastic phonon scattering and show that the only available to our knowledge
solution of Ref.~\cite{Fetter_scattering} suffers from the same problem: the non-canonicity of the variables overlooked there makes a non-trivial contribution to the scattering cross-section. The accurate result can be obtained from a finite number of diagrams presented here (Fig.~\ref{fig:inel_sc}), but in view of the length of the resulting expression and absence of its immediate application, we do not
derive it in a closed form here.

\textit{Vortex-phonon Hamiltonian in the plain-wave basis for
phonon fields.} For our purposes, it will be convenient to rewrite the vortex-phonon
Hamiltonian of Sec.~\ref{subsec:hydro} in the plane-wave basis for the
phonon fields explicitly obtaining the terms responsible for phonon
scattering:
\begin{eqnarray}
\eta(\mathbf{r}) = \sum_{\mathbf{q}} \sqrt{\omega_{\mathbf{q}}
\kappa/2V} \left[ \, e^{i \mathbf{q} \mathbf{r}} \, c_{\mathbf{q}} +
e^{-i \mathbf{q} \mathbf{r}}\, c^{\dagger}_{\mathbf{q}} \, \right]
\; , \nonumber
\\ \varphi(\mathbf{r}) = - i \sum_{\mathbf{q}} \sqrt{1 / 2 V\omega_{\mathbf{q}} \kappa }
\left[ \, e^{i \mathbf{q} \mathbf{r}} \, c_{\mathbf{q}} - e^{-i
\mathbf{q} \mathbf{r}}\, c^{\dagger}_{\mathbf{q}} \, \right]\; ,
\label{phonon_fields}
\end{eqnarray}
where $\omega_{\mathbf{q}}=c \, q$ and $V=L^3$ is
the system volume. The (harmonic part of) phonon Hamiltonian is given by Eq.~(\ref{phonons}) with $s$ labeling different wavenumbers $\mathbf{q}$.

A calculation analogous to that of Sec.~\ref{subsec:hydro} yields the relation for the canonical variables in the form
\begin{eqnarray}
a_k=\tilde{a}_k + \gamma_{k \, k_1 \, \mathbf{q}} \; \tilde{a}_{k_1}
[\tilde{c}_{\mathbf{q}} +
\tilde{c}^{\dagger}_{-\mathbf{q}}] + \mathrm{smaller \; terms}, \nonumber \\
\gamma_{k \, k_1 \, \mathbf{q}} = - \left(\frac{q}{8 V \, m_0 n \, c
}\right)^{1/2} \frac{q_x^2+q_y^2}{q^2} \; \delta_{k_1+q_z, k},
\label{transf.kelvon.plane}
\end{eqnarray}
(we omitted the terms that do not contain phonon operators and thus
result only in ``relativistic'' corrections to the kelvon spectrum
and kelvon-kelvon interactions) and for phonon operators
\begin{eqnarray}
c_\mathbf{q} = \tilde{c}_\mathbf{q} + \sigma^*_{\mathbf{q} \, k} \,
\tilde{a}_k + \sigma_{\mathbf{q} \, k} \, \tilde{a}^{\dagger}_{-k} %\nonumber \\
+ i \mu^*_{\mathbf{q} \, k_1 \, k_2}  \tilde{a}_{k_1}
\tilde{a}_{k_2} + i \mu_{\mathbf{q} \, k_1 \, k_2}
\tilde{a}^{\dagger}_{-k_1} \tilde{a}^{\dagger}_{-k_2}, \; + \;\;
\mathrm{smaller \; terms} \nonumber
\end{eqnarray}
\begin{eqnarray}
\sigma_{\mathbf{q} \, k} = i( \kappa \; q/c)^{1/2}
\frac{q_x+iq_y}{2 L q^2}\, \delta_{k,q_z}, \nonumber \\
\mu_{\mathbf{q} \, k_1 \, k_2} = -i
\frac{(q/m_0c)^{1/2}(q_x+iq_y)^2}{(32 \, n V)^{1/2}q^2} \,
\delta_{k_1+k_2, q_z}. \label{transf.phonon.plane}
\end{eqnarray}
Here and below we assume summation of products over repeating
indices.

As in the cylindrical case, up to neglected relativistic corrections, the variable transformations (\ref{transf.kelvon.plane}),
(\ref{transf.phonon.plane}) applied to Eqs.~(\ref{kelvons}),
(\ref{phonons}) do not change the spectrum of the elementary modes,
but generate a non-trivial contribution to the interaction
Hamiltonian in higher orders:
\begin{eqnarray}
H^{(1)}_{\mathrm{int}} = \Bigl[ \; S^*_{\mathbf{q} \, k}
\,\tilde{c}^{\dagger}_\mathbf{q} \, \tilde{a}_k + S_{\mathbf{q} \,
k}
\, \tilde{c}^{\dagger}_\mathbf{q} \, \tilde{a}^{\dagger}_{-k} %\nonumber \\
 + \; i M^*_{\mathbf{q} \, k_1 \, k_2} \,
\tilde{c}^{\dagger}_\mathbf{q} \, \tilde{a}_{k_1} \, \tilde{a}_{k_2}
\; + \; i M_{\mathbf{q} \, k_1 \, k_2} \,
\tilde{c}^{\dagger}_\mathbf{q} \, \tilde{a}^{\dagger}_{-k_1} \,
\tilde{a}^{\dagger}_{-k_2} \;\;
\nonumber \\  + \;  G_{k \, k_1 \, \mathbf{q}} \; \bigl( \; \tilde{a}^{\dagger}_{k}
\, \tilde{a}_{k_1} \, \tilde{c}_{\mathbf{q}} \; + \;
\tilde{a}^{\dagger}_{k} \, \tilde{a}_{k_1} \,
\tilde{c}^{\dagger}_{-\mathbf{q}} \; \bigr) \; \Bigr] \; + \;
\mathrm{H.c.}  + \; \mathrm{smaller \; terms} , \;\; \;\;\; \label{interaction_1}
\end{eqnarray}
where
\begin{eqnarray}
S_{\mathbf{q} \, k} \; = \;  \omega_\mathbf{q} \; \delta_{\mathbf{q},\mathbf{q'}} \; \sigma_{\mathbf{q'} \, k},  \nonumber \\
M_{\mathbf{q} \, k_1 \, k_2} \; = \; \omega_\mathbf{q} \; \delta_{\mathbf{q},\mathbf{q'}} \; \mu_{\mathbf{q'}
\, k_1
\, k_2}, \nonumber \\
G_{k \, k_1 \, \mathbf{q}} \; = \; \varepsilon_k \;\delta_{k,k'} \; \gamma_{k' \, k_1 \, \mathbf{q}}. \label{vertices}
\end{eqnarray}

Another relevant contribution to the interaction Hamiltonian comes
from the vortex-phonon coupling energy, Eq.~(\ref{H_int})
(equivalent to the one suggested by Hall and Vinen
\cite{Hall_Vinen}),
\begin{equation}
H^{(2)}_{\mathrm{int}} =  \int \! \! \! d \mathbf{r} \; \eta \nabla
\varphi \cdot \mathbf{v}_\mathrm{v}, \label{Energy.HV}
\end{equation}
where $\mathbf{v}_\mathrm{v} = \mathbf{v}_\mathrm{v}(\mathbf{r})$ is
the velocity field produced by the vortex line. Taking into account
this term alone leads to Fetter's effective interaction Hamiltonian \cite{Fetter_scattering}
(the terms $\propto \tilde{a}\tilde{c}\, \tilde{c}, \; \tilde{a}
\tilde{c}^{\dagger} \tilde{c}^{\dagger}$ are omitted)
\begin{eqnarray}
H^{(2)}_{\mathrm{int}} \, = \, \Bigl[ R_{\mathbf{q_1}\,\mathbf{q_2}}
\, \tilde{c}_{\mathbf{q_1}} \tilde{c}^{\dagger}_{\mathbf{q_2}} \; +
\; R^*_{\mathbf{q_1}\,-\mathbf{q_2}} \,
\tilde{c}_{\mathbf{q_1}} \tilde{c}_{\mathbf{q_2}} % \nonumber \\
+ \;  \; T_{k \, \mathbf{q_1}\,\mathbf{q_2}} \,
\tilde{a}^{\dagger}_k \, \tilde{c}_\mathbf{q_1}
\tilde{c}^{\dagger}_\mathbf{q_2} \;\Bigr] \nonumber \\
\; + \mathrm{H.c.} \; + \; \mathrm{smaller \; terms} ,
\label{interaction_2}
\end{eqnarray}
where
\begin{eqnarray}
R_{\mathbf{q_1}\,\mathbf{q_2}}\,=\,\frac{i\kappa}{2L^2}
\sqrt{\frac{\omega_\mathbf{q_1}}{\omega_\mathbf{q_2}}} \;
\frac{(\mathbf{q_1} \times \mathbf{q_2}\cdot
\hat{\mathbf{z}})}{|\mathbf{q_1}-\mathbf{q_2}|^2} \;
\delta_{q^{z}_{1}, q^{z}_{2}}, \nonumber \\
T_{k \, \mathbf{q_1}\,\mathbf{q_2}} \,=\, \frac{1}{2V}
\sqrt{\frac{\kappa L}{2 m_0 n}}
\mathbf{Q}_{\mathbf{q_1},\mathbf{q_2}} \cdot  \Bigl[ %\nonumber \\
\hat{\mathbf{z}}\bigl[
(\mathbf{q_2}-\mathbf{q_1})\cdot(\hat{\mathbf{x}}+i\hat{\mathbf{y}})\bigr]\,
 + k(\hat{\mathbf{x}}+i\hat{\mathbf{y}}) \,  \Bigr] \delta_{k+q^{z}_2, q^{z}_1}, \nonumber \\
 \mathbf{Q}_{\mathbf{q_1},\mathbf{q_2}}
= \frac{ \left(
\sqrt{\frac{\omega_{\mathbf{q_2}}}{\omega_{\mathbf{q_1}}}}
\mathbf{q_1} +
\sqrt{\frac{\omega_{\mathbf{q_1}}}{\omega_{\mathbf{q_2}}}}
\mathbf{q_2}\right) \times
(\mathbf{q_2}-\mathbf{q_1})}{|\mathbf{q_2}-\mathbf{q_1}|^2}. \;\;\;\;\;
\label{Q}
\end{eqnarray}
The complete interaction Hamiltonian up to the highest relevant
order in $\beta \ll 1$, $\alpha_k \ll 1$, and $|\eta|/n \ll 1$ is
given by $H_\mathrm{int}=H^{(1)}_\mathrm{int}+H^{(2)}_\mathrm{int}$.
Note that in this form the Hamiltonian conservers the momentum along
the vortex line (along the $z$-axis), but not the transverse
momentum. The change of transverse momentum during the scattering is
transferred to the vortex line as a whole (an analog of the
M\"{o}ssbauer effect), which results in a macroscopically small
displacement of the vortex line from its initial position. The
corresponding Goldstone mode can be straightforwardly taken into
account if necessary.

%%%%%%%%%%%%%%%%%%%%%%%%%%%%%%%% FIGURE geometry %%%%%%%%%%%%%
\begin{figure}
\includegraphics[width = 0.5\columnwidth,keepaspectratio=true]{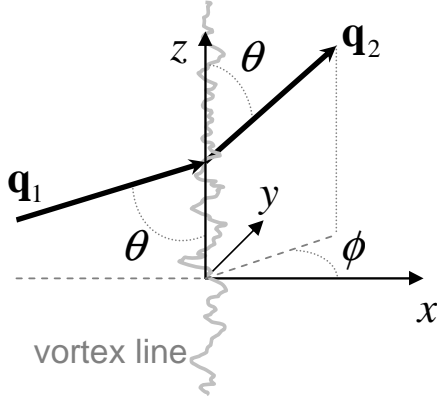}
\caption{Geometry of the elastic phonon scattering.}
\label{fig:geom}
\end{figure}

%%%%%%%%%%%%%%%%%%%%%%%%%%%%%%%% FIGURE elastic scattering %%%%%%%%%%%%%
\begin{figure}[htb]
\includegraphics[width = 0.7\columnwidth,keepaspectratio=true]{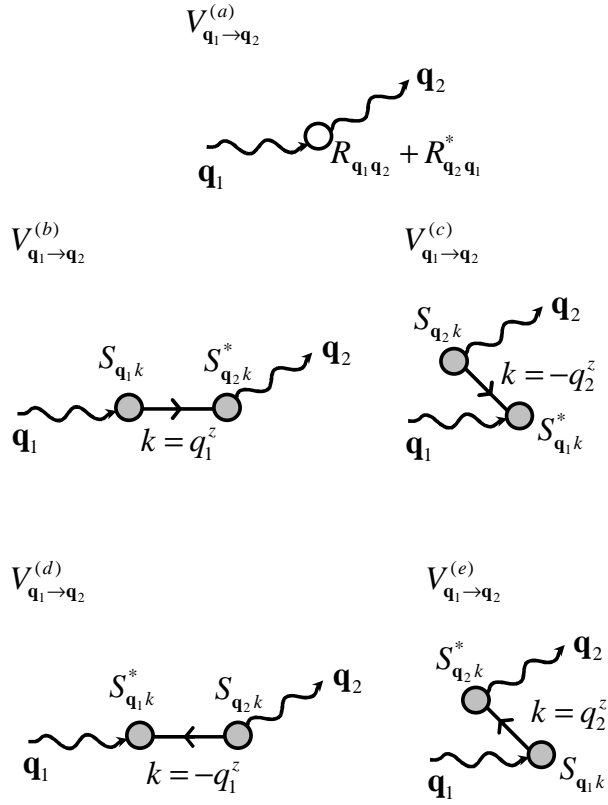}
\caption{Diagrammatic representation of the transition amplitude of
elastic phonon scattering on a vortex line. The wavy and straight
lines represent phonons and kelvons respectively.}
\label{fig:el_sc}
\end{figure}
%%%%%%%%%%%%%%%%%%%%%%%%%%%%%%%%%%%%%%%%%%%%%%%%%%%%%%%%%%%%%%%%

\textit{Elastic phonon scattering.} We consider the scattering geometry shown in Fig.~\ref{fig:geom}.
The vortex line has a distribution of kelvon occupation numbers $\{
n_k \}$. The elastic scattering differential cross-section can be
found as
\begin{equation}
\frac{d \sigma}{d \phi} = (V/c) \frac{d}{d \phi}\sum_{\mathbf{q}_2}
W_{\mathbf{q}_1 \rightarrow \mathbf{q}_2}, \label{diff.cross.}
\end{equation}
where $W_{\mathbf{q}_1 \rightarrow \mathbf{q}_2}$ is the probability
per unit time of a scattering event of a phonon with the wavenumber
$\mathbf{q}_1$ into the one with $\mathbf{q}_2$, such that
$\mathbf{q}_1 = q_1\,(\sin\theta, \, 0, \, \cos\theta)$,
$\mathbf{q}_2 = q_2\, (\sin\theta \cos\phi, \, \sin\theta \sin\phi
\, , \cos\theta)$. The probability $W_{\mathbf{q}_1 \rightarrow \mathbf{q}_2}$ is given by the Golden rule, $W_{\mathbf{q}_1
\rightarrow \mathbf{q}_2} = 2 \pi
\delta(\omega_{\mathbf{q}_1}-\omega_{\mathbf{q}_1}) |V_{\mathbf{q}_1
\rightarrow \mathbf{q}_2}|^2$. To the first (Born) approximation the
transition amplitude $V_{\mathbf{q}_1 \rightarrow \mathbf{q}_2}$ is
composed of the elementary processes shown diagrammatically in
Fig.~\ref{fig:el_sc}. Note that, in Ref.~\cite{Fetter_scattering} only
the term $V^{(a)}_{\mathbf{q}_1 \rightarrow \mathbf{q}_2}$ was taken
into account. The remaining terms describe a scattering process in
which the vortex line undergoes a virtual deformation (produced by a
kelvon with $k=q_{1z}$). It is easily seen that physically
$V^{(b-e)}_{\mathbf{q}_1}$ account for the oscillatory motion of the
vortex line induced during the scattering: if the
vortex were pinned by an external potential these terms would have
to vanish. The transition amplitudes in Fig.~\ref{fig:el_sc} are given by
\begin{eqnarray}
V^{(a)}_{\mathbf{q}_1 \rightarrow \mathbf{q}_2} =
R_{\mathbf{q_1}\,\mathbf{q_2}} + R^*_{\mathbf{q_2}\,\mathbf{q_1}} = \nonumber \\
\frac{i \kappa} {2L^2} \,
\frac{\sqrt{q_1q_2}(q_1+q_2)\sin^2\theta\sin\phi}{q_1^2+q_2^2
-2q_1q_2(\sin^2\theta\cos{\phi}+\cos^2\theta)} \,\delta_{q^{z}_{1},
\, q^{z}_{2}}, \label{V.a}
\end{eqnarray}
\begin{equation}
V^{(b)}_{\mathbf{q}_1 \rightarrow \mathbf{q}_2} =
\frac{S_{\mathbf{q}_1 \,k}\;S^*_{\mathbf{q}_2
\,k}}{\omega_q-\varepsilon_k} \, (n_k+1) , \label{V.b}
\end{equation}
\begin{equation}
V^{(c)}_{\mathbf{q}_1 \rightarrow \mathbf{q}_2} =
\frac{S_{\mathbf{q}_2 \,k}\;S^*_{\mathbf{q}_1
\,k}}{-\omega_q-\varepsilon_k} \, (n_{-k}+1), \label{V.c}
\end{equation}
\begin{equation}
V^{(d)}_{\mathbf{q}_1 \rightarrow \mathbf{q}_2} =
\frac{S_{\mathbf{q}_2 \,k}\;S^*_{\mathbf{q}_1
\,k}}{\omega_q+\varepsilon_k} \, n_{-k}, \label{V.d}
\end{equation}
\begin{equation}
V^{(e)}_{\mathbf{q}_1 \rightarrow \mathbf{q}_2} =
\frac{S_{\mathbf{q}_1 \,k}\;S^*_{\mathbf{q}_2
\,k}}{-\omega_q+\varepsilon_k} \, n_k. \label{V.e}
\end{equation}
Note, that at small scattering angles the amplitude
$V^{(a)}_{\mathbf{q}_1 \rightarrow \mathbf{q}_2}$ diverges and the
perturbation theory is not applicable. However, the small-angle scattering does not play an important role for the dissipative component of the mutual
friction force. Since $\varepsilon_k/\omega_q \sim a_0q_{1z}
\lesssim \beta \ll 1$ for $|k|=|q_{1z}|=|q_{2z}|$, we can neglect
the kelvon energy in the denominators of
Eqs.~(\ref{V.a})-(\ref{V.e}). Thus, the terms proportional to $\{
n_k \}$ cancel out in the total transition amplitude,
\begin{equation}
V_{\mathbf{q}_1 \rightarrow \mathbf{q}_2} =
R_{\mathbf{q_1}\,\mathbf{q_2}} \; \; + \Bigl[ \, S_{\mathbf{q}_1
\,k}\;S^*_{\mathbf{q}_2}\, - \, \mathrm{c.c.} \, \Bigr]/\omega_q,
\label{V.total}
\end{equation}
meaning that alternatively the amplitude can be obtained by
retaining only the terms of type $V^{(a-c)}_{\mathbf{q}_1
\rightarrow \mathbf{q}_2}$ but evaluating them in kelvon vacuum.
Substituting the vertex expressions into Eq.~(\ref{V.total}) and
doing the algebra yields
\begin{equation}
V_{\mathbf{q}_1 \rightarrow \mathbf{q}_2} = \frac{i \kappa}{2L^2} \,
\frac{\sin^2 \theta \sin\phi \cos\phi +\cos^2\theta
\sin\phi}{1-\cos\phi} \; \delta_{q_{1z}, q_{2z}}.\label{V.final}
\end{equation}
Finally, replacing the sum in Eq.~(\ref{diff.cross.}) by the
integral,
\begin{equation}
\frac{d \sigma}{d \phi} = (V/c) \int \frac{q_2 \,dq_2\, L^2}{2\pi}
\delta(\omega_{\mathbf{q}_1}-\omega_{\mathbf{q}_1}) |V_{\mathbf{q}_1
\rightarrow \mathbf{q}_2}|^2, \label{diff.cross.int}
\end{equation}
we arrive at
\begin{equation}
\frac{d \sigma}{d \phi} \; = \; L \, q \; \frac{\kappa^2}{8\pi c^2}
\; \frac{\bigl[ \sin^2 \theta \sin\phi \cos\phi +\cos^2\theta
\sin\phi \bigr]^2}{(1-\cos\phi)^2}. \label{diff.cross.final}
\end{equation}
For $\theta=\pi/2$ this result gives the differential cross-section
obtained in Ref.~\cite{Sonin}. Due to the divergency at $\phi \to 0$, this formula is only applicable in the cases where the small-angle scattering is irrelevant, as, e.g., in a calculation of the dissipative mutual friction force, which immediately follows from Eq.~(\ref{diff.cross.final}). In contrast, the calculation of the non-dissipative component is essentially non-perturbative and is discussed in Ref.~\cite{Sonin}.

%%%%%%%%%%%%%%%%%%%%%%%%%%%%%%%% FIGURE Inelastic scattering %%%%%%%%%%%%%
%%%%%%%%%%%%%%%%%%%%%%%%%%%%%%%%%%%%%%%%%%%%%%%%%%%%%%%%%%%%%%%%

\textit{Inelastic scattering.} Employing the same formalism, we describe the inelastic phonon
scattering, i.e. scattering accompanied by generation of Kelvin
waves. In the diagrammatic language the transition amplitude
$V_{\mathbf{q}_1 \rightarrow \mathbf{q}_2, k}$, where $k$ is the
kelvon wavenumber, is shown in Fig.~\ref{fig:inel_sc}. Note that the extra
diagrams with reversed propagators are dropped, which is
accounted for by evaluating the remaining ones over kelvon
vacuum. The scattering matrix element reported in
Ref.~\cite{Fetter_scattering} corresponds to the first term $\propto
T_{k \, \mathbf{q_1}\,\mathbf{q_2}}$ only.

%%%%%%%%%%%%%%%%%%%%%%%%%%%%%%
\section{Conclusions and discussion}
\label{sec:conclusions}

The zero-temperature limit of superfluid turbulence is non-trivial  and instructive.
A number of small and large parameters---Eqs.~(\ref{Lambda}), (\ref{ALPHA}), (\ref{beta:kelvon-phonon})---and (approximately) conserving quantities (energy, momentum, number of kelvons, LIA constants of motion), characterizing the phenomenon, bring about rich physics on one hand, and provide a solid basis for applicability of the analytic tools on the other.

The central part in the theory is played by the large parameter $\Lambda$, Eq.~(\ref{Lambda}). Implications of the condition $\Lambda \gg 1$ include (i) universality of the answers for all U(1)-type superfluids, (ii) applicability  of LIA, (iii) features following from (ii), of which the most important ones are (a) the integrability of LIA suppressing the efficiency of the pure Kelvin-wave cascade and (b) preemptive character of the self-induced motion of the vortices in the smallest Richardson-Kolmogorov eddies, responsible for cutting off the classical-fluid regime {\it before} it could experience a potential bottleneck associated with (a).

The energy and momentum conservation renders impossible the pure vortex ring cascade (originally proposed by Feynman) in which superfluid turbulence is supposed to decay into small rings in such a way  that each ring produces smaller ones independently of the rest of the vortex tangle. The absence of pure Feynman's cascade and suppressed efficiency of the pure Kelvin-wave cascade create specific circumstances under which there exists a range in the kelvon wavenumber space---with the size controlled by $\Lambda$---where the Kelvin-wave cascade is driven by this or that type of vortex line reconnections. The reconnections lift the integrability constraint of the LIA and push the vortex-line length to higher and higher wavenumbers.

In the theoretical limit of really large $\Lambda$ (to be taken with a grain of salt in view of realistic $\Lambda \lesssim15$ in $^4$He), we predict three distinct types of reconnection-driven Kelvin-wave cascades. The first two ones (in the down-the-cascade order)---(i) vortex-bundle-reconnection-based and (ii) neighboring-line-reconnection-based---apply only to the Richardson-Kolmogorov quasi-classical regime, while the third one, driven by local self-crossings, applies to non-structured tangles as well, including superfluid turbulence sustained by vibrating objects. At large-enough wave\-numbers, the local-self-crossings-driven cascade naturally crosses over to the pure Kelvin-wave cascade, as soon as the power of the latter becomes sufficient to compete with the former, eventually suppressing the reconnections by rendering the amplitude of Kelvin waves small compared to their wavelength.

At $T=0$, the inertial range of the pure Kelvin-wave cascade is controlled by the small parameter $\beta$, defined in Eq.~(\ref{beta:kelvon-phonon}), that guarantees irrelevance of compressibility and/or kelvon-phonon interaction before the cascade reaches the wavelength scale  $\lambda \sim  \lambda_{\rm ph}$, where phonon emission becomes dominant, cutting the cascade off. In a typical case, $1/\beta \gg \Lambda \sim \ln(1/\beta)$, implying  that the  inertial range of of the pure Kelvin-wave cascade is significantly larger than that of the reconnection-driven cascade(s).

Kinetics of the pure Kelvin-wave cascade are perturbative due to the asymptotically vanishing amplitude-to-wavelength ratio. This allows one to introduce a kinetic equation with the collision term associated with kelvon scattering. Because of the conservation of the total number of kelvons (which is nothing but the conservation of the angular momentum) and one-dimensional character of the problem, the leading kinetic process is the elastic three-kelvon scattering. As an effect of the LIA integrability constraint, the kinetics are entirely due to the interactions going beyond the LIA: the local contributions to the effective three-kelvon scattering amplitude exactly cancel each other leaving no order-$\Lambda$ collision terms and suppressing kinetic rates.

The small amplitude of Kelvin waves at the dissipative cutoff wavenumbers and weak kelvon-phonon coupling guaranteed by the parameter $\beta$ imply that the phonon emission is a perturbative process that can be naturally described within the Hamiltonian formalism.  Our derivation of the kelvon-phonon Hamiltonian revealed a subtlety of non-canonicity of the prime harmonic variables with respect to the interaction Hamiltonian. Adequately accounting for this fact at the level of hydrodynamic action leads to a proper kelvon-phonon interaction Hamiltonian, containing terms that would be missed otherwise. With this Hamiltonian the problem of the phonon cutoff of the Kelvin-wave cascade becomes rather straightforward. The leading process is two-kelvon inelastic scattering converting two kelvons with almost opposite momenta into one phonon. As a by-product, we resolved some long-standing controversies of the kelvon-phonon interaction, showing that these were due to the above-mentioned lacking terms in the interaction Hamiltonian. [We also revealed the role of a hidden phonon-vortex interaction in the theory of Berezinskii-Kosterlitz-Thouless transition \cite{KT_our}, which is not covered here.]

Having at hand a complete theoretical scenario for superfluid turbulence at $T=0$, it is important to check it against experiment and numeric simulations.  Discussing first the more simple case of non-structured turbulence, we note that a circumstantial evidence for the Kelvin-wave cascade---the absence of temperature dependence of the relaxation time superfluid turbulence in $^4$He at $T<70mK$---has been observed in both experiment \cite{Davis} and simulations \cite{Tsubota00}. The simulation of Ref.~\cite{Tsubota00}, performed within LIA (and after Ref.~\cite{Sv95} was published), could potentially identify  the  local self-crossings driven cascade. Moreover, the authors did observe production of vortex rings. However, they interpreted that as the pure Feynman's cascade. A direct numeric support for the local self-crossing scenario can be found in Ref.~\cite{sim_boll}, where the behavior of a vortex line attached to a vibrating object was studied within the LIA. It was observed that as a response to the external perturbation the vortex line gets kinky, after which frequent local self-crossings result in a systematic production of vortex rings.

Still very desirable is a simulation of the zero-temperature decay of non-structured turbulence within the full Biot-Savart description. Such a simulation could quantify our very rough order-of-magnitude estimate of the crossover between local-self-crossing-driven and pure Kelvin-wave cascades and answer the fundamental question of whether the inertial range of the cascade of self-crossings is large enough for realistic $\Lambda \lesssim 15$ to be taken seriously in the experimental context.

Turning now to the Richardson-Kolmogorov regime, we emphasize the recent experimental progress \cite{Golov} that allowed to probe the vortex line length as a function of temperature. As we discussed in Sec.~\ref{sec:crossover}, the experiment is in a very good qualitative agreement with our crossover scenario, the observed pronounced increase of the vortex line length at friction coefficient $\sim 10^{-3}$ being indicative of the line fractalization. Nevertheless, it is still a circumstantial rather than direct evidence. A direct experimental evidence for the local self-crossings driven scenario  might be an observation of small vortex rings emitted by spatially localized Richardson-Kolmogorov superfluid turbulence.

On the theoretical side, a simulation of the crossover from Richardson-Kolmo- \\ gorov to Kelvin-wave regime might be very instructive, especially given that numerically one can set $\Lambda$ to be arbitrarily large to render the role of this parameter more pronounced. Yet another missing theoretical piece---in view of only an order-of-magnitude estimate and a technical mistake in our Ref.~\cite{KS_04}, mentioned in Sec.~\ref{sec:pure}---is an accurate result for the numerical prefactor in the energy flux of the pure Kelvin-wave cascade, Eq.~(\ref{rel}).

Summarizing our results for the crossover from Richardson-Kolmogorov to Kelvin-wave regime, we cannot leave without a discussion the work by  L'vov,  Nazarenko, and Rudenko \cite{Lvov}, which triggered our interest to the problem. As discussed in Sec.~\ref{sec:crossover}, the authors of Ref.~\cite{Lvov} put forward an idea of bottleneck accumulation of energy in the wavenumber space between the classical-field hydrodynamic modes and Kelvin waves.  The notion
of bottleneck  in Ref.~\cite{Lvov} is unequivocally defined as being associated with the absence of {\it any} channel capable of accommodating the energy flux produced by the Richardson-Kolmogorov cascade, which implies {\it thermalization} of hydrodynamic modes in a certain range of length scales {\it above} the Kelvin-wave regime. In our Ref.~\cite{KS_crossover} (see also Sec.~\ref{sec:crossover}), we pointed out that this idea is inconsistent with the above-mentioned preemptive role of reconnections as an energy transport mechanism. While agreeing with us \cite{comment,Lvov2} on the importance of the self-induced motion of vortices in the smallest---of the size $\sim r_0$ defined in Eq.~(\ref{r_0-l_0})---Richardson-Kolmogorov eddies, and attempting to revise their theory accordingly in Ref.~\cite{Lvov2}, the same authors still insist on unimportance of reconnections and support the bottleneck idea. Note, however, that once we start from the grounds that the self-induced motion dominates vortex-line dynamics, reconnections are unavoidable---the lines simply do not interact with each other, and thus there is nothing to stop them from coming into a direct contact and reconnecting. We explicitly demonstrate in Sec.~\ref{sec:crossover} that these reconnections serve as an efficient channel of transferring the energy flux to lower scales thereby ruling out a possibility of the bottleneck.

In a certain loose sense, term `bottleneck' might be associated with the phenomenon of fractalization of vortex lines (cf.~Ref.~\cite{Golov}), since it is essentially an accumulation of energy in some region of the inertial range. We, however, see two reasons---one specific to the context of superfluid turbulence and, most importantly, a generic fundamental one---why doing so could be misleading. The specific reason is that the term has been already used in the context of the interface between hydrodynamic and Kelvin-wave modes, while fractalization of the lines is a purely Kelvin-wave effect, which happens deep (in the $\Lambda \to \infty$ limit) in the quantized regime. The fundamental reason is that the fractalization of the lines does not at all imply the absence of a channel(s) capable to accommodate the energy flux---the fractalization starting at a given wavelength scale does not result in any back action on larger scales, such as the development of a thermalized distribution above the bottleneck, which is a crucial ingredient of bottleneck scenarios. A more relevant term, putting the effect of vortex line fractalization in a broader context of exotic cascade phenomena, might be the {\it deposition} of the cascading constant of motion (e.g., energy in our case) within the cascade inertial range. While re-distributing the cascading quantity in a non-trivial way along the inertial range, the effect of deposition does not  plug the cascade, thus leaving its generic properties intact.


\begin{thebibliography}{99}

%1
\bibitem{Donnelly} R.J. Donnelly, {\it Quantized Vortices in He II}
(Cambridge University Press, Cambridge, 1991).

%2
\bibitem{Cambridge_workshop} C.F. Barenghi, R.J. Donnelly, and W.F. Vinen (eds.), {\it Quantized Vortex
Dynamics and Superfluid Turbulence}, Vol. 571 of Lecture Notes in
Physics, edited by (Springer-Verlag, Berlin, 2001).

%3
\bibitem{Vinen_Niemela} W. F. Vinen, J. J. Niemela, \textit{J. Low Temp. Phys.} \textbf{128}, 167 (2002).

\bibitem{Vinen06} W.F. Vinen, \textit{J. Low. Temp. Phys.} \textbf{145}, 7
(2006).

%%%%%%%%%%%%%%%%%%%%%%%%%%%%%%%%%%%%%%%%%%%%%%%%%%%%%%%%%%%%%%%%%%%%%%%%%%%%%%%%
 \bibitem{Sv95} B.V. Svistunov, \textit{Phys. Rev. B} {\bf 52}, 3647
(1995).

\bibitem{Nore} C. Nore, M. Abid, and M.E. Brachet, \textit{Phys. Rev. Lett.} {\bf 78}, 3896 (1997).

\bibitem{Davis} S.I. Davis, P.C. Henry, P.V.E. McClintock,
\textit{Physica B} {\bf 280}, 43 (2000).

\bibitem{Vinen2000} W.F. Vinen, \textit{Phys. Rev. B} {\bf 61}, 1410 (2000).

\bibitem{Tsubota00} M. Tsubota, T. Araki, and S.K.  Nemirovskii, \textit{Phys. Rev. B} {\bf 62}, 11751 (2000).

\bibitem{Vinen2001} W.F. Vinen, \textit{Phys. Rev. B} {\bf 64}, 134520 (2001).

\bibitem{Kivotides} D. Kivotides, J.C. Vassilicos, D.C. Samuels, C.F. Barenghi,
\textit{Phys. Rev. Lett.} {\bf 86}, 3080 (2001).

\bibitem{Vinen_2003} W.F. Vinen, M. Tsubota and A. Mitani, \textit{Phys.
Rev. Lett.} {\bf 91}, 135301 (2003).

\bibitem{KS_04} E.V. Kozik and B.V. Svistunov, \textit{Phys. Rev. Lett.} {\bf 92}, 035301 (2004).

%5
\bibitem{KS_05} E.V. Kozik and B.V. Svistunov, \textit{Phys. Rev. Lett.} {\bf 94}, 025301 (2005).

%6
\bibitem{KS_05_vortex_phonon} E. Kozik and B. Svistunov, \textit{Phys. Rev. B} {\bf 72}, 172505 (2005).

\bibitem{Bradley} D.I. Bradley, D.O. Clubb, S.N. Fisher, A.M. Gu\'enault, R. P. Haley, C. J. Matthews, G. R. Pickett, V. Tsepelin, and K. Zaki, \textit{Phys. Rev. Lett.} {\bf 96}, 035301 (2006).

\bibitem{sim_boll} R. Hanninen, M. Tsubota, W.F. Vinen, \textit{Phys. Rev. B} \textbf{75}, 064502 (2007).

\bibitem{Ladik} T.V. Chagovets, A.V. Gordeev, and L. Skrbek,
\textit{Phys. Rev. E} \textbf{76}, 027301 (2007).


\bibitem{Golov} P.M. Walmsley, A.I. Golov, H. E. Hall, A. A. Levchenko, and W. F. Vinen, \textit{Phys. Rev. Lett.} {\bf 99}, 265302 (2007).

\bibitem{Lvov} V.S. L'vov,   S.V. Nazarenko, and O. Rudenko,
\textit{Phys. Rev. B} {\bf 76}, 024520 (2007).

\bibitem {KS_crossover} E. Kozik and B. Svistunov, \textit{Phys. Rev. B} {\bf 77}, 060502, (2008).

\bibitem{Lvov2} V.S. L'vov,   S.V. Nazarenko, and O. Rudenko, \textit{J. Low Temp. Phys.} {\bf 153}, 140 (2008).

\bibitem {KS_scan} E. Kozik and B. Svistunov, \textit{Phys. Rev. Lett.} \textbf{100}, 195302 (2008).

\bibitem{Golov2} P.M. Walmsley and A.I. Golov, \textit{Phys. Rev. Lett.} \textbf{100}, 245301 (2008).


\bibitem{Barenghi} S.Z. Alamri, A.J. Youd, and C.F. Barenghi, \textit{Phys. Rev. Lett.} \textbf{101}, 215302 (2008).

\bibitem{vibr_review} M. Blažková, D. Schmoranzer, L. Skrbek, and W. F. Vinen, \textit{Phys. Rev. B} \textbf{79}, 054522 (2009).

%%%%%%%%%%%%%%%%%%%%%%%%%%%%%%%%%%%%%%%%%%%%%%%%%%%%%%%%%%%%%%%%%%%%%%%%%%%%%%%%%%%

%%%%%%%%%%%%% counterflow %%%%%%
%18
\bibitem {Vinen} W. F. Vinen, \textit{Proc. R. Soc. London, Ser. A} \textbf{240}, 114 (1957); \textit{ibid.} \textbf{242}, 493 (1957); \textit{ibid.} \textbf{243}, 400 (1957).
%19
\bibitem{Schwartz} K.W. Schwarz, \textit{Phys. Rev. B} {\bf 31}, 5782 (1985);
\textit{ibid.} {\bf 38}, 2398 (1988).
%%%%%%%%%%%%%%%%%%%%%%%%%%%%%%%%%

\bibitem{Maurer} J. Maurer, P. Tabeling,
\textit{Europhys. Lett.} \textbf{43}, 29 (1998).

%15
\bibitem{Stalp} S.R. Stalp, L. Skrbek, and R.J. Donnelly, \textit{Phys. Rev. Lett.} {\bf 82}, 4831 (1999).

\bibitem{Oregon} S.R. Stalp J.J. Niemela, W.F. Vinen, and R.J. Donnelly,
\textit{Phys. Fluids} \textbf{14}, 1377 (2002).


%20
\bibitem{Berloff} N.G. Berloff and B.V. Svistunov, \textit{Phys. Rev. A} {\bf 66}, 013603 (2002); and references therein.
%21
\bibitem{BEC_expt} C.N. Weiler, T.W. Neely, D.R. Scherer, A.S. Bradley, M.J. Davis, and B.P. Anderson, \textit{Nature} {\bf 455}, 948 (2008).
%22
\bibitem{Kibble-Zurek} T.W.B. Kibble, J. Phys. A {\bf 9}, 1387 (1976); W.H. Zurek, \textit{Nature} {\bf 317}, 505 (1985).

%23
\bibitem{Iordanskii} S.V. Iordanskii,
 \textit{Zh. Eksp. Teor. Fiz.} {\bf 49}, 225 (1965) [\textit{Sov. Phys. JETP} {\bf 22}, 160 (1966)].

%24
 \bibitem{Feynman} R.P. Feynman,  in {\it Progress in Low Temperature Physics}, edited by C.J. Gorter (Nort-Holland, Amsterdam, 1955), Vol. I, p. 17.


%27
\bibitem{Betchov} R. Betchov, \textit{J. Fluid Mech.} {\bf 22}, 471 (1965).

%28
\bibitem{Hasimoto} H. Hasimoto, \textit{J. Fluid Mech.} {\bf 51}, 477 (1972).

%29
\bibitem{Zakharov} V.E. Zakharov, S.V. Manakov, S.P. Novikov, and L.P. Pitaevskii, {\it Theory of Solitons} (Nauka, Moscow, 1980).

%30
\bibitem{Laurie} The authors acknowledge a discussion of this concern with Jason Laurie and Sergey Nazarenko.

\bibitem{Epstein_Baym} R.I. Epstein and G. Baym, Astrophys. J.
{\bf 387}, 276 (1992).


\bibitem{Sv91} B.V. Svistunov, \textit{J. Moscow Phys. Soc.}
{\bf 1}, 373 (1991).

\bibitem{Buttke} T.F. Buttke, \textit{J. of Comp. Phys.} {\bf 76}, 301 (1988).


\bibitem{Samuels_Donnelly} D.C. Samuels and R.J. Donnelly, \textit{Phys. Rev. Lett.} \textbf{65}, 187 (1990).

\bibitem{Fetter_scattering} A.L. Fetter, \textit{Phys. Rev.} {\bf 186}, 128
(1969), and references therein.

\bibitem{Bretin} V. Bretin, P. Rosenbush, F. Chevy,
G.V. Shlyapnikov, and J. Dalibard, \textit{Phys. Rev. Lett.} {\bf 90},
100403 (2003).

\bibitem{Mizushima} T. Mizushima, M. Ichioka, and K. Machida,
\textit{Phys. Rev. Lett.} {\bf 90}, 180401 (2003).

\bibitem{Stoof} J.-P. Martikainen and H.T.C. Stoof, \textit{Phys. Rev. A}
{\bf 69}, 053617 (2004).

\bibitem{LLStatMech2} E.M. Lifshitz and L.P. Pitaevskii,
\textit{Statistical Mechanics}, Part 2, Pergamon Press, New York,
1980.

\bibitem{Popov} V.N. Popov, \textit{Functional Integrals in Quantum Field The-
ory and Statistical Physics}, Reidel, Dordrecht, 1983.

\bibitem{LLHydrodynamics} E.M. Lifshitz and L.P. Pitaevskii,
\textit{Fluid Mechanics}, Pergamon Press, New York, 1987.

\bibitem{soliton1} G. Reinisch and J.C. Fernandez, Phys. Rev. B {\bf
25}, 7352 (1982); Kazumi Maki, \textit{Phys. Rev. B} {\bf 26}, 2181 (1982).

\bibitem{Sonin} E.B. Sonin, \textit{Phys. Rev. B} \textbf{55}, 485 (1997).

\bibitem{Pitaevskii} L.P. Pitaevski, \textit{Zh. Eksp.
Teor. Fiz.} {\bf 35}, 1271 (1958) [\textit{Sov. Phys. JETP} {\bf 8}, 888
(1959)].

\bibitem{Demircan} E. Demircan, P. Ao, and Q. Niu, \textit{Phys. Rev. B} \textbf{52},
476 (1995).

\bibitem{Hall_Vinen} H.E. Hall and W.F. Vinen, \textit{Proc. Roy. Soc. Ser. A}, \textbf{238}, 204 (1956); \textit{ibid} \textbf{238}, 215
(1956).

\bibitem{comment} In the revised version of their manuscript, the authors of Ref.~\cite{Lvov} admit potential importance
of the self-induced motion, referring to the preprint of our Ref.~\cite{KS_crossover}.


\bibitem{KT_our} E. Kozik, N. Prokof'ev, and B. Svistunov,
%\textit{Vortex-phonon interaction in the Kosterlitz-Thouless theory},
\textit{Phys. Rev. B} \textbf{73}, 092501 (2006).

\end{thebibliography}
\end{document}